# Signal Design Impacts on GNSS and LEO-PNT L1 Band Acquisition Complexity: A Case Study with Xona Pulsar

Jérôme Leclère, Rajnesh Joshi, Thyagaraja Marathe, Tyler G. R. Reid

*Xona Space Systems*

**Biographies**

**Dr. Jérôme Leclère** is a GNSS FPGA Designer at Xona in Montreal, Canada. He holds an engineering degree in Electronics and Signal Processing from ENSEEIHT (France, 2008) and a Ph.D. in GNSS from EPFL (Switzerland, 2014). Dr. Leclère has worked on hardware-based GNSS receivers and GNSS acquisition, as well as on signal processing algorithms for lidar and satellite phased antenna arrays. Currently, his work focuses on Pulsar signals and receiver development.

**Rajnesh Joshi** is a Firmware Engineer at Xona in Montreal, Canada. He holds an engineering degree in Computer Engineering (2024) from Simon Fraser University in British Columbia, and has had previous internship experiences in embedded and FPGA design in both industry and research including at a telecommunications, and a biomedical engineering firm, as well as a reconfigurable computing research lab.

**Dr. Thyagaraja Marathe** is a Principal Engineer at Xona, specializing in PNT using LEO satellite constellations. He previously worked at Rx Networks and Accord Software on GNSS corrections and receiver development. He holds a Ph.D. in Geomatics Engineering from the University of Calgary, where he also completed postdoctoral research. His interests include LEO PNT, high-accuracy positioning, and GNSS receiver design. He also holds degrees from BITS Pilani and VTU, India.

**Dr. Tyler G. R. Reid** is a co-founder and CTO of Xona whose focus is commercial satellite navigation services from Low Earth Orbit. Tyler previously worked as a Research Engineer at Ford Motor Company in localization and mapping for self-driving cars. He was also a Software Engineer at Google and a lecturer at Stanford University where he co-taught the course on GPS. Tyler received his Ph.D. ('17) and M.Sc. ('12) in Aeronautics and Astronautics from Stanford where he worked in the GPS Research Lab and his B.Eng in Mechanical Engineering from McGill ('10). He is a recipient of the RTCA Jackson Award.

**Abstract**

GNSS signal acquisition is the process of detecting and estimating signal parameters by exploiting correlation operations that provide processing gain and enable the detection of signals below the noise floor. The design of an acquisition engine requires balancing achievable sensitivity, computational complexity, memory requirements, power consumption, and implementation complexity. In particular, increasing coherent and non-coherent integration times improves sensitivity but can significantly increase receiver processing requirements, limiting its applicability in low-power and resource-constrained receivers.

This paper presents a comparative analysis of L1-band signal acquisition complexity by evaluating the required integration time and corresponding implementation requirements for different signal designs under equivalent acquisition performance targets. The analysis considers the impact of signal characteristics, including spreading codes, secondary codes, modulation, and signal bandwidth, on coherent and non-coherent acquisition performance. For a given attenuation level and target probability of false alarm and detection, the required integration times are derived and translated into acquisition complexity metrics, including arithmetic operations, processing time, and memory requirements.

The analysis is applied to Pulsar X1, a LEO-enabled PNT signal, and representative GNSS L1 signals to determine under which attenuation, assistance, and integration conditions the higher received power of LEO-PNT translates into reduced receiver acquisition workload. By relating the required integration to arithmetic operations, processing time, and memory requirements, the study provides a receiver-level comparison relevant to both conventional and snapshot-based receivers, where reducing acquisition processing can directly contribute to lower energy consumption.

Published at ION GNSS+ 2026, Orlando, FL, USA, Sept. 14–18, 2026. DOI:10.33012/2026.20902

## 1. INTRODUCTION

Global Navigation Satellite System (GNSS) receivers rely on signal acquisition to detect visible satellites and obtain coarse estimates of their code delay and Doppler frequency before tracking the signal (Ward, 2017). Since acquisition requires searching a large uncertainty space while detecting signals that are often well below the thermal noise floor, it is one of the most computationally demanding stages of a receiver. High-sensitivity receivers therefore rely on coherent and non-coherent integration to provide sufficient processing gain for reliable detection under weak signal conditions (van Diggelen, 2021). The corresponding integration time directly impacts receiver complexity, processing time, memory requirements, and energy consumption.

These considerations have become increasingly important with the emergence of low-power GNSS receivers and snapshot-based positioning techniques, where only a short segment of the received signal is collected before acquisition and navigation processing are performed locally or remotely (van Diggelen, 2009; van Dierendonck & Al-Fanek, 2018; Beuchert & Rogers, 2021). In such applications, minimizing the amount of processing required during acquisition is essential to reduce power consumption to meet stringent constraints, and improve overall system efficiency. Similar considerations have motivated the development of high-sensitivity assisted GNSS receivers, where acquisition performance plays a dominant role in receiver operation.

Although all GNSS receivers rely on the same fundamental acquisition principles, the required processing effort depends strongly on the characteristics of the transmitted signal. Received signal power, spreading code length, primary code period, pilot/data power allocation, secondary codes, modulation, occupied bandwidth, and acquisition search space all influence the amount of processing gain that must be generated by the receiver to satisfy a given probability of detection and probability of false alarm (van Diggelen, 2021). At the same time, these characteristics also affect the practical implementation of the acquisition engine through their impact on arithmetic complexity, memory requirements, and processing time.

Previous work has extensively investigated acquisition algorithms for weak-signal GNSS receivers (Psiaki, 2001; Ziedan, 2006), focusing on improving acquisition sensitivity (van Diggelen, 2009, 2021), or reducing computational complexity (Akopian, 2005; Foucras et al., 2013; Leclère et al., 2015, 2018a; Zhou et al., 2022). More recently, energy-efficient acquisition techniques have also attracted significant attention in the context of low-power and IoT GNSS receivers (Morales-Ferre et al., 2021). A receiver-complexity-based comparison methodology was previously applied to GPS L1 and L5 acquisition (Leclère et al., 2018b). The present work significantly extends this approach by considering six representative L1-band GNSS and LEO-PNT signals, incorporating signal-specific implementation losses and search-space characteristics, and evaluating both acquisition sensitivity and implementation complexity under identical acquisition performance requirements.

Extending the coherent integration time is a well-established approach for improving acquisition sensitivity under weak signal conditions. However, the associated implementation cost is not solely determined by the integration duration but also by the signal structure. Signals with different primary code periods, spreading code lengths, secondary codes, or modulation schemes require different acquisition architectures and therefore different arithmetic complexity, memory resources, and processing time to achieve the same integration time. For example, the legacy GPS L1 C/A signal employs a short 1 ms primary code that repeats within a 20 ms navigation data bit, allowing coherent integration to be extended over several code periods with relatively limited implementation overhead. In contrast, most modern GNSS signals employ longer primary code periods together with secondary (overlay) codes, such that extending coherent integration generally requires longer coherent processing intervals and additional signal processing. Consequently, identical integration times do not necessarily imply identical receiver implementation complexity.

This paper proposes a unified methodology to compare the acquisition complexity of representative L1-band navigation signals. To ensure a fair comparison, all signals are evaluated under identical acquisition performance requirements, defined by a common probability of detection and probability of false alarm. A common acquisition sensitivity framework is first established by fixing the target probabilities of detection and false alarm and determining the integration time required for each signal under different attenuation conditions. The analysis explicitly accounts for signal-specific characteristics, including received signal power, effective carrier-to-noise density ratio, pilot/data power allocation, modulation losses, implementation losses, and acquisition search space. The resulting integration times are subsequently used to evaluate representative receiver implementations in terms of arithmetic complexity, memory requirements, and processing time.

The proposed methodology is applied to Pulsar X1, GPS L1 C/A, GPS L1C, Galileo E1, BeiDou B1I, and BeiDou B1C. Beyond comparing acquisition sensitivity, the paper quantifies how differences in signal design translate into practical receiver implementation cost. The results provide insight into the processing and potential energy savings offered by different signal designs for both conventional and snapshot-based GNSS receivers.

## 2. SIGNAL CHARACTERISTICS IMPACTING ACQUISITION COMPLEXITY

GNSS signal acquisition consists of detecting the presence of a navigation signal and estimating its code delay and Doppler frequency by searching for the maximum of the cross-correlation between the received signal and a locally generated replica. Although all GNSS acquisition algorithms rely on the same correlation principle, the characteristics of the transmitted signal largely determine the integration strategy required to achieve a given acquisition performance and, consequently, the complexity of the receiver implementation.

Several signal properties directly influence acquisition performance. The received signal power determines the achievable signal-to-noise ratio (SNR) for a given integration time, while the spreading code structure, secondary codes, modulation, and signal bandwidth constrain the integration strategy and affect the computational and memory requirements of the acquisition engine. This section reviews the signal characteristics that are relevant to the comparative analysis presented in this paper.

### 2.1. Received signal power and $C/N_0$

The received carrier-to-noise density ratio ($C/N_0$) is the primary metric used to characterize the strength of a navigation signal at the receiver and is the main driver of acquisition sensitivity. For a given acquisition algorithm and target probability of detection and false alarm, a higher effective $C/N_0$ reduces the integration time required to detect the signal.

The effective $C/N_0$ available for acquisition depends on several factors beyond the transmitted signal power. The received carrier power is determined by the transmitted power, antenna gains, propagation losses, and receiver characteristics. Additional implementation-dependent losses may also occur depending on how the receiver processes the signal.

Many modern GNSS signals separate their transmitted power between pilot and data components. Since acquisition is often performed using only the pilot component, or only the data component, only a fraction of the total transmitted signal power contributes to the correlation process, resulting in a reduced effective $C/N_0$. Likewise, receivers may intentionally simplify the acquisition processing by using an approximate local signal replica instead of the nominal modulation. For example, Galileo E1 may be acquired using a BPSK approximation rather than a full BOC(1,1) local replica, reducing implementation complexity at the expense of correlation loss. These implementation losses are incorporated into the analysis presented in this paper.

Unlike traditional MEO GNSS constellations, LEO-PNT systems benefit from a reduced free-space path loss, resulting in a higher received signal power at the user receiver. However, this increased received power is partially offset by the aggregate interference generated by the LEO constellation itself, which contributes to the effective noise density experienced by the receiver (Betz, 2015). This effect is not limited to high-power LEO constellations, as MEO constellations are also affected, although to a lesser extent (Paonni et al., 2026). Because the desired RNSS signals and aggregate RNSS interference are attenuated together while the receiver thermal noise remains unchanged, the relative contribution of aggregate interference decreases with attenuation. It therefore becomes less significant under attenuated-signal conditions.

As a result, the effective $C/N_0$ improvement depends not only on the received signal power but also on the aggregate interference assumptions adopted for the constellation. In this work, the received signal powers and effective noise densities are determined for both LEO and MEO constellations and derived from the compatibility analyses presented in (International Telecommunication Union, 2015), (Reid et al., 2025).

To provide a consistent comparison across all signals, this paper considers the effective $C/N_0$ available after accounting for signal power allocation, receiver implementation losses, and the background noise assumptions. Several attenuation levels are then applied to determine the integration time required to achieve identical acquisition performance for all signals.

### 2.2. Primary spreading code characteristics

The primary spreading code defines the periodic sequence used to spread the navigation signal and forms the basis of the acquisition process. Its main characteristics include the chip rate, code length, and code period, which together determine the duration of one coherent correlation interval.

The code period directly influences the acquisition strategy. Typically, receivers perform coherent integration over one or more complete code periods before combining successive results through non-coherent accumulation. Consequently, the primary code period establishes the minimum coherent integration interval naturally supported by the signal.

The code length and chip rate also influence the implementation complexity of the acquisition engine. Longer codes require larger local code replicas and larger correlation buffers, while higher chip rates increase the sampling rate required to preserve correlation performance. These characteristics affect the number of arithmetic operations, processing time, and memory requirements that will be evaluated in Section 4.

### 2.3. Secondary codes and navigation data

Although longer coherent integration generally improves acquisition sensitivity, it is limited by unknown sign transitions present in the transmitted signal. These transitions originate either from the navigation data or from secondary (overlay) spreading codes applied to the primary code.

When an unknown transition occurs during coherent integration, the accumulated signal energy may be partially or completely cancelled, reducing the acquisition performance. To avoid this degradation, coherent integration is typically limited to intervals over which the transmitted signal remains constant, or techniques such as multiple hypotheses or secondary code correlation must be evaluated to compensate for the unknown transitions of the secondary code.

The signals considered in this work differ significantly in this respect. GPS L1 C/A transmits navigation data at 50 bit/s, allowing coherent integration over multiple primary code periods before a potential data bit transition occurs. In contrast, modern GNSS signals generally employ secondary codes whose chip duration is equal to the primary code period. As a result, coherent integration beyond a single primary code period requires accounting for the secondary code sequence.

To isolate the impact of signal design on acquisition complexity, this paper considers three representative integration strategies. The first limits coherent integration to a single primary code period and achieves additional sensitivity through non-coherent accumulation of successive correlation magnitudes. The second performs coherent integration over the entire required integration interval, assuming that any data or secondary-code transitions are properly handled. The third combines both approaches by extending coherent integration beyond one primary code period and then applying non-coherent integration, which we call hybrid integration. These three strategies represent common implementation approaches and provide a consistent basis for comparing different signal designs.

### 2.4. Modulation and signal bandwidth

The modulation and occupied bandwidth of a navigation signal influence both the received signal power distribution and the correlation properties observed by the receiver. Different modulation schemes, such as BPSK, BOC, CBOC, and EFQPSK, produce different autocorrelation functions and distribute the signal energy differently across the available spectrum.

The receiver front-end bandwidth also determines the amount of signal energy and thermal noise captured during acquisition. Depending on the modulation, part of the transmitted signal energy may fall outside the receiver bandwidth, reducing the effective received $C/N_0$ available for acquisition.

These effects are incorporated into the link budget and acquisition sensitivity analysis presented in Section 3. In particular, the impact of the EFQPSK modulation employed by Pulsar X1 is explicitly evaluated to quantify the corresponding acquisition performance.

### 2.5. Summary of analyzed signals

**TABLE 2.1** summarizes the characteristics of the L1-band signals considered throughout this work, following (Xona, 2026; U.S. Department of Defense, 2022a, 2022b; European Union, 2025; China Satellite Navigation Office, 2017, 2019). The selected parameters correspond to those that directly influence acquisition performance and implementation complexity.

Received signal power determines the integration time required to achieve a target acquisition performance, while the spreading code, secondary code, modulation, and signal bandwidth influence the receiver architecture needed to realize that integration. These characteristics form the basis of the comparative analysis developed in the following sections, where each signal is evaluated under identical acquisition performance requirements to quantify the resulting implementation complexity.

Note that since the target is low-complexity receivers, we consider only a single lobe of the BOC-modulated signals, processed as BPSK, also called single side band (SSB) processing (otherwise the sampling rate would increase).

**TABLE 2.1** Relevant parameters for signals analyzed

| Signal | Pulsar X1 | GPS L1 C/A | GPS L1C (SSB) | Galileo E1 (SSB) | BeiDou B1I | BeiDou B1C (SSB) |
|---|---|---|---|---|---|---|
| Minimum Received Power (dBW) | −148.2 | −158.5 | −158.25 | −157.25 | −163 | −159 |
| Double-sided Bandwidth[1] (MHz) | 1.77 | 2.046 | 2.046 | 4.092 | 4.092 | 2.046 |

| Signal | Pulsar X1 | GPS L1 C/A | GPS L1C (SSB) | Galileo E1 (SSB) | BeiDou B1I | BeiDou B1C (SSB) |
|---|---|---|---|---|---|---|
| Primary Code Chip Rate (Mchip/s) | 1.023 | 1.023 | 1.023 | 1.023 | 2.046 | 1.023 |
| Primary Code Chip Length (chip) | 1023 | 1023 | 10230 | 4092 | 2046 | 10230 |
| Primary Code Duration (ms) | 1 | 1 | 10 | 4 | 1 | 10 |
| Secondary Code Chip Rate (chip/s) | 1000 | N/A | 100 | 250 | 1000 | 100 |
| Secondary Code Chip Length (chip) | 100 | N/A | 1800 | 25 | 20 | 1800 |
| Secondary Code Chip Length (ms) | 100 | N/A | 18000 | 100 | 20 | 18000 |
| Navigation Frame Data Rate (bit/s) | 1000 | 50 | 100 | 250 | 50 | 100 |
| Doppler Frequency Search Space (kHz)[2] | ±33.6 | ±4.47 | ±4.47 | ±3.52 | ±3.75 | ±3.75 |
| Data and Pilot | Yes, both. | No, only data. | Yes, both. | Yes, both. | No, only data. | Yes, both. |

[1] Null-to-null bandwidth for all signals except Pulsar where it is the bandwidth containing 99.5 % of the signal power (Xona, 2026)
[2] Doppler for actual constellations (van Diggelen, 2021) except Pulsar where it is the nominal constellation (Xona, 2026)

The acquisition complexity of each signal results from the combination of the factors summarized in **TABLE 2.2** rather than from any individual characteristic. Some factors, such as a higher received signal power, reduce the amount of processing gain required for successful acquisition, whereas others, including pilot/data power sharing, modulation approximation losses, longer primary code periods, or a larger acquisition search space, increase the receiver processing requirements. The methodology presented in the following section combines these effects into a common acquisition sensitivity analysis, allowing all signals to be compared under identical acquisition performance requirements.

**TABLE 2.2** Summary of the dominant factors affecting acquisition complexity for signals analyzed

| Signal | Link budget factors | Signal structure factors | Acquisition search factors |
|---|---|---|---|
| Pulsar X1 | ++ Highest received power<br>− Higher self-interference (issue only at low attenuation)<br>− Pilot/data power split | + Short 1 ms code<br>− Watermark & modulation loss | ○ Moderate search space (due to 8× larger Doppler) |
| GPS L1 C/A | + No pilot/data power split | + Short 1 ms code<br>+ 20 PRN codes per data bit | + Smaller search space<br>+ Easier to extend coherent integration up to a certain amount<br>− Long coherent integration difficult with data bits (no pilot channel) |
| GPS L1C (SSB) | − Pilot/data power split<br>− BOC/BPSK loss | − Long 10 ms primary code | ○ Moderate search space (due to 10× longer code) |
| Galileo E1 (SSB) | − Pilot/data power split<br>− BOC/BPSK loss | − 4 ms primary code | ○ Moderate search space (due to 4× longer code) |

| Signal | Link budget factors | Signal structure factors | Acquisition search factors |
|---|---|---|---|
| BeiDou B1I | + No pilot/data power split<br>− Low power | + Short 1 ms primary code<br>− Higher chip rate<br>− Overlay code and data on same channel | + Small to moderate search space (due to 2× longer code) |
| BeiDou B1C (SSB) | − Pilot/data power split<br>− BOC/BPSK loss | − Long 10 ms primary code | ○ Moderate search space (due to 10× longer code) |

## 3. ACQUISITION SENSITIVITY ANALYSIS

The objective of this section is to determine the integration time required by each signal to achieve identical acquisition performance under different attenuations. Rather than comparing signals based solely on their nominal received power, a common acquisition performance model is adopted so that all signals are evaluated using the same probability of detection and probability of false alarm requirements.

The analysis follows the same general methodology as the link budget approach presented in (van Diggelen, 2009, 2021) for GNSS receiver sensitivity analysis. However, it is extended to explicitly account for signal-specific characteristics that affect acquisition performance, including pilot/data power allocation, secondary codes, modulation losses, receiver implementation assumptions, and the effective noise density associated with LEO-PNT compatibility. The resulting integration times are subsequently used in Section 4 to evaluate receiver implementation complexity.

### 3.1. Methodology

The objective of the proposed methodology is to determine the amount of integration required for each signal to satisfy identical acquisition performance requirements. This is achieved by comparing the SNR available after the minimum coherent integration interval with the SNR required to satisfy the target probabilities of detection and false alarm. The difference between these two quantities determines the additional coherent or non-coherent integration required for successful acquisition.

To ensure a fair comparison, the same acquisition performance criteria and receiver assumptions are applied to every analyzed signal. Consequently, differences in the required integration time reflect the signal-specific characteristics and search-space assumptions, rather than differences in the target acquisition performance.

An example of a complete worksheet for both coherent and non-coherent integration is provided in Appendix A.

#### *3.1.1. Acquisition performance requirements*

##### *3.1.1.1. Detection Probability and False Alarm Probability*

Signal acquisition is formulated as a binary hypothesis test in which each acquisition cell corresponds to one code delay and Doppler hypothesis. The acquisition threshold is selected to satisfy a target probability of false alarm while ensuring a specified probability of detection for the correct acquisition cell. These two metrics define the acquisition performance requirements and directly determine the integration time required to detect a signal.

In this work, the acquisition performance targets are selected to reflect the requirements of low-power GNSS receivers, including snapshot-based receivers, where only a limited amount of signal is collected before the acquisition results are transmitted to a cloud processor for position computation. In this operating mode, the detected correlation peak may be used directly to estimate the pseudorange before frame synchronization has been achieved. As discussed by van Diggelen (2009, Section 6.8.4.2), a false acquisition can therefore result in a pseudorange error approaching one full code period (approximately 300 km for a 1 ms code). However, van Diggelen also notes that very stringent false-alarm requirements are mainly necessary when no subsequent means of validating the measurements is available. In a snapshot architecture, the cloud processor can exploit measurements from multiple satellites, together with available constellation or assistance information, to perform consistency checks and reject erroneous measurements. Consequently, occasional isolated false acquisitions can be tolerated, allowing a less stringent false-alarm requirement and therefore improved acquisition sensitivity.

For this reason, a global probability of false alarm $P_{\mathrm{FA,G}} = 10^{-2}$ is adopted throughout this work, where $P_{\mathrm{FA,G}}$ denotes the probability of at least one false alarm during the complete acquisition search for one satellite. This value is selected as a representative compromise rather than as a strict system-level requirement. For example, assuming 12 independent satellite searches with $P_{\mathrm{FA,G}} = 10^{-2}$, the probability of obtaining at least one false acquisition is approximately 11.4 %, while the probabilities of obtaining two or more and three or more false acquisitions are only approximately 0.62 % and 0.021 %, respectively. Thus, false acquisitions may occasionally occur, as expected in an architecture relying on subsequent measurement validation, while multiple simultaneous false acquisitions remain unlikely. Since acquisition evaluates a large number of code-delay and Doppler hypotheses, $P_{\mathrm{FA,G}}$ is translated into the corresponding probability of false alarm for an individual acquisition cell using the total number of searched cells. **Figure 3.1** illustrates this relationship, and Section 3.1.1.3 provides the corresponding expression.

The target cell probability of detection $P_{\mathrm{D,C}}$ is set to 0.85. Although this value corresponds to the successful acquisition of a single satellite, it results in a very high probability of acquiring a sufficient number of satellites for positioning. As shown by (van Diggelen, 2009), assuming ten satellites are visible, a per-satellite detection probability of 0.85 yields a probability greater than 0.9999 of detecting at least four satellites, which is sufficient to compute a navigation solution. This value therefore represents a practical compromise between acquisition sensitivity and receiver complexity.

Once the cell probability of false alarm has been determined, the corresponding detection threshold is obtained from the statistical distribution of the acquisition metric under the noise-only hypothesis (discussed in the next section). The minimum signal-to-noise ratio required to achieve the target probability of detection is then obtained from the statistical distribution of the acquisition metric under the signal-present hypothesis. Finally this will drive the number of non-coherent or coherent accumulations required. These steps are illustrated in **Figure 3.1** and **Figure 3.2**, and constitute the first stage of the acquisition sensitivity analysis. Note that the coherent accumulations impact the search space, therefore, several iterations may be required to reach the desired constraints.

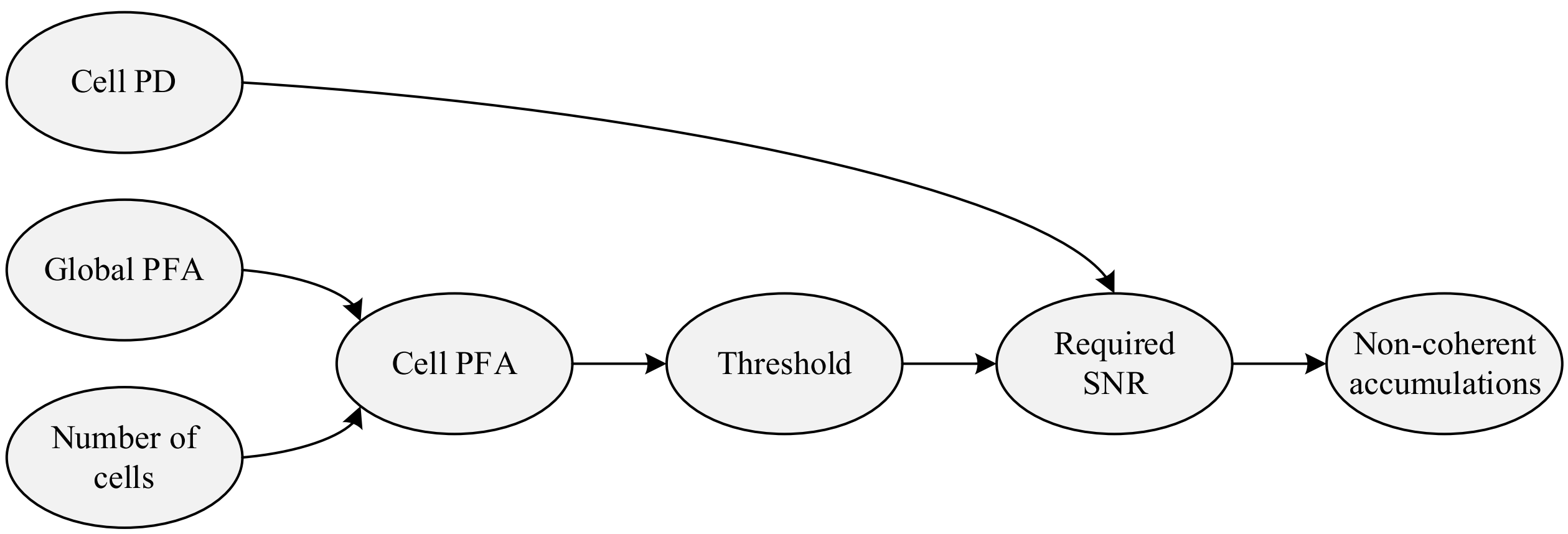

**Figure 3.1** Relation between the different parameters for non-coherent integration only

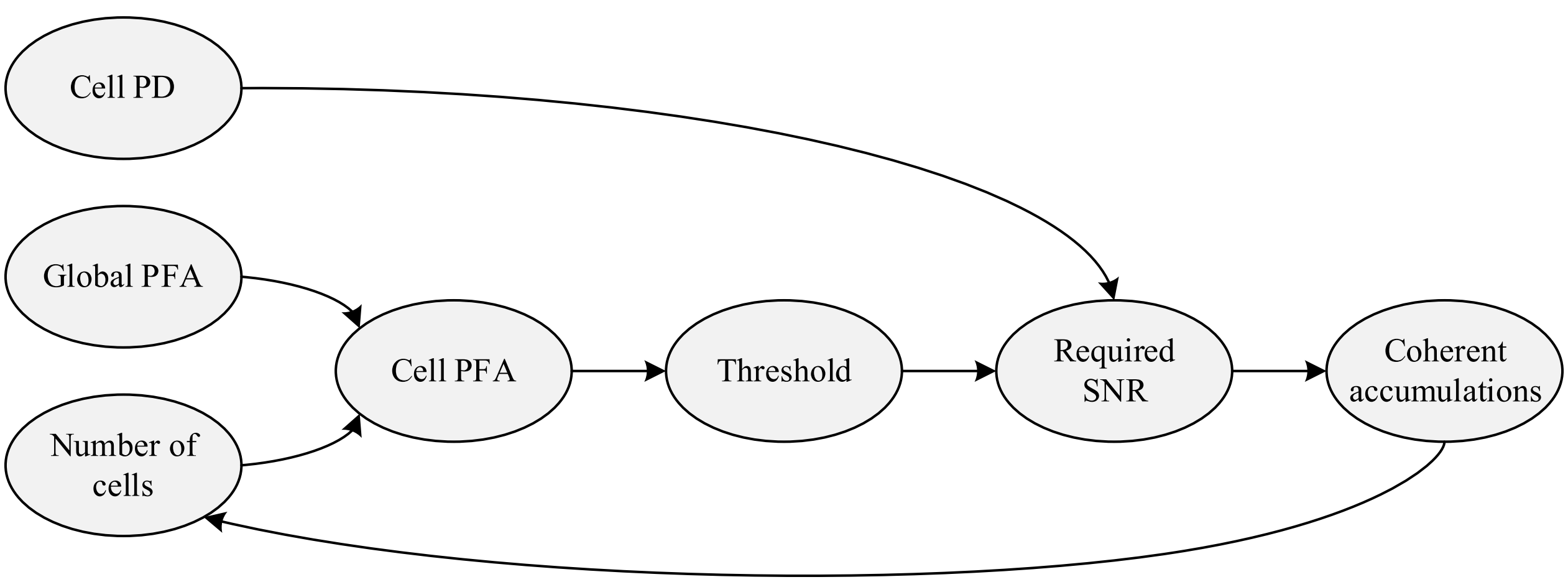

**Figure 3.2** Relation between the different parameters for coherent integration only

*3.1.1.2. Connecting SNR to Probabilities*

The relation between the SNR and PFA and PD depends on the integration strategy.

For coherent integration, the in-phase and quadrature correlator outputs follow Gaussian distributions whose mean is determined by the received signal power and whose variance depends on the receiver noise. The magnitude of the complex-valued correlation follows a Rice distribution in the presence of a signal and a Rayleigh distribution in the absence of a signal (van Diggelen, 2009). The required post-correlation SNR is therefore obtained from the corresponding probability of detection and probability of false alarm expressions. The probability of false alarm is given by $\mathrm{e}^{-T^2/2}$ and the probability of detection is given by $Q_1\left(\sqrt{2SNR}, T\right)$, where $T$ is the threshold and $Q_1$ is the first-order Marcum Q-function.

For non-coherent integration, successive coherent correlation magnitudes are accumulated. The resulting acquisition statistic follows a non-central chi-square distribution under the signal-present hypothesis and a central chi-square distribution under the noise-only hypothesis. However, in this analysis it is approximated as Gaussian using the central limit theorem. This approximation simplifies the worksheet-based comparison and provides a consistent analytical framework across all signals. Since some cases involve a small number of non-coherent accumulations, the approximation may introduce a residual error; a direct central/non-central chi-square treatment will be considered in future refinements.

*3.1.1.3. Impact of search space*

The global PFA is related to the cell PFA as

$$P_{\mathrm{FA,G}} = 1 - \left(1 - P_{\mathrm{FA,C}}\right)^{N_{\mathrm{CB}} N_{\mathrm{FB}}}, \tag{1}$$

where $N_{\mathrm{CB}}$ is the number of code bins, and $N_{\mathrm{FB}}$ is the number of frequency bins (van Diggelen, 2009, Section 6.8.4.1).

The number of code bins depends on the number of chips in the PRN code and on the number of samples per chip (assume equal to 2 in this analysis for all the signals). Therefore, codes with longer PRN codes will need a lower cell PFA for a same global PFA.

The number of frequency bins depends on the frequency search space $\Delta f$ and on frequency step $\delta f$ as $N_{\mathrm{FB}} = \Delta f / \delta f$. The frequency step $\delta f$ is usually inversely proportional to the coherent integration time $T_{\mathrm{C}}$. Common values are $1/(2T_{\mathrm{C}})$ or $2/(3T_C)$. The larger the step, the smaller the number of frequency bins, but the higher the implementation loss (which may itself increase the required $T_{\mathrm{C}}$, therefore, it is a trade-off); in this study, we have considered a step of $1/(2T_{\mathrm{C}})$, which leads to an average loss of 0.30 dB (van Diggelen, 2009).

As for the frequency search space, it includes the Doppler offset due to the satellite's movements, the Doppler induced by the oscillator offset, and to a lesser extent to the Doppler offset due to the user movements. For this parameter, it makes sense to differentiate the cases with or without assistance, because the values change drastically.

Without assistance: the actual maximum Doppler can be found in (van Diggelen, 2021; Table 17.1); and a typical TCXO offset is 3 ppm (van Diggelen, 2021; Section 17.2.1), which converts to about 4700 Hz. **TABLE 3.1** summarizes the values considered for all the signals analyzed.

**TABLE 3.1** Frequency search space without assistance

| Signal | Pulsar X1 | GPS L1 C/A | GPS L1C | Galileo E1 | BeiDou B1I | BeiDou B1C |
|---|---|---|---|---|---|---|
| Maximum oscillator offset (Hz) | 4700 | | | | | |
| Maximum satellite Doppler (Hz) | 33600 | 4470 | 4470 | 3520 | 3750 | 3750 |
| Maximum frequency offset (Hz) | 38300 | 9170 | 9170 | 8220 | 8450 | 8450 |
| Considered search space $\Delta f$ (Hz) | 76600 | 18340 | 18340 | 16440 | 16900 | 16900 |

With assistance, the actual Doppler residual depends on the knowledge of time and position. The offset due to the time directly comes from the Doppler rate, whereas the offset due to the position is proportional to the ratio $\frac{\mathrm{speed}}{\mathrm{range}}$(van Diggelen, 2021; Section 17.6). For this analysis, we assumed a 1 s time offset and a 50 km position uncertainty. As for the oscillator, which can be estimated using assistance, a typical residual is 100 ppb (van Diggelen, 2021), which leads to an offset of 160 Hz. **TABLE 3.2** summarizes the values considered for all the signals analyzed.

**TABLE 3.2** Frequency search space with assistance

| Signal | Pulsar X1 | GPS L1 C/A | GPS L1C | Galileo E1 | BeiDou B1I | BeiDou B1C |
|---|---|---|---|---|---|---|
| Oscillator offset residual (Hz) | 160 | | | | | |
| Time offset (s) | 1 | | | | | |
| Doppler vs time error (Hz/s) | 230 | 0.84 | 0.84 | 0.58 | 0.62 | 0.62 |
| Doppler due to time offset (Hz) | 230 | 0.84 | 0.84 | 0.58 | 0.62 | 0.62 |
| Position offset (km) | 50 | | | | | |
| Doppler vs position error (Hz/km) | 6.85 | 0.88 | 0.88 | 0.7 | 0.76 | 0.76 |
| Doppler due to position offset (Hz) | 342.59 | 44.00 | 44.00 | 35.00 | 38.00 | 38.00 |
| Maximum frequency offset (Hz) | 733 | 205 | 205 | 196 | 199 | 199 |

The search-space model considered in this work includes code delay and Doppler frequency but does not introduce Doppler rate as an additional acquisition dimension. This approximation is negligible for the analyzed MEO signals because of their comparatively small Doppler rates, and it remains appropriate for short coherent integration intervals of X1. For longer X1 coherent integrations, however, the higher LEO dynamics may require Doppler-rate compensation and/or associated loss in the link budget. In an assisted receiver, the Doppler rate can be predicted from the same orbital, time, and approximate user-state information used to predict the Doppler frequency, substantially reducing the residual uncertainty (Choi et al., 2026). Even without assistance, Doppler frequency and Doppler rate are jointly constrained by the satellite-user geometry (Leclère et al., 2025) and therefore need not be searched as two completely independent dimensions. The present complexity analysis does not include additional Doppler-rate hypotheses and should therefore be regarded as increasingly optimistic for the longest coherent X1 integration intervals.

#### *3.1.2. Sensitivity analysis methodology*

Starting from the nominal received signal strength, the effective $C/N_0$ available to the receiver is determined by accounting for the receiver front-end characteristics and signal-specific corrections. The corresponding SNR after the minimum coherent integration interval is then computed, after which additional coherent or non-coherent accumulation is applied until the acquisition performance requirements defined in Section 3.1.1 are satisfied.

##### *3.1.2.1. Received signal power*

The analysis begins with the nominal received signal strength specified for each navigation signal. Those are listed in **TABLE 2.1**. Additional attenuation is then introduced to emulate degraded propagation conditions and to evaluate acquisition performance over a range of received signal levels.

Several signal-dependent losses are subsequently applied to determine the effective received power available to the acquisition process. Modern GNSS signals often divide the transmitted power between pilot and data components. Since low-complexity acquisition is generally performed using the pilot channel only, the available received power is reduced accordingly. Likewise, receivers may intentionally simplify the acquisition process by processing only part of the transmitted signal spectrum. Low-power receivers may want to process wideband BOC-modulated signals using a single main lobe to reduce front-end filter bandwidth and sampling rate requirements.

Adopting this practical constraint, this study considers acquisition of GPS L1C, Galileo E1, and BeiDou B1C signals using SSB processing. This trade-off drastically lowers implementation complexity at the expense of a manageable power loss, which is taken into consideration. **TABLE 3.3** outlines the link budget that shows the final signal strength received considered for the acquisition sensitivity analysis.

**TABLE 3.3** Signal strength and losses considered before front end

| Signal | Pulsar X1 | GPS L1 C/A | GPS L1C (SSB) | Galileo E1 (SSB) | BeiDou B1I | BeiDou B1C (SSB) |
|---|---|---|---|---|---|---|
| Nominal Signal Strength (dBm) | −118.2 | −128.5 | −128.25 | −127.25 | −133 | −129 |
| Data/Pilot Channel Splitting Loss (dB) | −3.01 | 0 | −1.25 | −3.01 | 0 | −1.25 |
| BOC/BPSK Loss (dB) | 0 | 0 | −3.57 | −3.42 | 0 | −3.57 |
| Received Signal strength (dBm) | −121.21 | −128.50 | -133.07 | −133.68 | −133.00 | −133.82 |

*3.1.2.2. Effective noise power spectral density*

Following the receiver sensitivity methodology described by van Diggelen (2009; 2021), the receiver front end is characterized by its effective noise temperature and equivalent noise bandwidth. An effective temperature of 290 K is considered, corresponding to a thermal noise power spectral density of $N_0 = kT_{\mathrm{eff}} - 204\ \mathrm{dBW/Hz}$.

The ratio between the front-end bandwidth and the signal chip rate is chosen to be the same as in van Diggelen (2009; 2021). This assumption is retained in the subsequent link-budget calculations, since the front-end bandwidth determines the integrated noise power.

In addition to thermal noise, the acquisition process is affected by multiple-access interference generated by other RNSS signals. As discussed by Betz (2015), the aggregate interference from navigation signals can be represented as an equivalent contribution to the noise power spectral density through the spectral separation between the desired and interfering signals. Although this effect is particularly significant for high-power LEO-PNT systems, it is also present for conventional MEO GNSS (Paonni et al., 2026), and is therefore considered consistently for all signals in this study.

Following the compatibility framework of (International Telecommunication Union, 2015) and the application to Pulsar presented by Reid et al. (2025), the RNSS interference is separated into three contributions: $I_{\mathrm{ref}}$, representing interference generated by the reference RNSS system; $I_{\mathrm{alt}}$, representing interference from the alternate system considered in the coordination analysis; and $I_{\mathrm{rem}}$, representing interference from the remaining RNSS systems. These terms are treated as equivalent post-correlation noise power spectral densities and are combined with the thermal noise to obtain the effective noise density experienced by the receiver.

Non-RNSS external interference, denoted $I_{\mathrm{ext}}$ in (International Telecommunication Union, 2015), is not included in the nominal analysis. Such interference is strongly dependent on the local operating environment, and no generally representative average value was identified.

For an excess attenuation $A$ expressed in dB, assumed to affect the desired and interfering RNSS signals equally, the total RNSS interference density is

$$I_{\mathrm{RNSS}}(A) = 10\log_{10}\left(10^{\frac{I_{\mathrm{ref}}-A}{10}} + 10^{\frac{I_{\mathrm{rem}}-A}{10}} + 10^{\frac{I_{\mathrm{alt}}-A}{10}}\right). \tag{2}$$

Since the same attenuation is applied to all three contributions, $I_{\mathrm{RNSS}}(A) = I_{\mathrm{RNSS}}(0) - A$.

The effective noise power spectral density used in the sensitivity analysis is then

$$\begin{aligned} N_{0,\mathrm{eff}}(A) &= 10\log_{10}\left(10^{\frac{N_0}{10}} + 10^{\frac{I_{\mathrm{RNSS}}(A)}{10}}\right) \\ &= 10\log_{10}\left(10^{\frac{N_0}{10}} + 10^{\frac{I_{\mathrm{ref}}-A}{10}} + 10^{\frac{I_{\mathrm{rem}}-A}{10}} + 10^{\frac{I_{\mathrm{alt}}-A}{10}}\right), \end{aligned} \tag{3}$$

with all power spectral densities expressed in dBW/Hz. The corresponding increase relative to the thermal noise floor is

$$\Delta N_0(A) = N_{0,\mathrm{eff}}(A) - N_0. \tag{4}$$

**TABLE 3.4** summarizes the interference densities adopted for each signal. For GPS L1 C/A, GPS L1C, and Galileo E1C, the interference densities are derived from the corresponding GPS-Pulsar and Galileo-Pulsar coordination analyses. Since these studies are intended for compatibility assessment and use maximum aggregate interference conditions, whereas the present link-budget analysis targets representative average conditions, the reported values are reduced by 2.45 dB. This correction corresponds to the difference between maximum and mean aggregate gain obtained from Pulsar constellation simulations using the latest satellite antenna pattern. In the absence of equivalent mean aggregate-gain simulations for the MEO constellations, the same difference is assumed for GPS and Galileo.

Values for BeiDou B1I and B1C are estimated from signals with similar modulation characteristics. B1I is derived from GPS L1 C/A because both signals use BPSK modulation. Its chip rate is twice that of L1 C/A, resulting in an approximately 3 dB reduction in SSC. For $I_{\text{ref}}$, the 4.5 dB difference in nominal received signal power is additionally included. B1C is derived from Galileo E1C because both employ non-time-multiplexed MBOC-type modulation; the same $I_{\text{rem}}$ and $I_{\text{alt}}$ values are therefore assumed, while $I_{\text{ref}}$ is reduced by 1.75 dB to account for the difference in received signal power.

**TABLE 3.4** Interference densities adopted for each signal

| Signal | $I_{\text{ref}}$ (dBW/Hz) | $I_{\text{rem}}$ (dBW/Hz) | $I_{\text{alt}}$ (dBW/Hz) | $I_{\text{RNSS}}(0)$ (dBW/Hz) |
|---|---|---|---|---|
| GPS L1 C/A | −208.14 | −207.49 | −226.19 | −204.76 |
| GPS L1C | −210.09 | −208.40 | −215.21 | −205.64 |
| Galileo E1C | −211.73 | −210.86 | −215.21 | −207.46 |
| BeiDou B1I | −215.64 | −210.49 | −229.19 | −209.29 |
| BeiDou B1C | −213.48 | −210.86 | −215.21 | −208.04 |
| Pulsar X1 | −192.31 | −207.49 | — | −192.18 |

For Pulsar X1, the reference-system interference is calculated directly as

$$I_{\text{ref,X1}} = P_{\text{max}} + G_{\text{agg,mean}} + SSC. \quad (5)$$

with $P_{\text{max}} = -139.1$ dBW, an average aggregate gain $G_{\text{agg,mean}} = 7.49$ dB obtained by constellation simulation using the latest satellite antenna pattern, and $SSC = -60.70$ dB/Hz, resulting in $I_{\text{ref,X1}} = -192.31$ dBW/Hz. The remaining-system interference is approximated using the GPS L1 C/A value because X1 has the same BPSK chip rate. When X1 is the reference signal, no separate alternate-system contribution is introduced; interference from the existing RNSS constellations is included in $I_{\text{rem}}$.

The resulting increase in effective noise density is shown in **TABLE 3.5** for the attenuation levels considered in this study. Multiple-access interference therefore has its largest effect under strong-signal conditions. For the MEO signals, the increase in noise density is between approximately 1 and 3 dB under nominal conditions and falls below 0.4 dB after 10 dB of attenuation. The effect is considerably larger for Pulsar X1 because of its substantially higher received signal power, increasing the nominal noise floor by approximately 12 dB. However, the desired signal and RNSS interference are attenuated together whereas the receiver thermal noise remains unchanged. Consequently, the interference contribution progressively becomes negligible as attenuation increases, with the X1 noise-density penalty decreasing to approximately 1.7 dB at 15 dB attenuation and 0.6 dB at 20 dB. This behavior is particularly relevant to the acquisition comparison, since the high received power of X1 is primarily beneficial under the attenuated conditions where RNSS self-interference becomes less dominant.

It should be noted that the link budget combines the specified minimum desired-signal powers with representative aggregate-interference levels. These conditions are not expected to occur simultaneously for all satellites in a particular constellation geometry: a geometry producing generally weaker received signals would also tend to reduce the aggregate RNSS interference. The resulting acquisition requirements should therefore be interpreted as conservative sensitivity-oriented values rather than as a prediction of a specific instantaneous constellation realization.

From the effective $N_0$, the corresponding noise power and IF SNR are computed. These quantities represent the available signal quality before the minimum coherent integration interval.

**TABLE 3.5** Increase in effective noise density for different attenuation levels for each signal (dB)

| Signal | 0 dB | 5 dB | 10 dB | 15 dB | 20 dB | 25 dB | 30 dB |
|---|---|---|---|---|---|---|---|
| GPS L1 C/A | 2.65 | 1.02 | 0.35 | 0.11 | 0.04 | 0.01 | 0.00 |
| GPS L1C | 2.27 | 0.85 | 0.29 | 0.09 | 0.03 | 0.01 | 0.00 |
| Galileo E1 | 1.62 | 0.58 | 0.19 | 0.06 | 0.02 | 0.01 | 0.00 |
| BeiDou B1I | 1.13 | 0.39 | 0.13 | 0.04 | 0.01 | 0.00 | 0.00 |
| BeiDou B1C | 1.44 | 0.51 | 0.17 | 0.05 | 0.02 | 0.01 | 0.00 |
| Pulsar X1 | 12.10 | 7.64 | 4.01 | 1.70 | 0.61 | 0.20 | 0.07 |

*3.1.2.3. Minimum coherent integration and implementation losses*

The IF SNR obtained in the previous section is first converted into the signal-to-noise ratio corresponding to the minimum coherent integration interval supported by the considered signal. Throughout this work, the minimum coherent interval is defined as one primary code period, which represents the shortest coherent integration naturally performed by the acquisition engine. The corresponding number of accumulated samples depends on both the primary code duration and the receiver sampling frequency, assumed to be twice the signal chip rate.

The ideal coherent processing gain is then obtained from the coherent integration time. In practice, however, several implementation effects reduce the achievable gain. These include front-end filtering, finite quantization resolution, residual carrier frequency mismatch, code alignment errors resulting from the acquisition search grid, and signal-specific effects such as the Pulsar watermarking and EFQPSK modulation losses. The total implementation loss is subtracted from the ideal coherent gain to obtain the effective coherent processing gain.

For the resolution, a 2-bit analog-to-digital converter (ADC) is considered, leading to a loss of 1.01 dB (van Diggelen, 2021). For the frequency mismatch, we have considered a step of $1/(2T_C)$, which leads to an average loss of 0.30 dB, as mentioned in Section 3.1.1.3. For the code alignment, since the sampling rate is twice the chip rate and the code step is assumed to be one sample, it is half a chip, which leads to an average loss of 1.20 dB.

Pulsar watermarking inverts 21 chips among the 1023 transmitted (Anderson, 2025), introducing a loss of $20\log_{10}\left((1023 - 2 \times 21)/1023\right) \approx 0.36$ dB. Pulsar EFQPSK modulation when processed as BPSK introduces a loss of about 0.50 dB, based on simulation results.

Finally, the effective coherent gain is applied to the input SNR to determine the coherent SNR after the minimum coherent integration interval. At this stage, the signal conditions available after one primary code period have been completely characterized. The following step of the methodology determines the additional coherent or non-coherent integration required to reach the target probability of detection while satisfying the probability of false alarm established in Section 3.1.1.

**3.2. Required integration time**

Using the methodology described in Section 3.1, the required integration time was computed for each analyzed signal under identical acquisition performance requirements. Results are presented for non-coherent-only, coherent-only, and hybrid acquisition strategies, considering assisted and non-assisted acquisition scenarios.

For the hybrid strategy, coherent integration is limited to representative maximum values, after which any additional processing gain required to satisfy the acquisition-performance target is obtained through non-coherent integration. Several coherent-integration constraints are considered (see **TABLE 3.6**) to represent the different practical integration regimes evaluated later in Section 4. When the coherent-only integration requirement is smaller than the imposed limit, no non-coherent accumulation is required and the hybrid result reduces to the coherent-only result.

These limits are selected to correspond to the acquisition architectures investigated in Section 4.1. Their implementation significance and associated processing-memory trade-offs are discussed there; in the present section, only their effect on the required total integration time is considered.

**TABLE 3.6** Coherent integration constraint for hybrid integration

| Signal type | Hybrid case | Coherent integration constraint |
|---|---|---|
| GPS L1 C/A | Case 1 | $N_{\mathrm{C}}^{\max} = 10$ |
| GPS L1 C/A | Case 2 | $N_{\mathrm{C}}^{\max} = 19$, including the average bit-transition loss |
| GPS L1 C/A | Case 3 | $N_{\mathrm{C}}^{\max} = 20$ |
| Modern pilot signal | Case 1 | $N_{\mathrm{C}}^{\max} = 5$ |
| Modern pilot signal | Case 2 | $N_{\mathrm{C}}^{\max} = 10$ |
| Modern pilot signal | Case 3 | $T_{\mathrm{C}}^{\max} = 100$ ms |

### *3.2.1. Non-assisted acquisition*

*3.2.1.1. Noncoherent-only integration*

**Figure 3.3** and **Figure 3.4** show the noncoherent integration length and time required for each signal at different attenuations to reach the required probability of detection, and global probability of false alarm, without assistance.

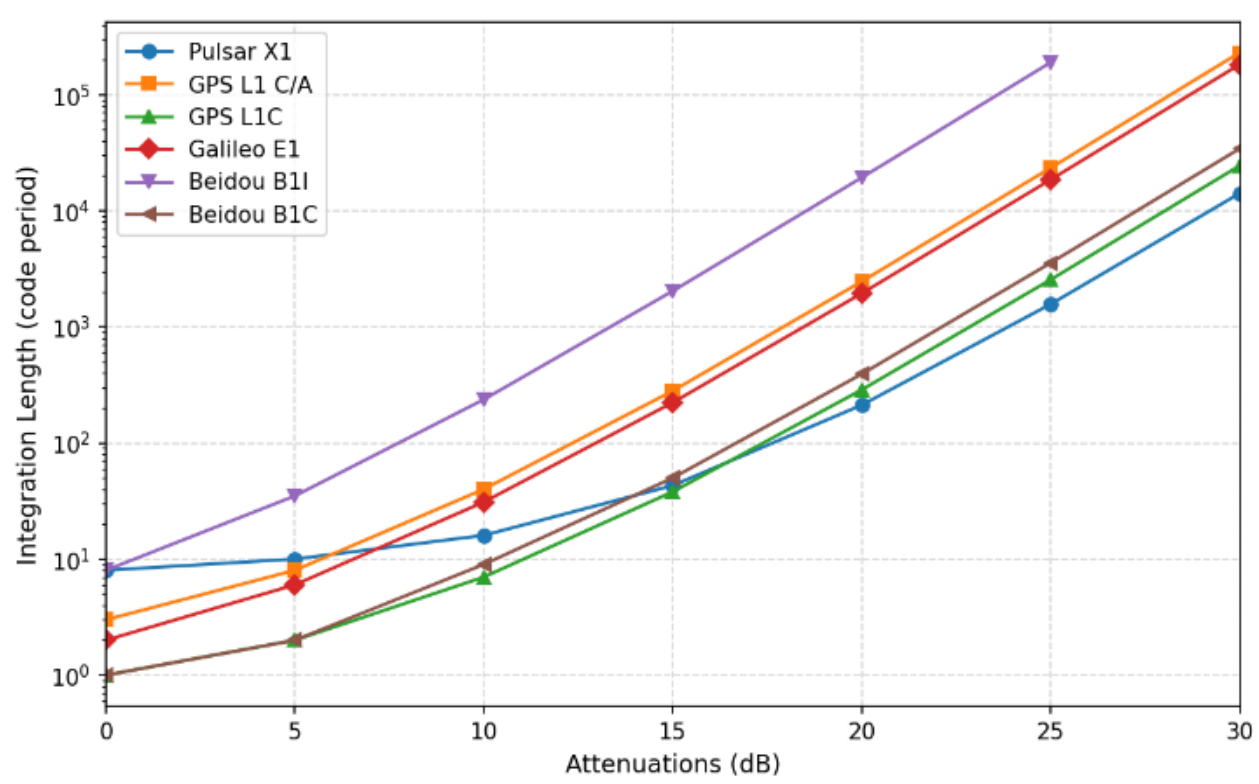

**Figure 3.3** Required noncoherent integration length without assistance

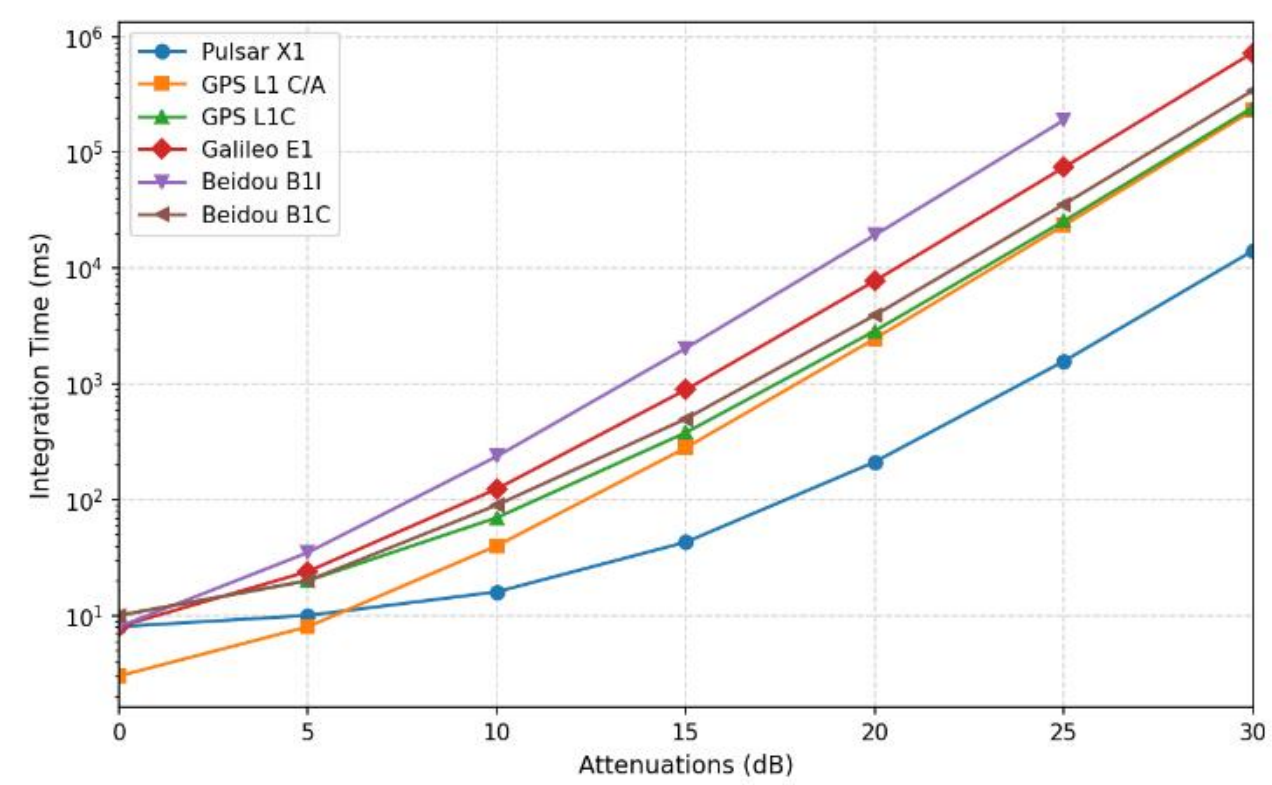

**Figure 3.4** Required noncoherent integration time without assistance

*3.2.1.2. Coherent-only integration*

**Figure 3.5** and **Figure 3.6** show the coherent integration length and time required for each signal at different attenuations to reach the required probability of detection, and global probability of false alarm, without assistance.

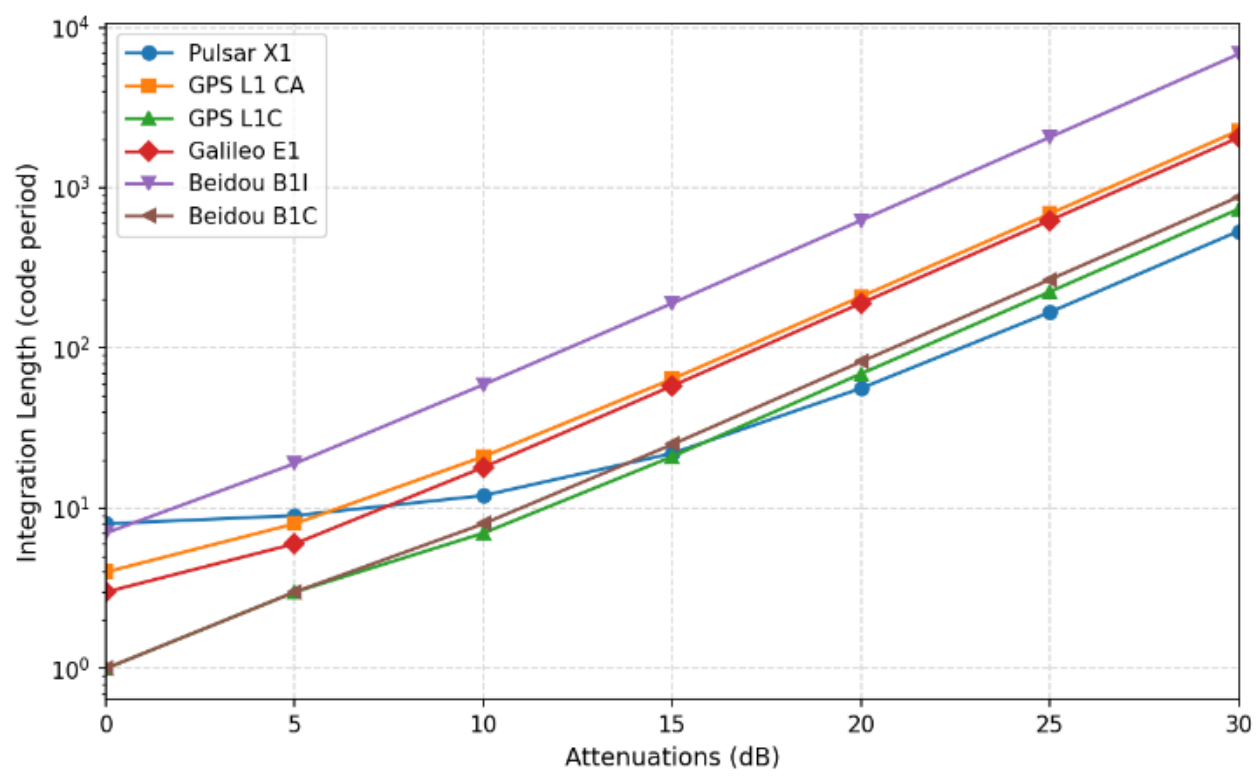

**Figure 3.5** Required coherent integration length without assistance

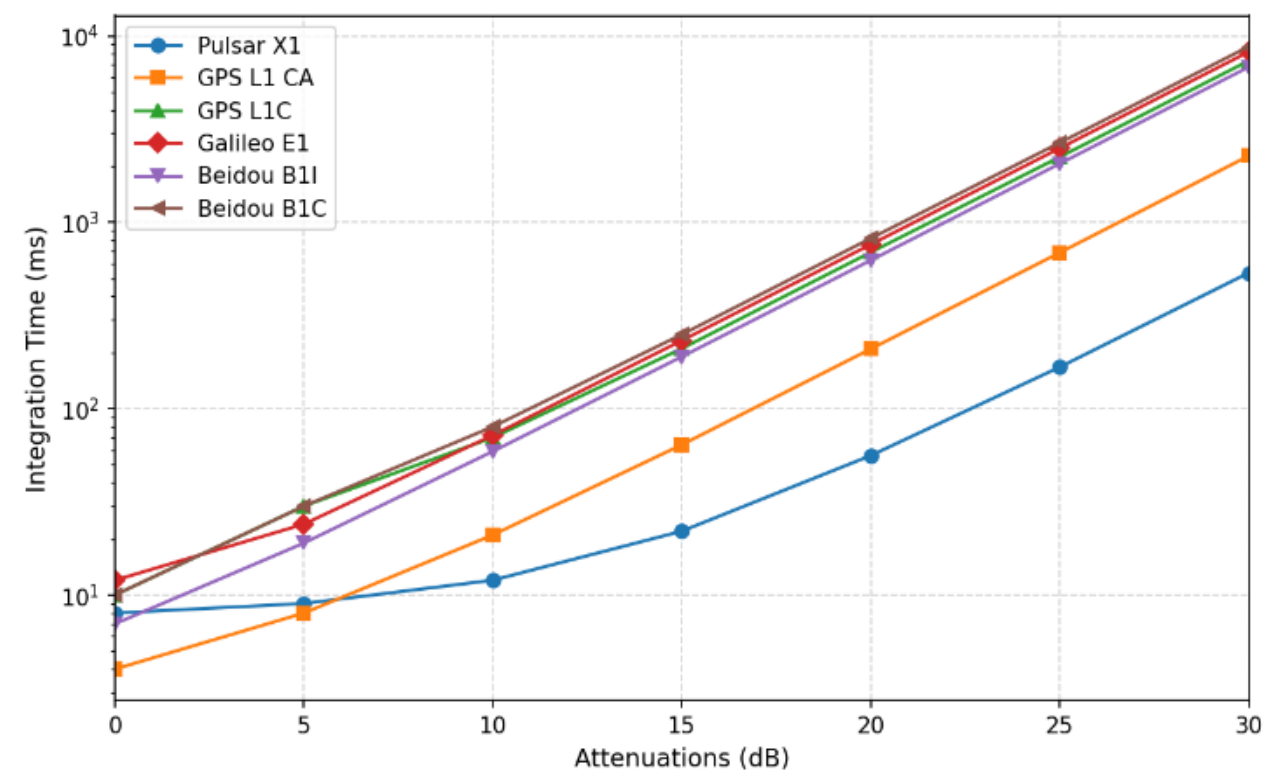

**Figure 3.6** Required coherent integration time without assistance

*3.2.1.3. Hybrid integration*

**Figure 3.7** shows the total integration time for each signal at different attenuations to reach the required probability of detection, and global probability of false alarm, without assistance.

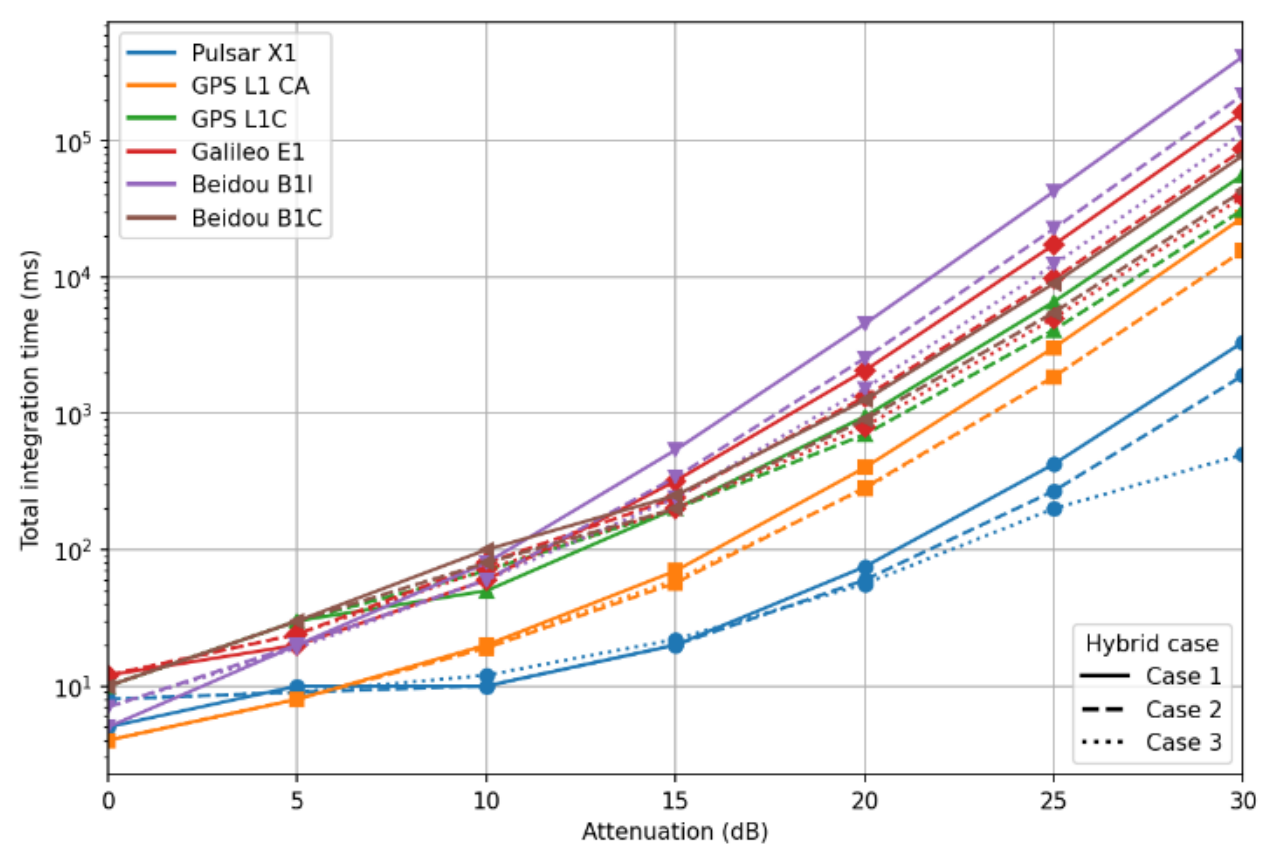

**Figure 3.7** Required coherent integration time without assistance

***3.2.2. Assisted acquisition***

*3.2.2.1. Non-coherent-only integration*

**Figure 3.8** and **Figure 3.9** show the noncoherent integration length and time required for each signal at different attenuations to reach the required probability of detection, and global probability of false alarm, with assistance.

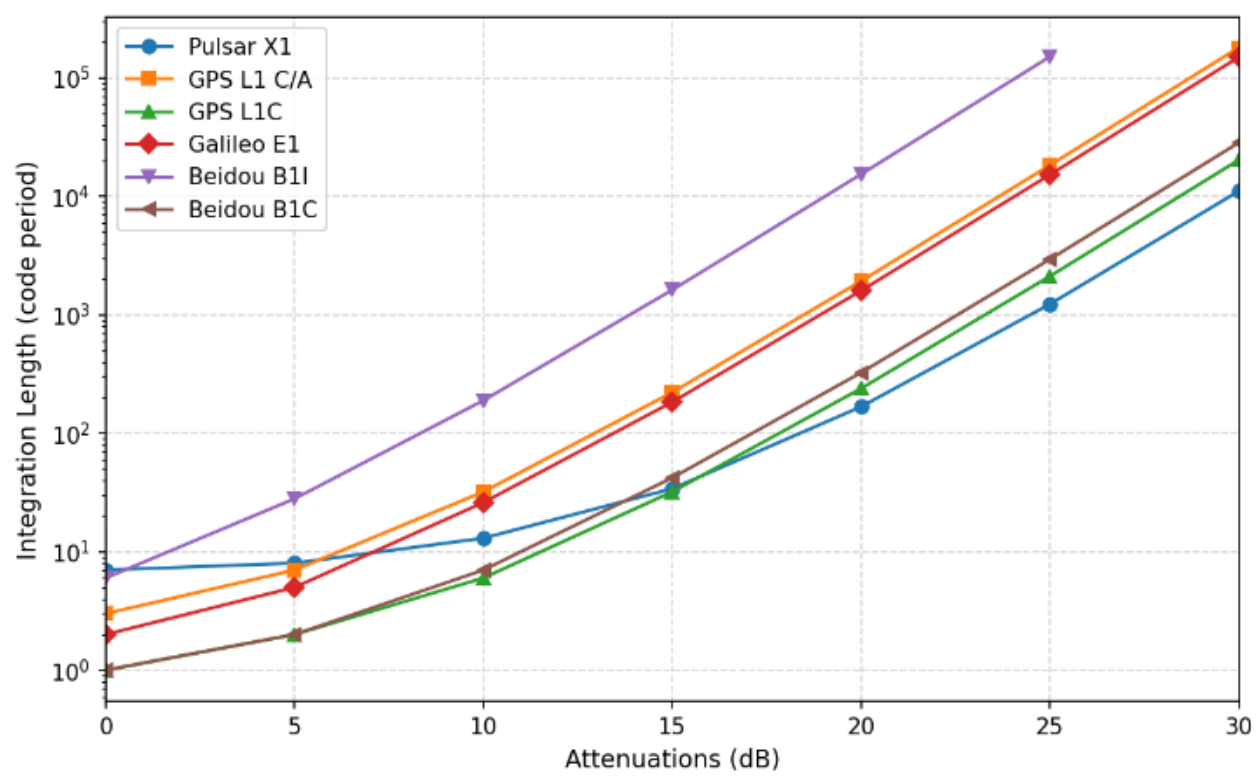

**Figure 3.8** Required noncoherent integration length with assistance

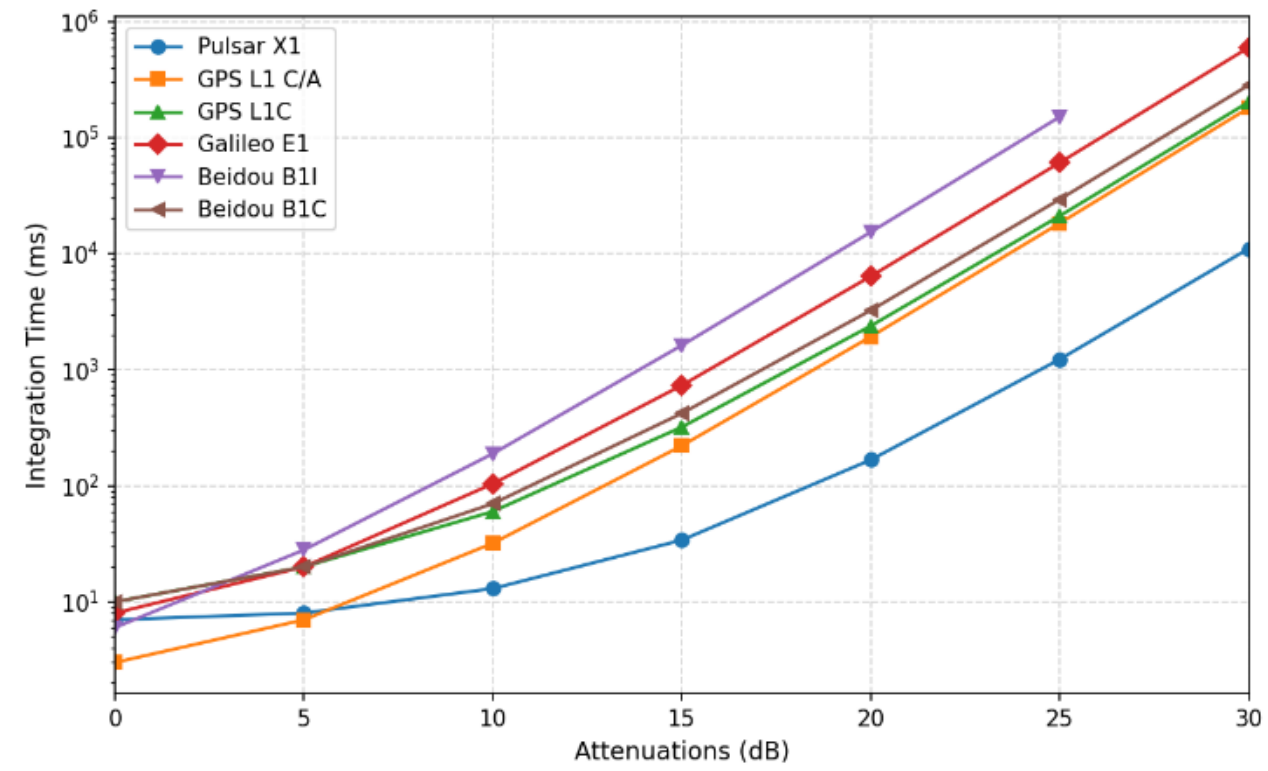

**Figure 3.9** Required noncoherent integration time with assistance

*3.2.2.2. Coherent-only integration*

**Figure 3.10** and **Figure 3.11** show the coherent integration length and time required for each signal at different attenuations to reach the required probability of detection, and global probability of false alarm, with assistance.

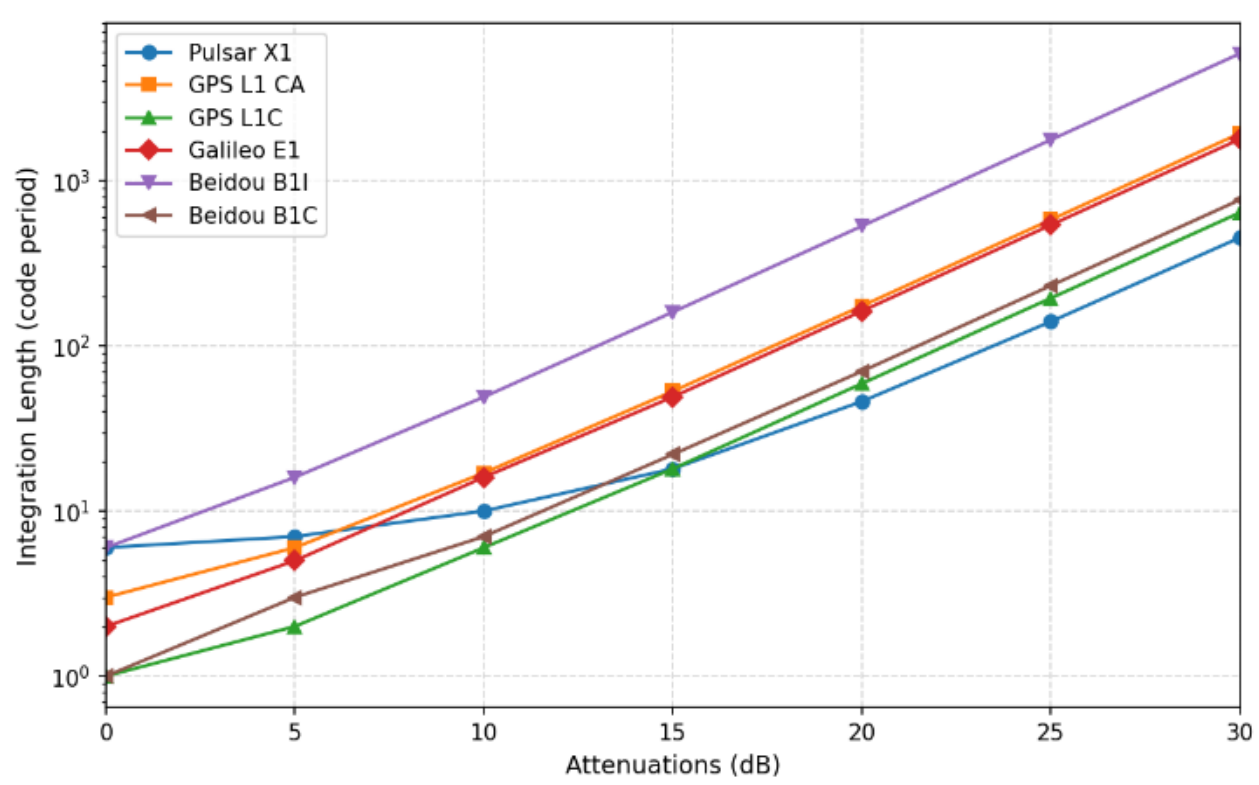

**Figure 3.10** Required coherent integration length with assistance

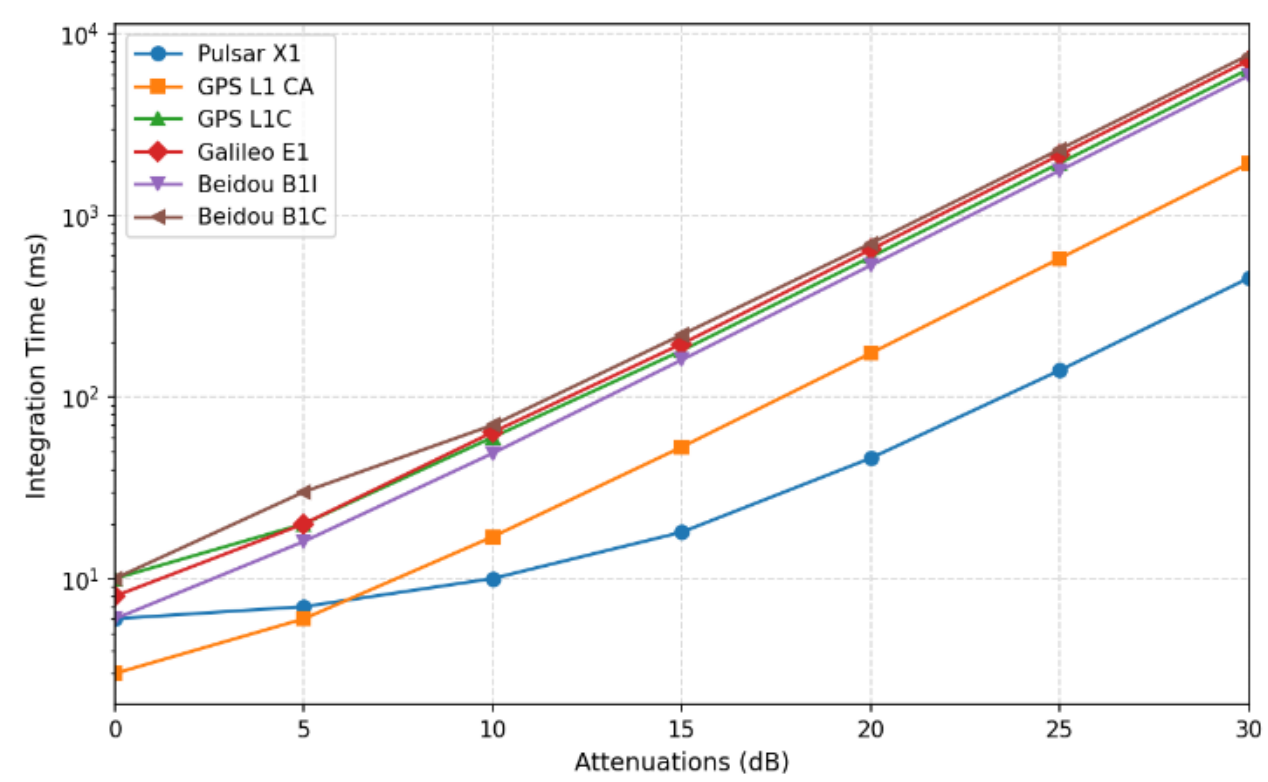

**Figure 3.11** Required coherent integration time with assistance

*3.2.2.3. Hybrid integration*

**Figure 3.12** shows the total integration time for each signal at different attenuations to reach the required probability of detection, and global probability of false alarm, with assistance.

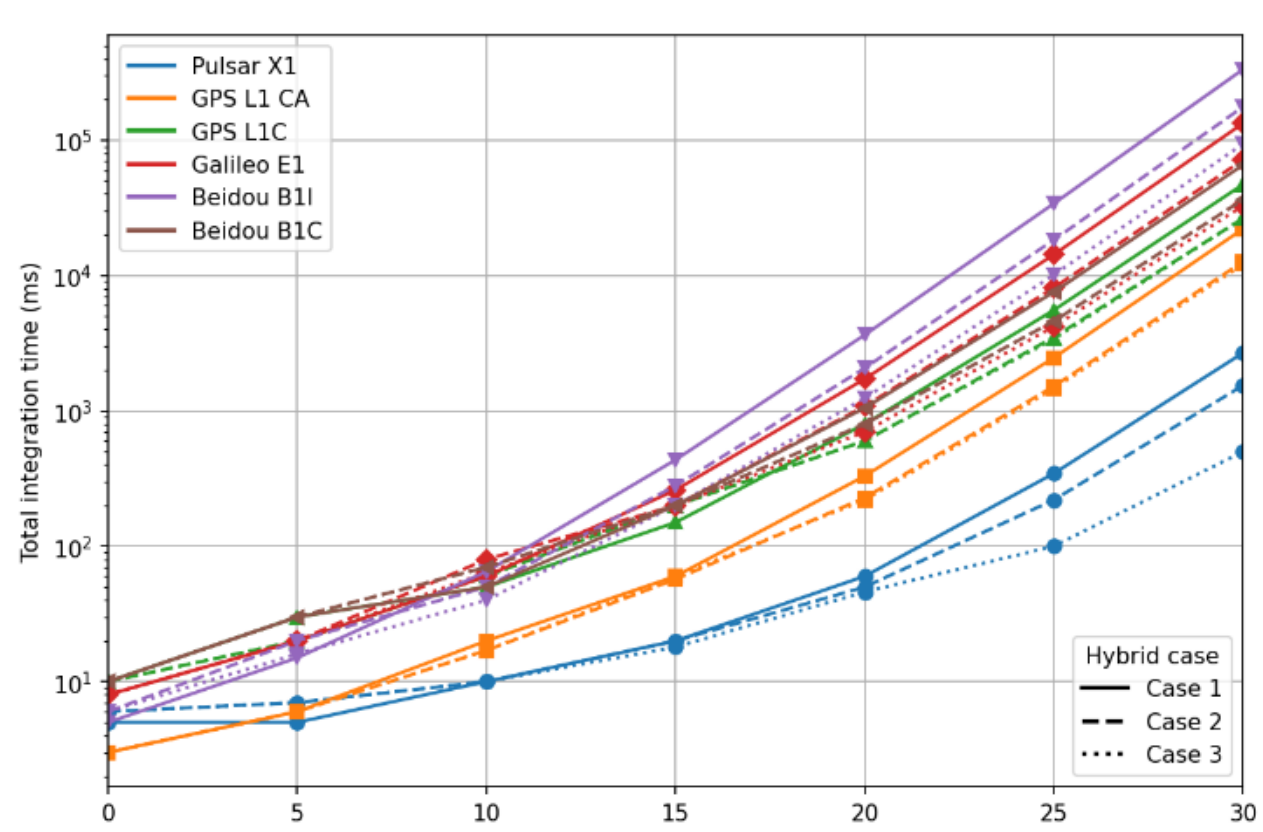

**Figure 3.12** Required coherent integration time with assistance

### *3.2.3. General observations*

Three main outcomes can be deduced from the previous tables.

First, as attenuation increases, Pulsar X1 requires the shortest coherent and non-coherent integration times among the analyzed signals. This behavior is primarily explained by the higher received signal power of the LEO constellation, which reduces the amount of processing gain that must be generated by the receiver to satisfy the target acquisition performance. The advantage becomes increasingly significant as the attenuation increases, where the required processing gain is dominated by receiver sensitivity.

Second, at low attenuation levels, Pulsar X1 does not always provide the shortest integration time because aggregate RNSS interference partially offsets its higher received signal power. As shown in **TABLE 3.5** this penalty decreases rapidly with attenuation, allowing the link-budget advantage of X1 to become increasingly dominant under weaker-signal conditions.

Among the MEO signals, GPS L1 C/A consistently requires the shortest integration times. This is primarily due to its favorable link budget, as the full transmitted power is available for acquisition without pilot/data power sharing or BOC approximation losses. In addition, its short 1023-chip primary code minimizes the code search space, requiring a smaller final SNR for the same detection probabilities. Together, these characteristics make GPS L1 C/A the least demanding MEO signal from an acquisition perspective. Note that this is consistent with previous observations from (van Diggelen, 2021; Section 18.7, van Diggelen, 2014).

The results presented in this section quantify the integration time required for each signal to satisfy identical acquisition performance requirements. Although integration time provides an important indicator of acquisition sensitivity, it does not directly reflect the implementation cost of the receiver. Signals requiring the same integration time may nevertheless exhibit significantly different arithmetic complexity, memory requirements, and processing time due to differences in signal structure, such as primary code length, chip rate, modulation, and the adopted acquisition strategy.

Consequently, the required integration times obtained in this section constitute only the first step of the comparison. In the following section, these values are used as inputs to representative receiver implementations in order to evaluate the corresponding computational complexity, memory requirements, and processing time. This enables a fair comparison of the practical implementation cost associated with acquiring each analyzed signal.

## 4. RECEIVER IMPLEMENTATION COMPLEXITY

Section 3 quantified the amount of integration required for each signal to satisfy identical acquisition performance requirements. This section evaluates the implementation cost associated with achieving these integration times. Representative FFT-based acquisition architectures are considered for non-coherent integration only, coherent integration only, and hybrid involving both integrations. Then, their arithmetic complexity (for software receivers), processing time (for hardware receivers), and memory requirements are analyzed. Unless otherwise stated, the integration times determined in Section 3 are used directly as inputs to the complexity analysis.

### 4.1. Receiver architectures considered

The acquisition can be split into two parts: primary code correlation, i.e. computing a correlation whose length is the primary code; and extra integration, where we combine consecutive primary code correlation results to improve sensitivity. The primary code correlation considered for this study is described in Section 4.1.1. After the primary-code correlation has been performed, the acquisition process diverges depending on the selected integration strategy: non-coherent vs coherent vs hybrid.

Non-coherent integration extends the observation time by accumulating the magnitudes of successive primary-code correlations, whereas coherent integration combines the complex correlation results before the magnitude operation. Non-coherent integration is simpler to implement, does not depend on the signal structure, and does not increase the number of Doppler bins to test, but it brings less gain than coherent integration. The number of non-coherent accumulations is denoted $N_{\mathrm{NC}}$. The non-coherent integration is discussed in Section 4.1.2.

Extending the coherent integration interval is considerably more dependent on the signal structure than non-coherent integration. Since coherent accumulation is performed before the magnitude operation, the receiver must preserve the sign relationship between successive primary code correlations. Consequently, the implementation strategy depends on both the signal structure and the required number of coherent accumulations, $N_{\mathrm{C}}$. Since the GPS L1 C/A signal has a different structure than modern GNSS signals (several primary code periods within one data bit vs. single primary code period within one data bit or one secondary code chip), the architecture to extend the coherent integration for both is clearly different, and Sections 4.1.3 and 4.1.4 discuss them separately.

Finally, both types of integration can be combined, where coherent integration is followed by non-coherent integration. This is discussed as well in Sections 4.1.3 and 4.1.4.

BeiDou B1I is retained in the non-coherent baseline but omitted from the coherent and hybrid comparisons. Its data-only structure combines multiple primary-code periods per navigation bit with an overlay code, requiring additional transition handling compared with GPS L1 C/A. Together with its lower received power and higher chip rate, B1I provides no structural or link-budget advantage over L1 C/A for the coherent integration architectures considered here.

The architectures described below are selected according to the signal structure and required coherent integration depth; they are not all intended to be used for every operating condition.

### *4.1.1. FFT-based primary code correlation*

To ensure a fair comparison between the analyzed signals, a common FFT-based acquisition architecture is adopted throughout this work. FFT-based acquisition provides an efficient implementation of the parallel code phase search and is representative of modern GNSS receivers (van Diggelen, 2021, Section 18.7.1). Since the objective of this paper is to compare the implementation complexity imposed by different signal structures rather than different acquisition algorithms, the same acquisition architecture is used for all signals. FFT-based acquisition is particularly attractive in the assisted scenario considered here, because assistance substantially reduces the Doppler search space while the code-delay search space remains unchanged; parallel code-phase search therefore preserves its main advantage.

The architecture considered is illustrated in **Figure 4.1**. The received signal is first buffered in memory, allowing the same block of samples to be reused for every Doppler hypothesis without repeatedly accessing the front-end. This approach is common in modern receivers (van Diggelen, 2021, Section 18.7.1) and is particularly well suited to snapshot-based receivers and enables efficient evaluation of all code-delay and Doppler hypotheses. For each Doppler hypothesis, the input signal is mixed with a local carrier replica, transformed into the frequency domain using an FFT, multiplied by the conjugated FFT of the local spreading code, and transformed back to the time domain using an inverse FFT. The resulting correlation vector simultaneously provides the correlation values for all code-delay hypotheses associated with the selected Doppler bin.

For a primary code period containing $N_{\mathrm{P}}$ samples, the correlation is performed using double-block zero padding (DBZP). Consequently, the input block spans two primary code periods, and the FFT length $N_{\mathrm{FFT,P}}$ is selected as the smallest power of two greater than or equal to $2N_{\mathrm{P}}$. This approach prevents circular correlation losses caused by navigation data or overlay-code transitions while also satisfying the FFT length requirements (Leclère, 2014). More optimized zero-padding techniques have been proposed, particularly for signals whose primary code length is not close to a power of two (Leclère, 2015), but these optimizations are not considered here in order to maintain a consistent comparison across all analyzed signals.

**TABLE 4.1** summarizes the implementation parameters for each signal. The longer primary code periods of modern GNSS signals lead to significantly larger FFT sizes than legacy GPS L1 C/A or BeiDou B1I, which directly impacts both the software processing workload and the hardware resources required for implementation.

The FFT of the local spreading code is independent of both the received data and the Doppler hypothesis and therefore needs to be computed only once for each acquisition attempt. For hardware receivers, it can either be computed using a dedicated FFT block or reuse the FFT employed for the input signal during the acquisition process. Unlike the remaining operations, this computation is not repeated for every Doppler hypothesis. Nevertheless, it is still included in the implementation complexity analysis, as its relative contribution becomes more significant when the acquisition search space is reduced, particularly in assisted acquisition scenarios.

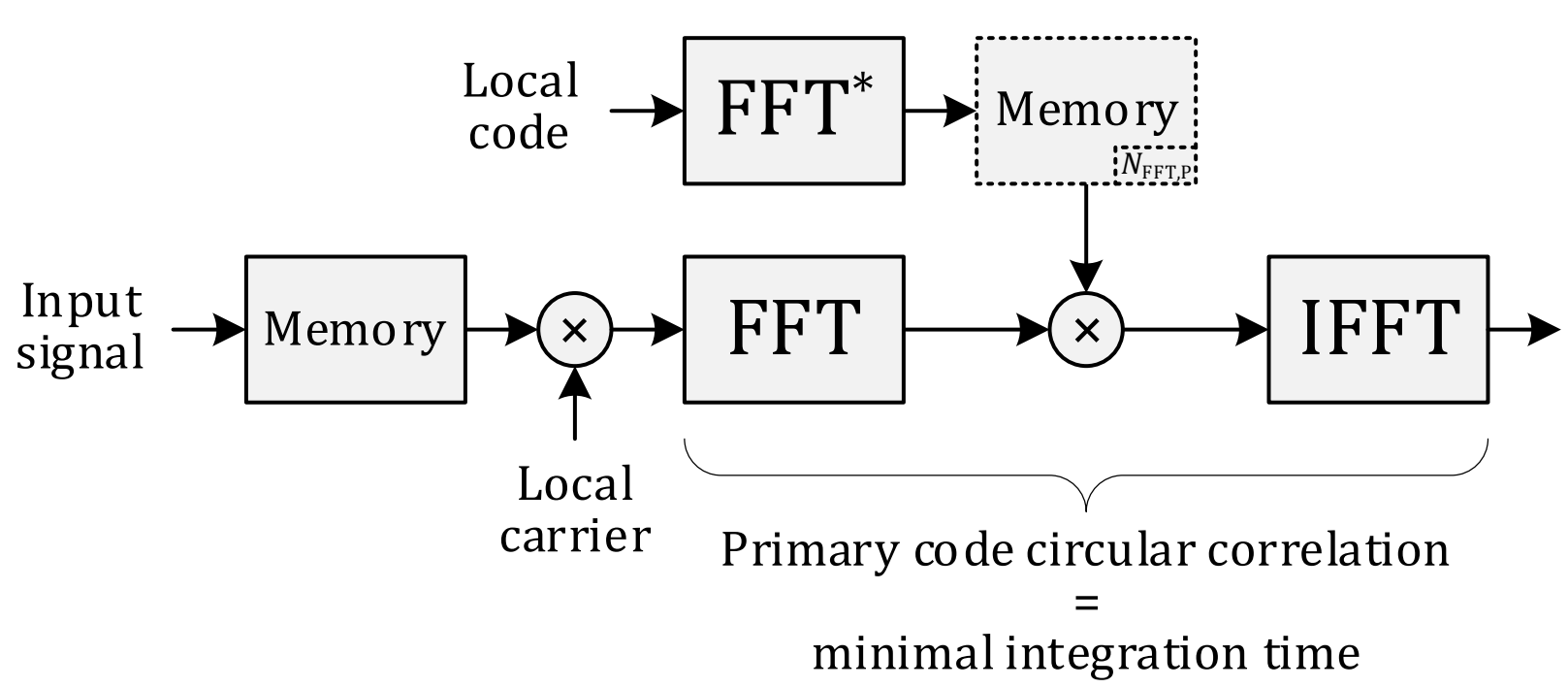

**Figure 4.1** Architecture considered for primary-code correlation

**TABLE 4.1** Representative implementation parameters related to FFT for each signal

| Signal | X1 | L1 C/A | L1C | E1 | B1I | B1C |
|---|---|---|---|---|---|---|
| Primary code length in samples ($N_{\mathrm{P}}$) | 2046 | 2046 | 20460 | 8184 | 4092 | 20460 |
| Number of samples in two primary code periods ($2N_{\mathrm{P}}$) | 4092 | 4092 | 40920 | 16368 | 8184 | 40920 |
| Required FFT length ($N_{\mathrm{FFT,P}}$) | 4096 | 4096 | 65536 | 16384 | 8192 | 65536 |
| Number of BRAMs required[1] | 10 | 10 | 117 | 32 | 19 | 117 |
| Number of bits required ($M_{\mathrm{FFT,P}}$)[2] | 360 Kibit | 360 Kibit | 4212 Kibit | 1152 Kibit | 684 Kibit | 4212 Kibit |

[1] Number of BRAMs required is an estimate provided by AMD Vivado v2022.2 for the Fast Fourier Transform LogicCORE IP, using a 12-bit Pipelined I/O configuration and the xczu47dr device
[2] 1 Kibit = 1024 bits

### *4.1.2. Non-coherent integration*

In the non-coherent acquisition strategy, the coherent integration duration is limited to a single primary code period. After each FFT-based correlation, the magnitude of the complex correlation vector is computed, producing one real-valued correlation output for every code-delay hypothesis. These outputs are accumulated over multiple primary code periods until the required non-coherent integration length determined in Section 3.2 is reached.

This approach is straightforward to implement since it only requires storing one accumulation value for each code-delay hypothesis. For every additional primary code period, a new FFT-based correlation is performed, and its magnitude is added to the previously accumulated result. The implementation therefore scales linearly with the number of non-coherent accumulations, making this strategy applicable to any navigation signal regardless of its data or secondary-code structure.

The architecture is illustrated in **Figure 4.2**. There are $(N_{\mathrm{NC}}+1)N_{\mathrm{P}}$ input samples stored (value shown in the bottom right of the input memory box). The + 1 in the expression is because the primary code correlations are performed over two code periods, therefore, even if $N_{\mathrm{NC}}$ correlations are computed, the final correlation needs an extra period. The non-coherent integration is implemented with a memory storing the accumulated magnitudes using a simple accumulation stage (Leclère, 2017), and this memory stores $N_{\mathrm{P}}$ samples (value shown in the bottom right of the non-coherent integration box).

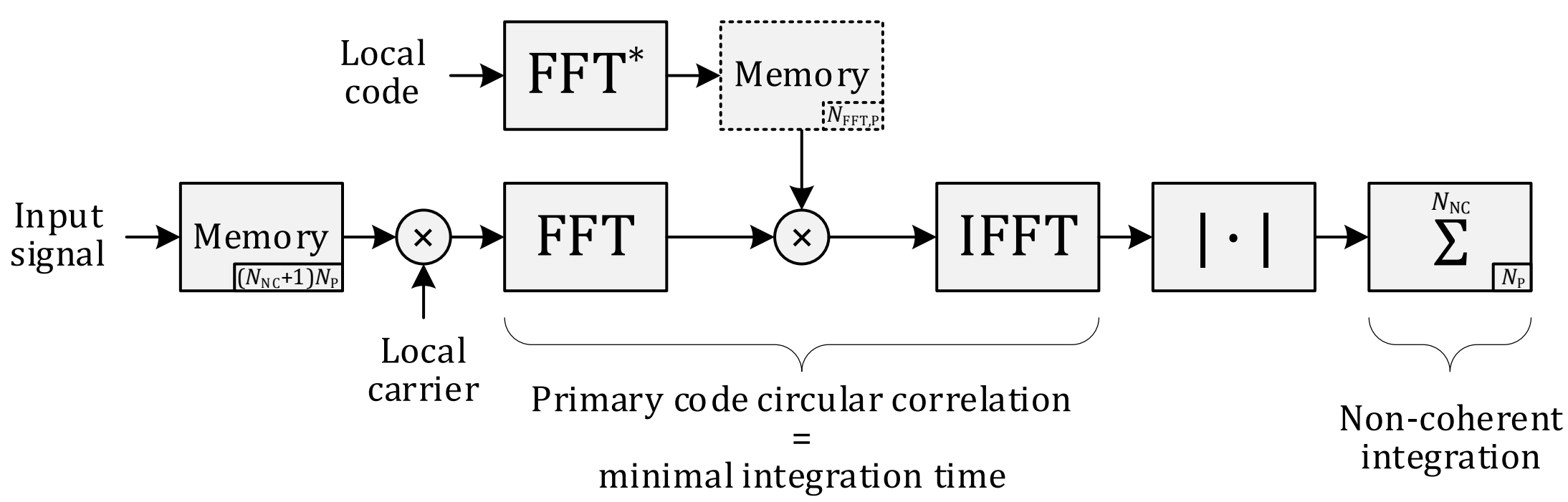

**Figure 4.2** Representative architecture for non-coherent integration

### *4.1.3. Extending coherent integration for GPS L1 C/A signal*

The GPS L1 C/A signal is a particular case because its 1 ms primary code repeats 20 times within a single navigation data bit. Consequently, coherent integration can be extended beyond one primary code period using several implementation strategies, each providing a different trade-off between complexity and detection performance.

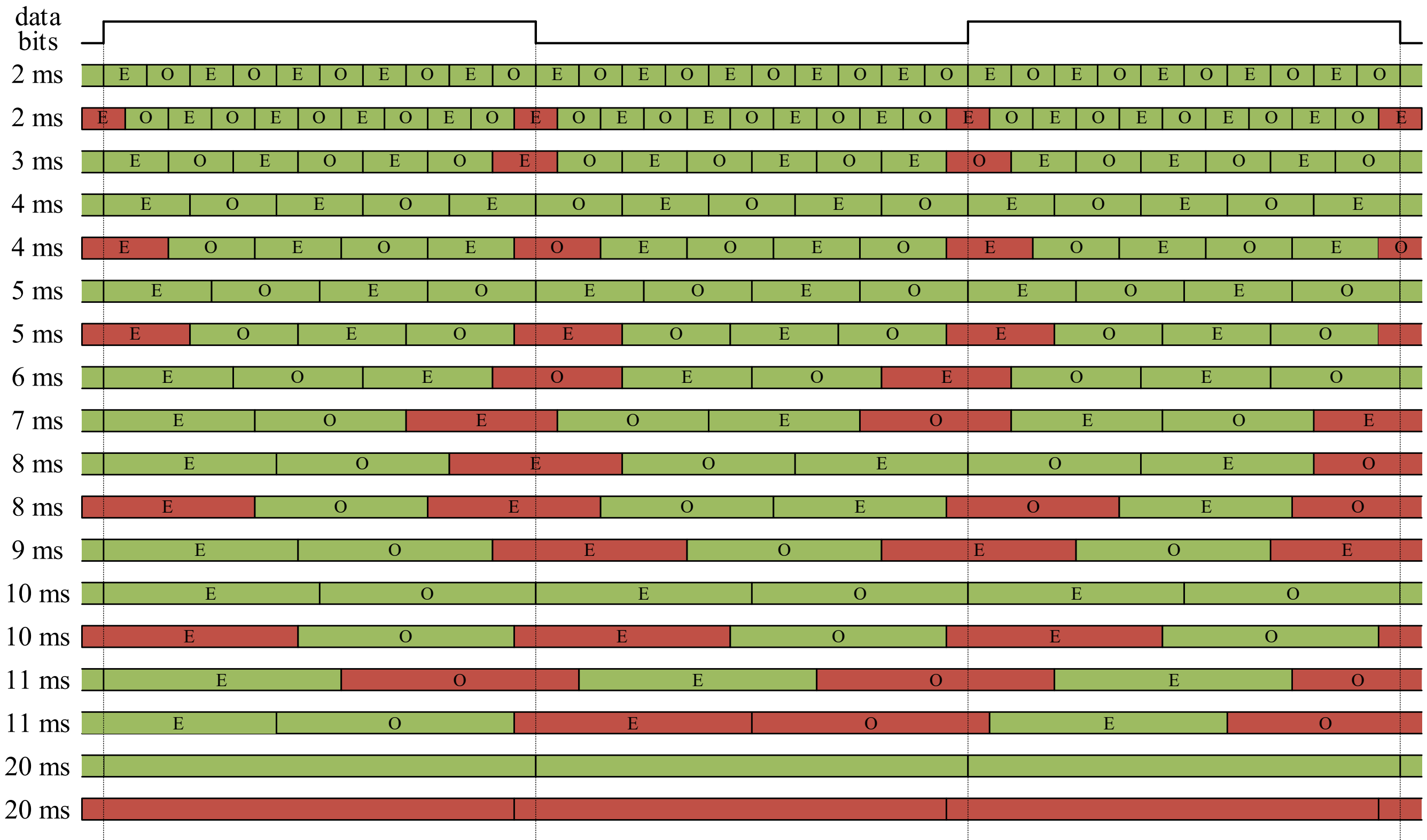

**Figure 4.3** Illustration of GPS L1 C/A data bit crossing for different coherent integration times and starting points. E = even, O = odd

*4.1.3.1. Alternate block method*

For coherent integration times up to 10 ms, the alternate block method proposed by (Psiaki, 2001) can be employed to extend the coherent integration time without any loss (method called alternate half-bit method in (Psiaki, 2001) because it is typically used with an integration time of 10 ms to maximize the coherent integration time, but it can be used for any duration between 1 ms and 10 ms, and in this study we focus on the minimum processing required to reach a target performance, therefore it is useful to use less than 10 ms if it is enough). In this method, two consecutive data blocks, each spanning the desired coherent integration interval, are processed separately as shown in **Figure 4.5**. Since the duration covered by both blocks is not more than the 20 ms data bit duration, one of the two blocks is guaranteed to be free of navigation-bit transitions as shown in **Figure 4.3** and therefore achieves the full coherent processing gain. **Figure 4.3** also shows that from 11 ms of coherent integration time, there can be two consecutive blocks each crossing a data bit, hence the 10 ms limit of this method to guarantee to not have any loss (or then more consecutives blocks should be processed, but this reduces the efficiency of the method, and is thus not considered here).

If only coherent integration is performed, the integration time can be freely selected between 1 ms and 10 ms. When coherent integration is followed by non-coherent integration, however, all coherently integrated results accumulated along at least one path should be free of data-bit transitions to avoid any associated loss. When $20/N_\mathrm{C}$ is an even integer where $N_\mathrm{C}$ is the coherent integration time in milliseconds greater than 1, i.e., when $N_\mathrm{C} = 2$, 5, or 10, the coherent results can simply be separated into even and odd paths. Since successive data-bit transitions are then separated by an even number of coherent blocks, they always affect blocks of the same parity, guaranteeing that the other path remains transition-free. For $N_\mathrm{C} = 4$ ms, $20/N_\mathrm{C} = 5$ is odd, so successive transitions alternate between even and odd blocks; consequently, ordinary even/odd splitting no longer guarantees a transition-free path. A dedicated 5-block-periodic routing can still confine all transition-affected blocks to the same path, but this results in an unbalanced distribution, with one path accumulating 40 % of the blocks and the other 60 %. Finally, when $N_\mathrm{C}$ is not a divisor of 20, successive data-bit transitions do not occur at a fixed spacing in terms of coherent blocks, and no nontrivial fixed two-path routing can guarantee that all transition-affected blocks remain confined to one path. Therefore, when non-coherent integration is used, only $N_\mathrm{C} = 2$, 5, or 10 are considered in this study.

The main drawback of this method is that is requires roughly twice the amount of input samples to be stored and to be processed, since for example if $N_\mathrm{C} = 10$ and $N_\mathrm{NC} = 1$, then 21 ms of input samples must be stored and processed.

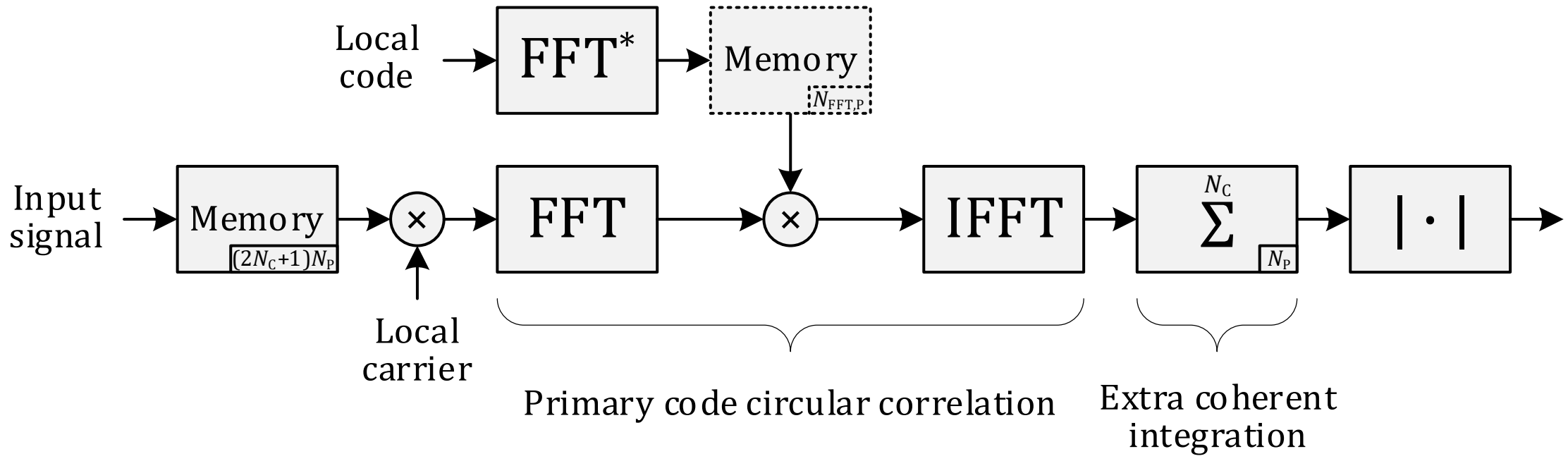

**Figure 4.4** Architecture for extended coherent integration for GPS L1 C/A via alternate block method without non-coherent integration

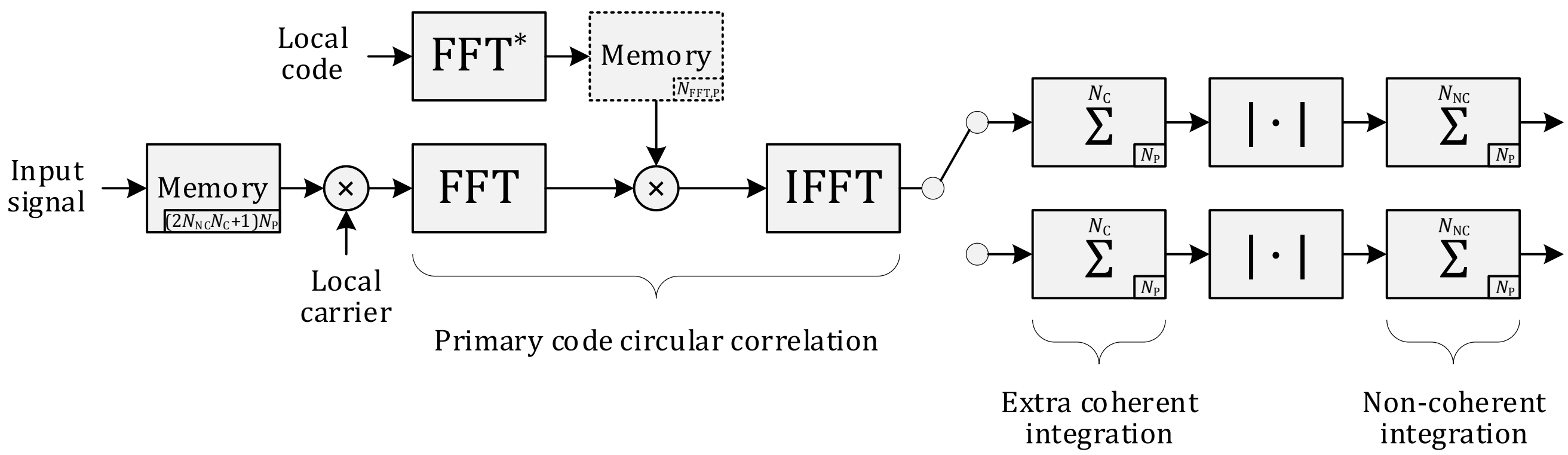

**Figure 4.5** Architecture for extended coherent integration for GPS L1 C/A via alternate block method with non-coherent integration

*4.1.3.2. Full-bits method with* $N_C \leq 20$

To extend the coherent integration beyond 10 ms without loss due to data-bit transitions, the integration time must be increased directly to the full data-bit duration of 20 ms, using the full-bits method proposed by Psiaki (2001). In this approach, 20 parallel branches start their coherent accumulations at 1 ms-spaced offsets and each integrates over 20 ms, as shown in **Figure 4.6**. Among the 20 branches, one is guaranteed to have its integration boundaries aligned with the data-bit boundaries and is therefore free of transitions, while all other branches cross a bit boundary, as illustrated in **Figure 4.3**.

The resulting coherent outputs from each branch can then be accumulated non-coherently over successive 20 ms intervals if required, as shown in **Figure 4.6**.

Note that it is also possible to use this method for $N_\mathrm{C}$ lower than 20, where $N_\mathrm{C}$ accumulators are enough. Without non-coherent integration, any $N_\mathrm{C}$ can be used, with non-coherent integration, $N_\mathrm{C}$ is limited to divisors of 20, i.e., 1, 2, 4, 5, 10.

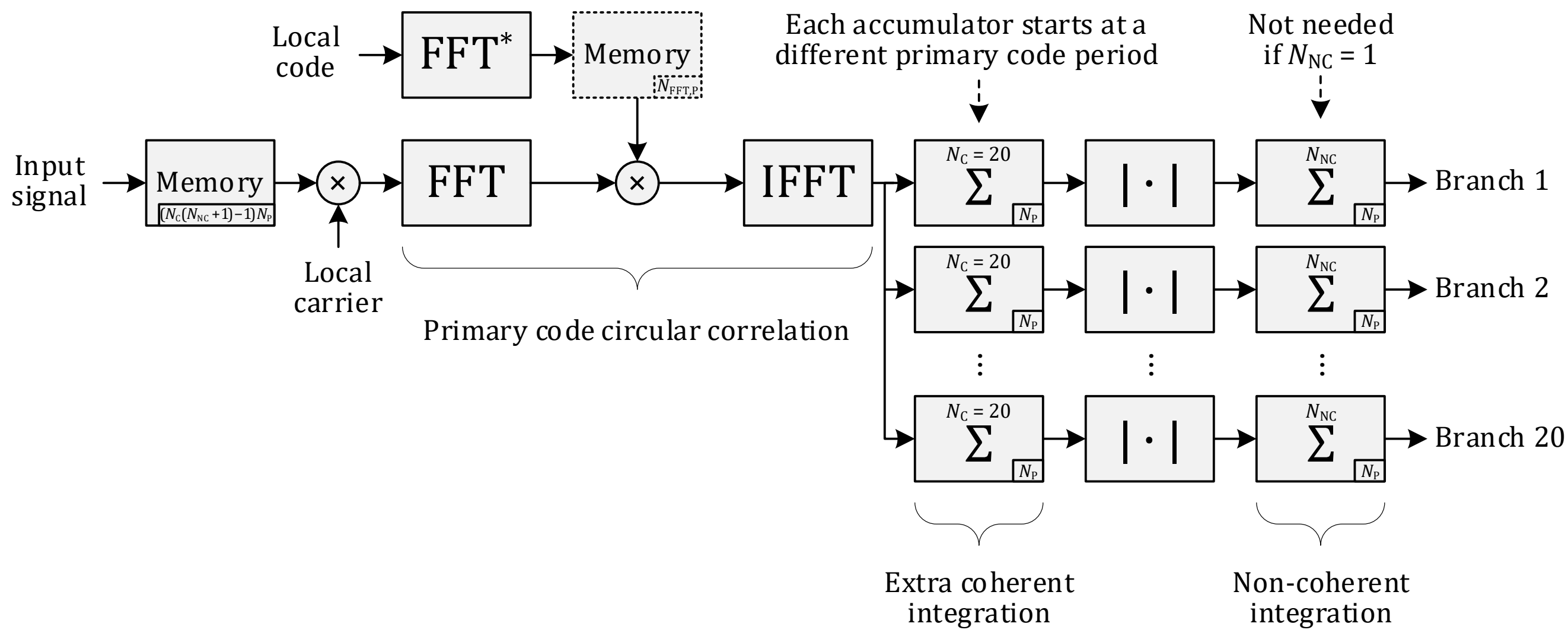

**Figure 4.6** Architecture for extended coherent integration for GPS L1 C/A via full-bits method with $N_\mathrm{C} = 20$

### *4.1.3.3. Full-bits method with $N_C > 20$*

The coherent integration time can be extended beyond the data bit, but since the sign of the 20 ms correlation results is unknown, all the combinations must be tested. This grows extremely rapidly since to integrate coherent over $N_\mathrm{B}$ bits (which would give a coherent integration time of 20 $N_\mathrm{B}$ ms), there are $2^{N_\mathrm{B}-1}$ combinations. The raw approach to this would be either to have $2^{N_\mathrm{B}-1}$ accumulators for each of the 20 branches as shown in **Figure 4.7**, which is the fastest approach but would require a lot of memory ($20 \times 2^{N_\mathrm{B}-1}$ accumulators); or to have a single accumulator for each of the 20 branches and test the $2^{N_\mathrm{B}-1}$ combinations one after each other, which reduces the amount of memory but increases the processing time. In this latter case, the $N_\mathrm{B}$ outputs from the first stage of extra coherent integration should be either stored (requiring 20 memories) as shown in **Figure 4.8**, or recomputed as shown in **Figure 4.9**. Note that in the serial case, a more efficient way is to compute a new combination from a previous one using Gray code order, as explained in Section 4.1.4.1.

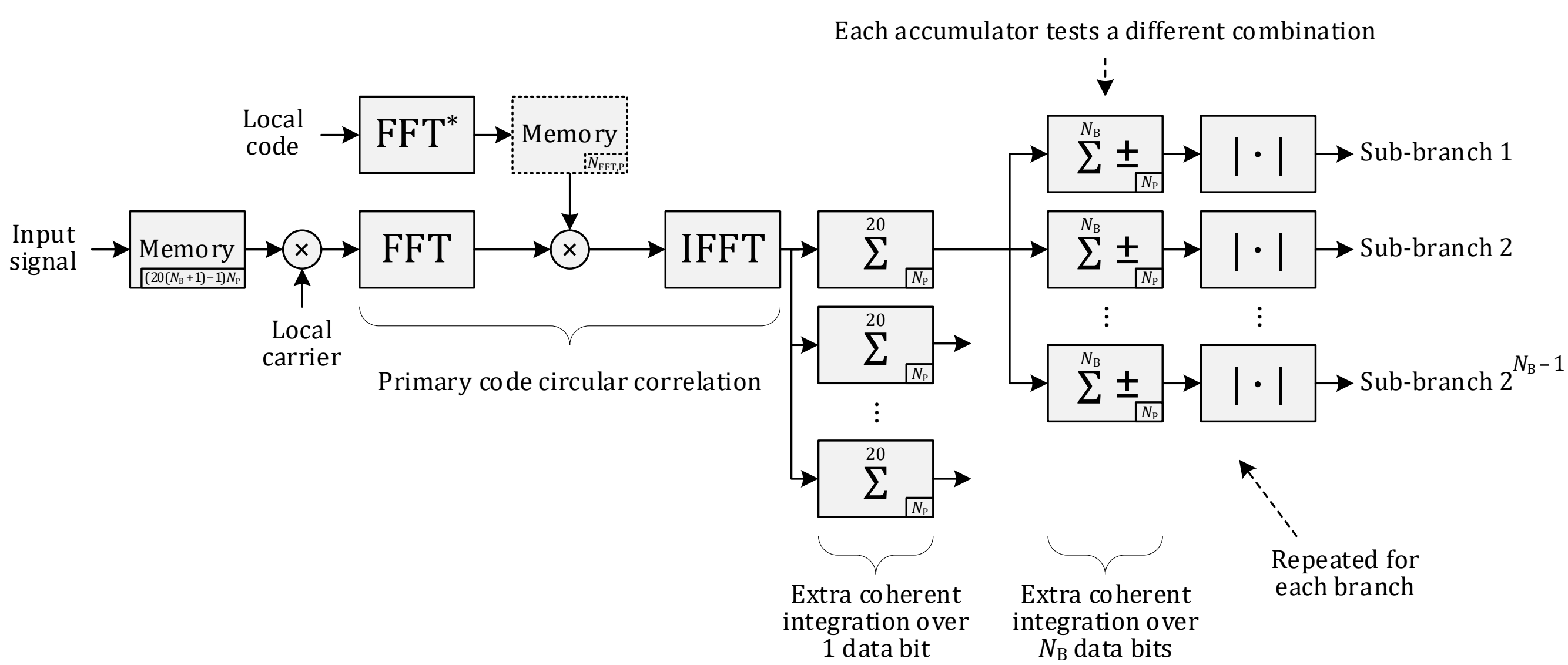

**Figure 4.7** Architecture for extended coherent integration for GPS L1 C/A via full-bits method with $N_\mathrm{C} > 20$ and parallel extra coherent integration

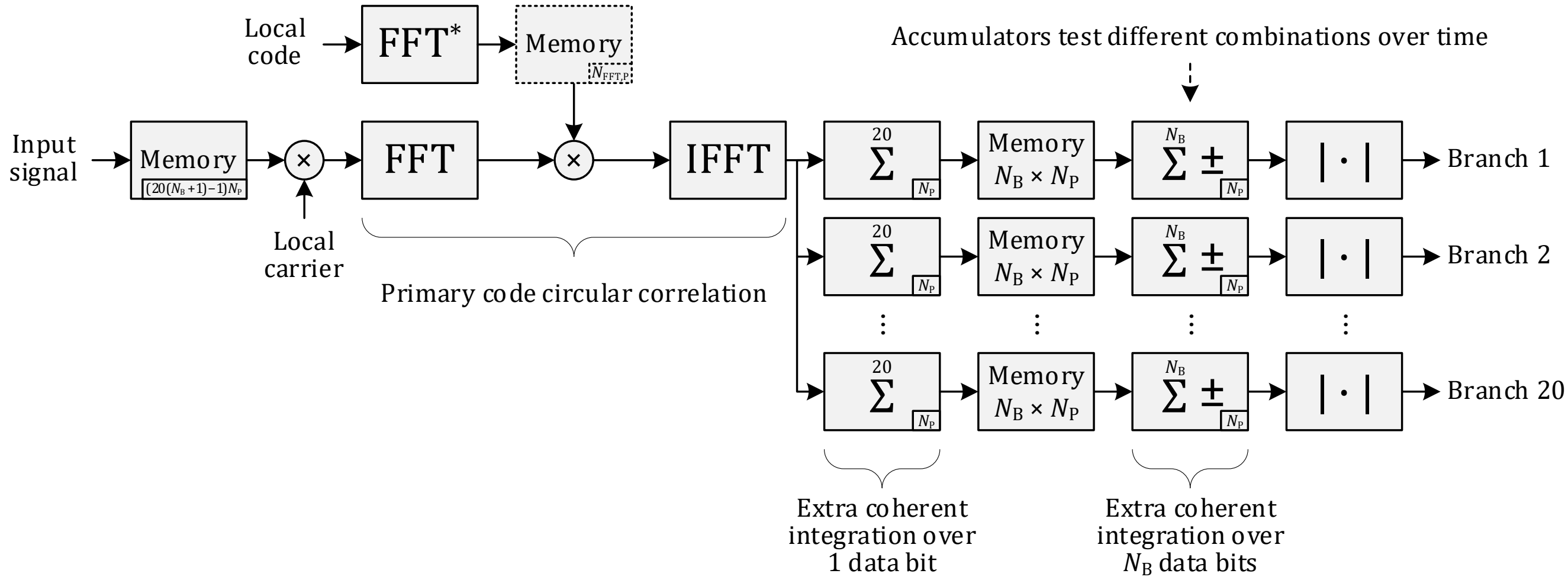

**Figure 4.8** Architecture for extended coherent integration for GPS L1 C/A via full-bits method with $N_C > 20$ and serial extra coherent integration with intermediate memory

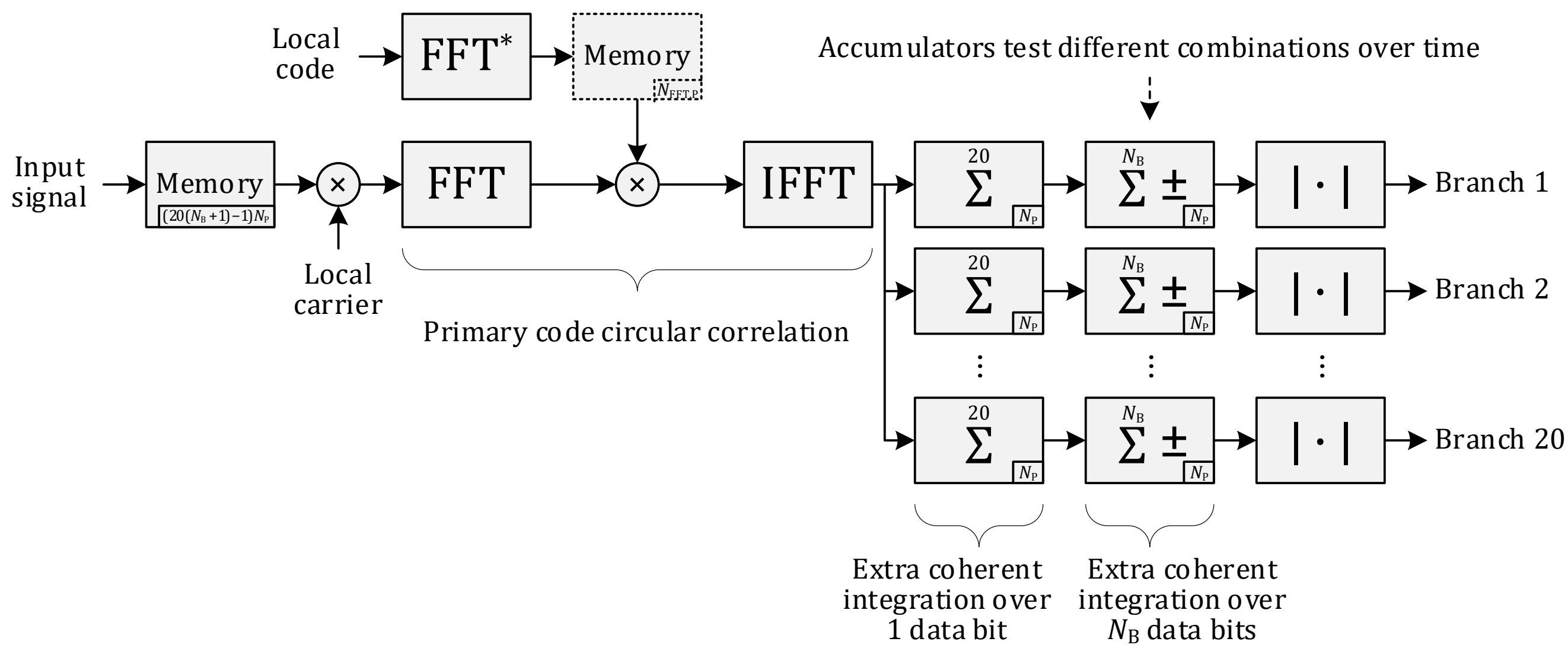

**Figure 4.9** Architecture for extended coherent integration for GPS L1 C/A via full-bits method with $N_C > 20$ and serial extra coherent integration without intermediate memory

*4.1.3.4. Average bit-transition loss method*

Finally, when a sufficiently long non-coherent integration is performed, another possibility is to use an odd coherent integration time that is not a divisor of the 20 ms navigation-bit duration (i.e., $N_C = 3$, 7, 9, 11, 13, 17, or 19) and perform non-coherent integration on all results, as shown in **Figure 4.10**. In this case, the position of the navigation-bit transition varies across successive coherent integrations as shown in **Figure 4.3**, such that its effect can be characterized statistically over a sufficiently large number of non-coherent accumulations. Rather than explicitly identifying or avoiding these transitions, the resulting degradation can be modeled as an average implementation loss, as described by (van Diggelen, 2009, Section 6.6.2). This approach provides flexibility in the selection of the coherent integration time and the simple implementation, at the expense of a longer total integration time compared to the other methods due to an additional loss. The corresponding average losses for the coherent integration times considered in this work are summarized in **TABLE 4.2**.

Note that it is possible to select a coherent integration time that is a divisor of 20, but in this case the average loss is not predictable in advance because the bit transition can happen at different times in the coherent integration window (leading to different possible losses) but it will happen always at the same time during a processing (see e.g. the rows 4 ms in **Figure 4.3**, where one case has always no loss while the other has another loss). Therefore, they could be considered if desired selecting a loss in the link budget, but we do not consider them in this study.

Note also that the non-coherent integration is not required, performing the detection just on single window, but in this case we may or may not have a loss, and the detection could not happen at all (but it can be a strategy to tolerate to miss few satellites in order to reduce the complexity).

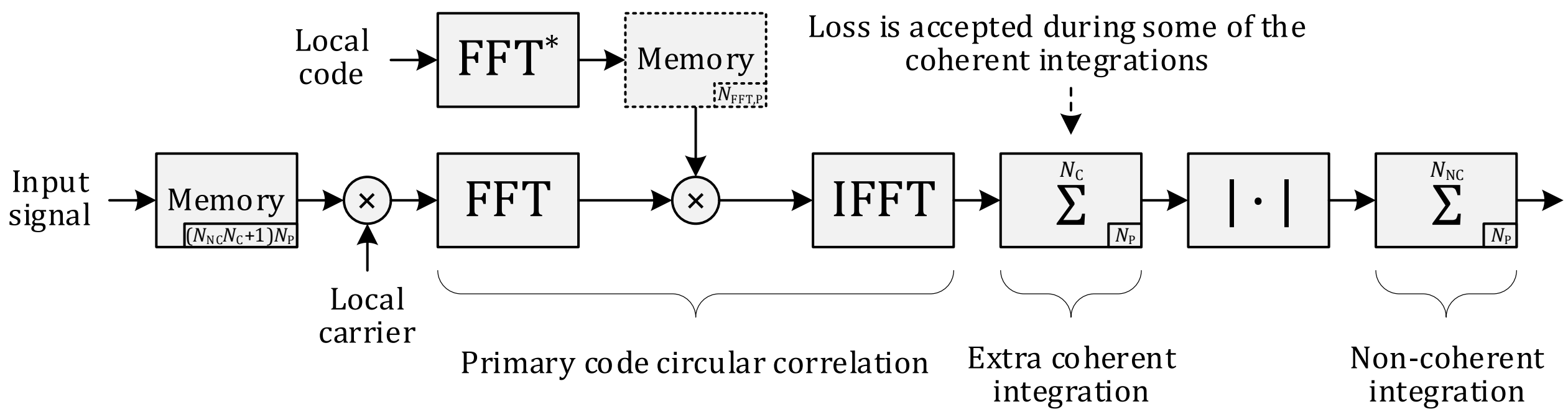

**Figure 4.10** Architecture for extended coherent integration for GPS L1 C/A via average bit-transition method

**TABLE 4.2** Average bit alignment loss as a function of the coherent integration time for GPS L1 C/A

| Coherent integration time (ms) | 3 | 7 | 9 | 11 | 13 | 17 | 19 |
|---|---|---|---|---|---|---|---|
| Effective integration | 58/60 | 128/140 | 160/180 | 190/220 | 218/260 | 268/340 | 290/380 |
| Average bit alignment loss (dB) | −0.29 | −0.78 | −1.02 | −1.27 | −1.53 | −2.07 | −2.35 |

### *4.1.4. Extending coherent integration for modern signals (pilot channel + overlay code)*

Modern GNSS signals differ fundamentally from GPS L1 C/A because one primary code period generally coincides with one navigation data bit or one secondary code chip. Consequently, every additional coherent accumulation introduces a new unknown sign (Borio et al., 2008), which must be compensated properly for full accumulation, as illustrated in **Figure 4.11**. The strategy for the extra coherent integration depends on the signal structure and on the number of coherent accumulations $N_C$ desired. The main strategies are explained hereafter.

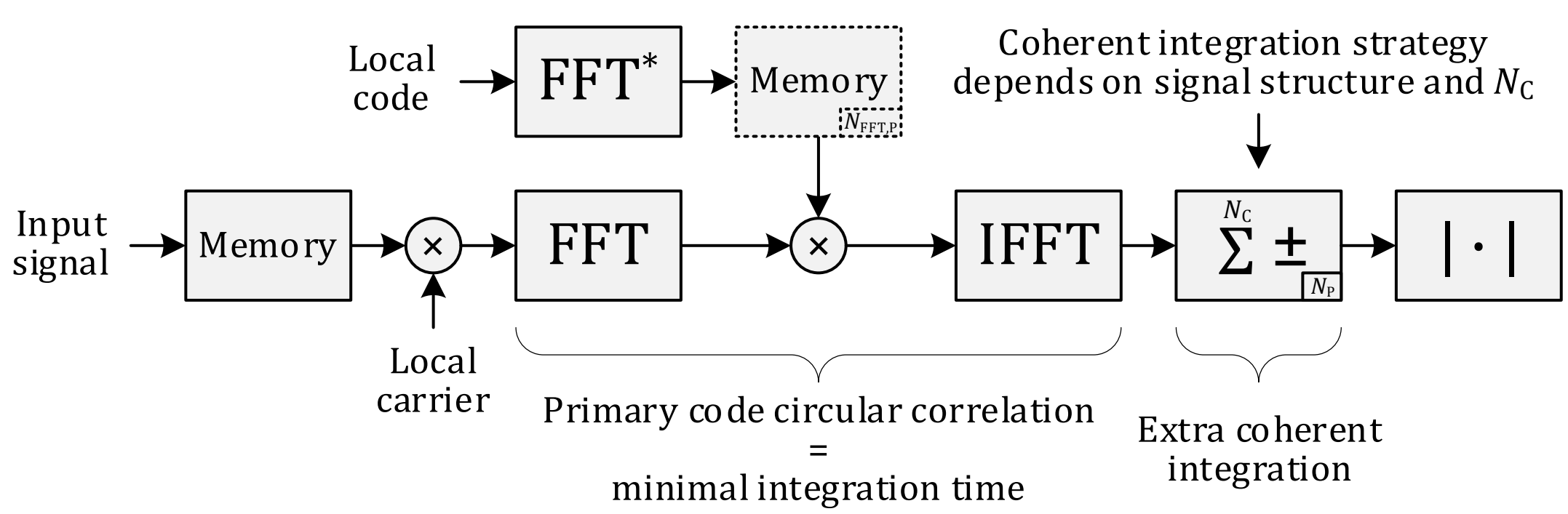

**Figure 4.11** Generic architecture for extended coherent integration

#### *4.1.4.1. Multi-hypothesis approach for small $N_C$*

For short coherent integration intervals, all possible sign combinations may be tested, leading to an exponential growth of the number of combinations of $2^{N_C-1}$, where $N_C$ is the number of coherently accumulated primary code periods. Therefore, this approach remains practical only for a limited number of coherent accumulations, but is particularly well suited to signals having a very strong power, such as Pulsar X1, where few extra coherent accumulations of such strong signal can be equivalent to long coherent integration time for MEO GNSS signals; or to signals employing long primary code periods, such as GPS L1C and BeiDou B1C with their 10 ms code period, where relatively long integration time can be achieved with only few extra coherent accumulations (Galileo E1 also benefits from this approach to a lesser extent owing to its intermediate 4 ms code period).

The raw approach to this would be either to have $2^{N_\mathrm{C}-1}$ accumulators as shown in **Figure 4.12**, which is the fastest approach but requires significant amount of memory; or to have a single accumulator and test the $2^{N_\mathrm{C}-1}$ combinations one after another, which reduces the amount of memory but increases the processing time. In this latter case, the $N_\mathrm{C}$ outputs from the primary code correlation should be either stored (requiring $N_\mathrm{C}$ memories) as shown in **Figure 4.13**, or recomputed as shown in **Figure 4.14**.

In the serial case, while a raw approach would require $N_\mathrm{C}$ accesses per combination for $2^{N_\mathrm{C}-1}$ combinations, i.e. a total of $N_\mathrm{C}\,2^{N_\mathrm{C}-1}$ accesses, there exists a recursive approach that uses the previous result to compute the next one. The first combination required $N_\mathrm{C}$ accesses, then selecting a combination that differs by only one bit (e.g. from $+++$ to $++-$, then to $+--$, then to $+-+$, with $N_\mathrm{C}=3$), only one additional access is required for each of the $2^{N_\mathrm{C}-1}-1$ remaining combinations, which gives a total of $N_\mathrm{C}+2^{N_\mathrm{C}-1}-1$ accesses. Such recursive algorithm exploits Gray code order, and is discussed in detail in (Wang et al., 2014; Leclère et al., 2026).

Up to this point, only coherent integration has been considered. If additional sensitivity is required, non-coherent integration can be applied across successive coherently integrated blocks. Because the unknown sign pattern may differ from one coherent block to the next, the maximizing sign combination must be selected independently for each block. A direct implementation could evaluate all possible combinations across successive blocks before detection, but this is unnecessary because the maximization is separable. For each code-delay hypothesis, the maximum magnitude across all sign combinations is therefore selected independently for each coherent block, and these per-block maxima are then accumulated non-coherently. For example, for two consecutive blocks $a$ and $b$, $\max_{i,j}\left(|a_i|+|b_j|\right)=\max_i(|a_i|)+\max_j\left(|b_j|\right)$. This avoids an additional combinatorial growth across non-coherent blocks and leads to the hybrid architectures shown in **Figure 4.15** through **Figure 4.17** (Qiu et al., 2022).

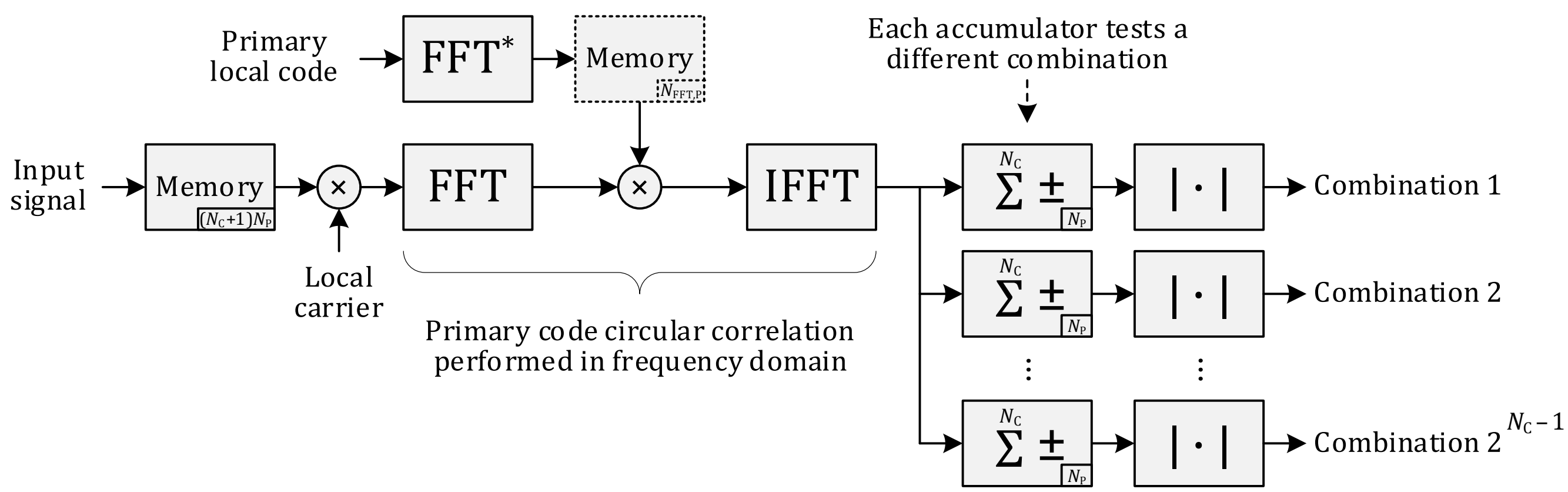

**Figure 4.12** Architecture for extended coherent integration for other GNSS signals via multi-hypothesis evaluation and parallel extra coherent integration (small $N_\mathrm{C}$)

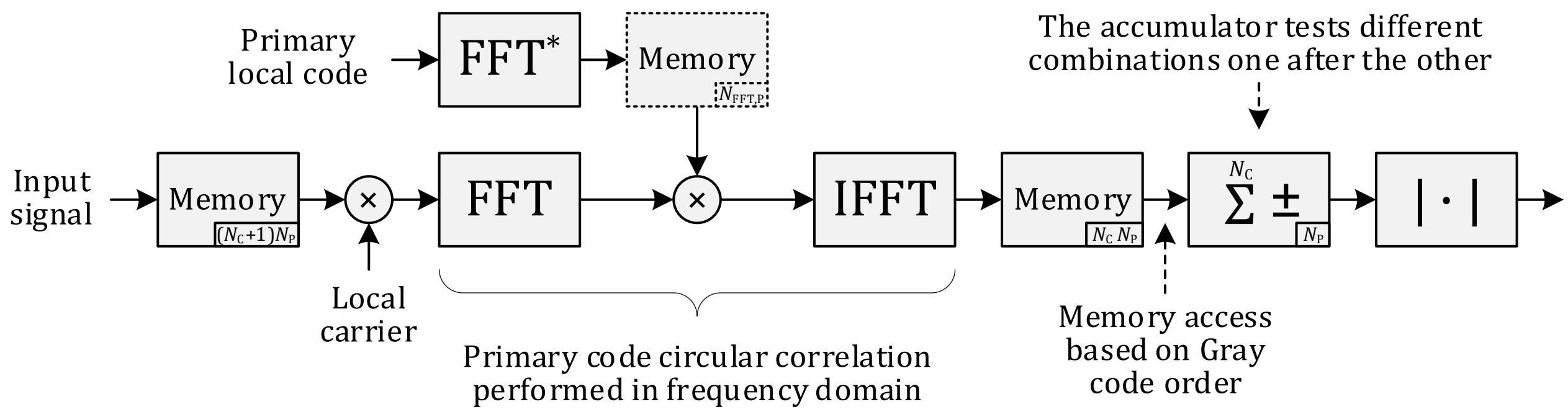

**Figure 4.13** Architecture for extended coherent integration for other GNSS signals via multi-hypothesis evaluation and serial extra coherent integration with intermediate memory (small $N_\mathrm{C}$)

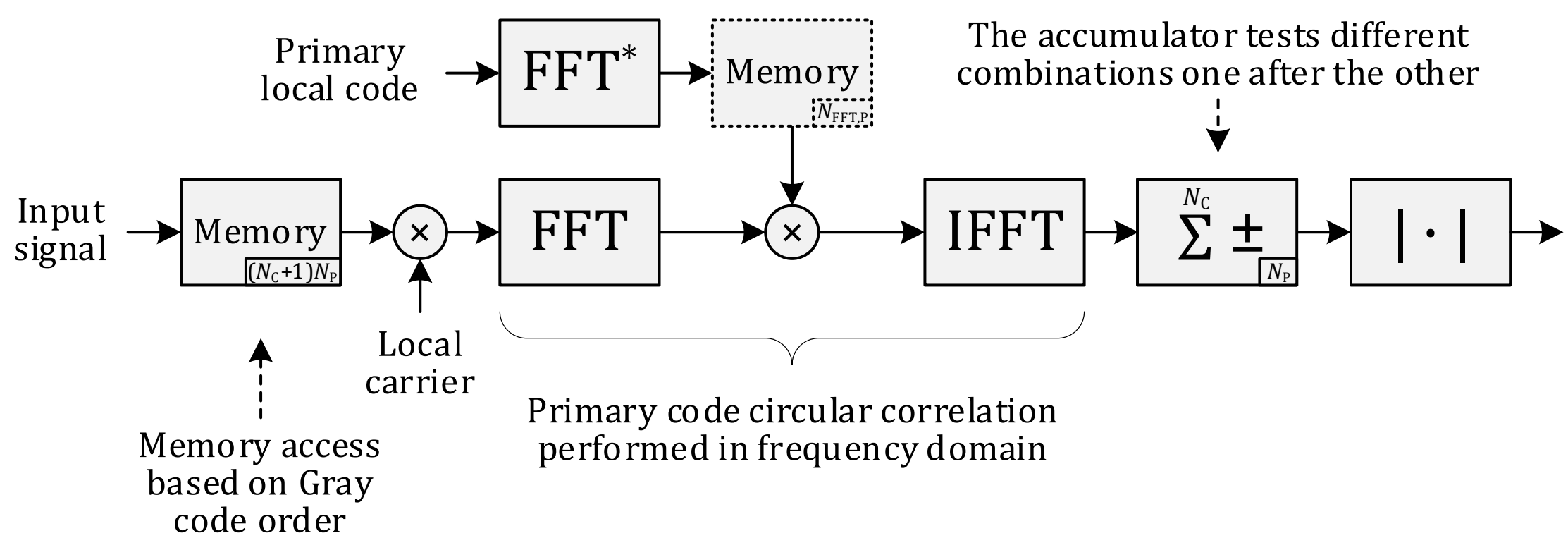

**Figure 4.14** Architecture for extended coherent integration for other GNSS signals via multi-hypothesis evaluation and serial extra coherent integration without intermediate memory (small $N_\mathrm{C}$)

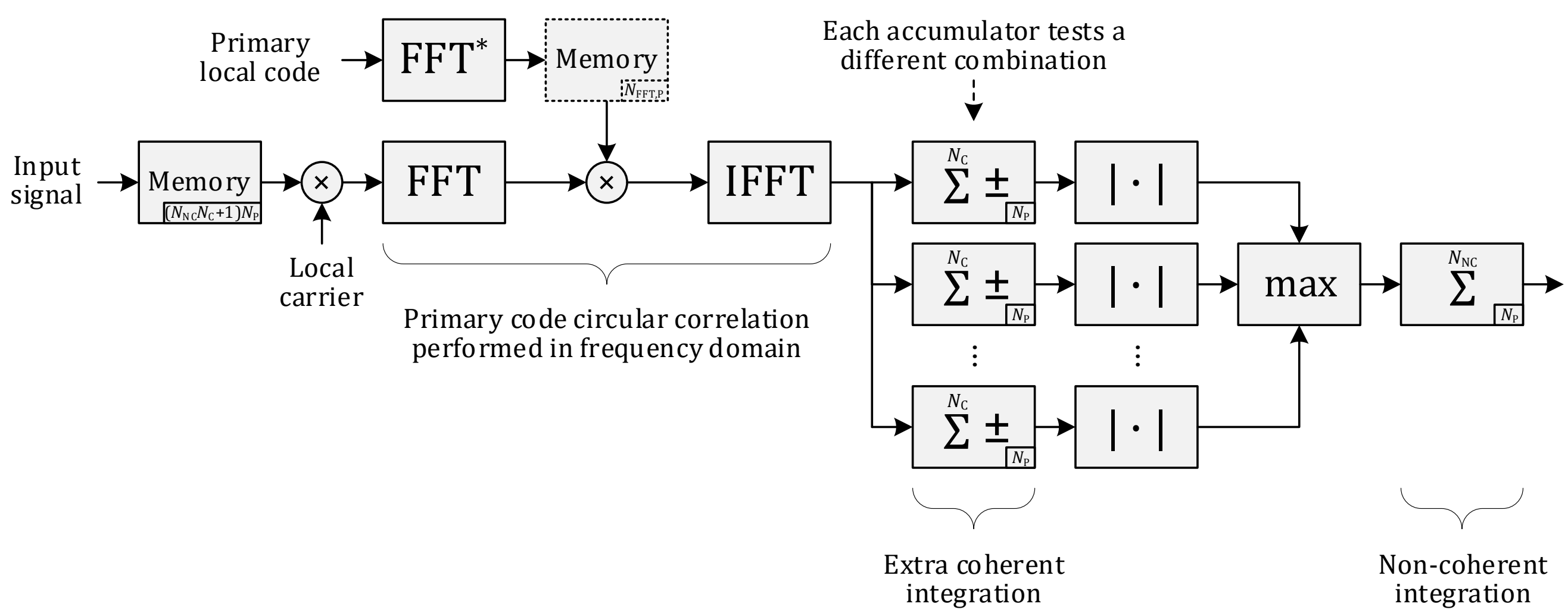

**Figure 4.15** Architecture for extended coherent integration for other GNSS signals via multi-hypothesis evaluation, parallel extra coherent integration and non-coherent integration (small $N_\mathrm{C}$)

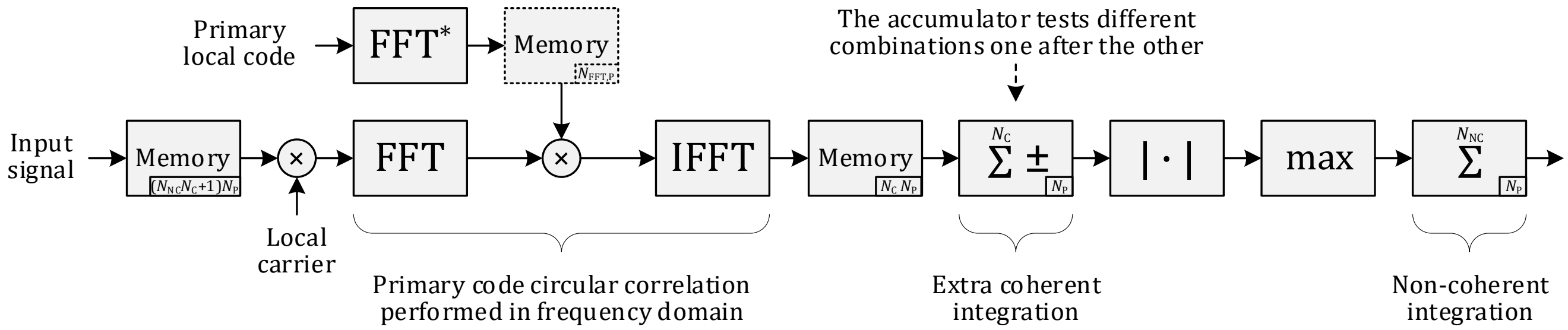

**Figure 4.16** Architecture for extended coherent integration for other GNSS signals via multi-hypothesis evaluation, serial extra coherent integration with intermediate memory, and non-coherent integration (small $N_C$)

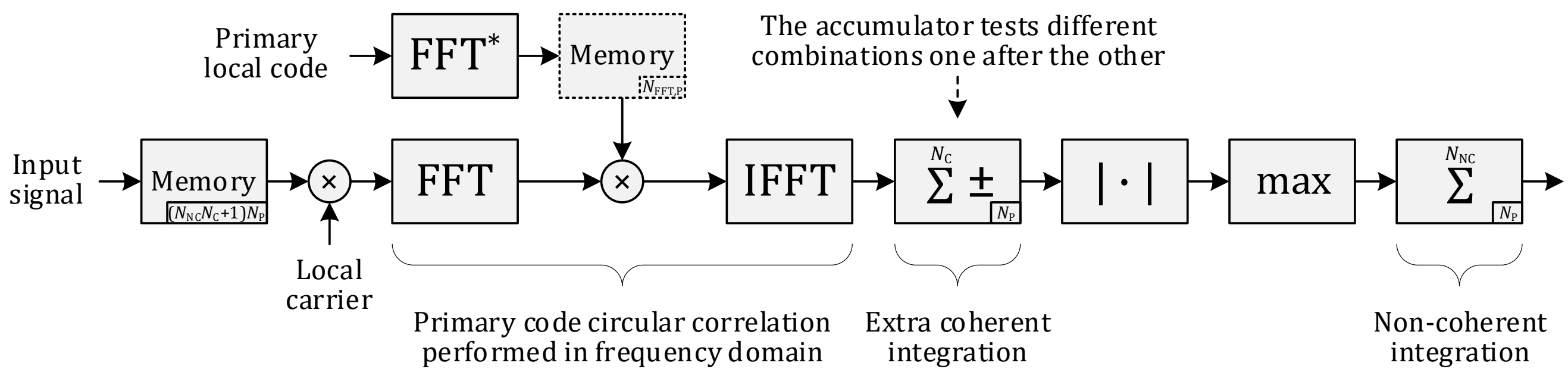

**Figure 4.17** Architecture for extended coherent integration for other GNSS signals via multi-hypothesis evaluation, serial extra coherent integration without intermediate memory, and non-coherent integration (small $N_C$)

*4.1.4.2. Secondary code correlation for intermediate and large $N_C$*

As seen in the previous section, when the required coherent integration interval increases, exhaustive hypothesis testing rapidly becomes impractical due to the exponential growth that will become larger than the number of possible delays. In this case, the receiver should instead correlate the successive coherent correlations with the known secondary code, either partially or completely.

A full secondary code correlation uses $N_S$ primary code periods or more and still tests the $N_S$ delays; whereas a partial secondary code correlation uses $N_C$ primary code periods (with $N_C < N_S$), and still test the $N_S$ delays. As with the previous architectures, this can be implemented in a parallel way duplicating branches, as shown in **Figure 4.18** and **Figure 4.19**, in a serial way, as shown in **Figure 4.20** and **Figure 4.21**, or in another parallel way using a frequency-domain implementation of the secondary code correlation, as shown in **Figure 4.22** and **Figure 4.23**.

Previous work showed that frequency-domain secondary-code correlation introduces non-negligible hardware resources for relatively short secondary codes, for which parallel or serial time-domain correlation can be more efficient (Leclère, 2017). Storing the intermediate primary-code correlation vectors can also incur a substantial memory cost, making recomputation attractive in memory-constrained hardware implementations. Conversely, software receivers typically have more abundant memory, making frequency-domain processing and intermediate buffering more attractive. Therefore, **Figure 4.18** to **Figure 4.21** are considered for the hardware model, whereas **Figure 4.22** and **Figure 4.23** are considered for the software model.

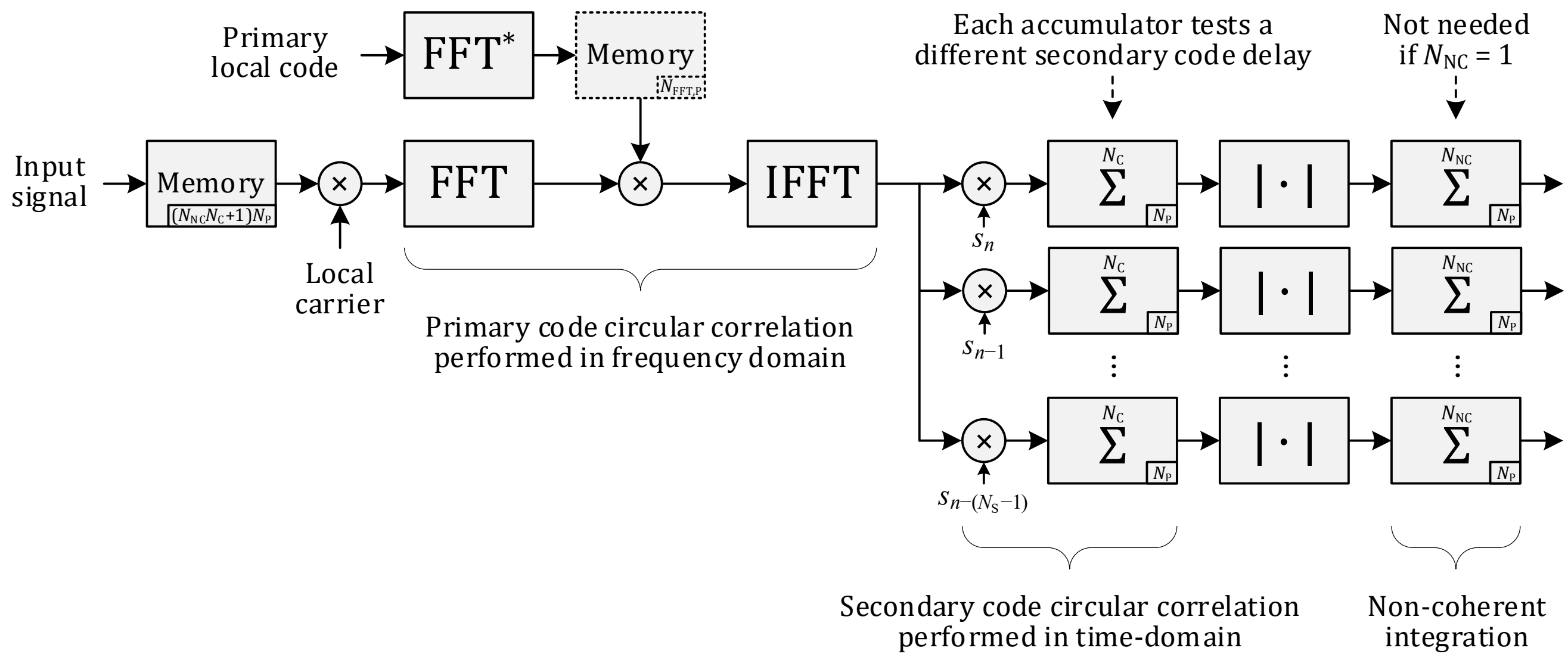

**Figure 4.18** Architecture for extended coherent integration for other GNSS signals via time-domain parallel partial secondary code correlation (intermediate $N_C$)

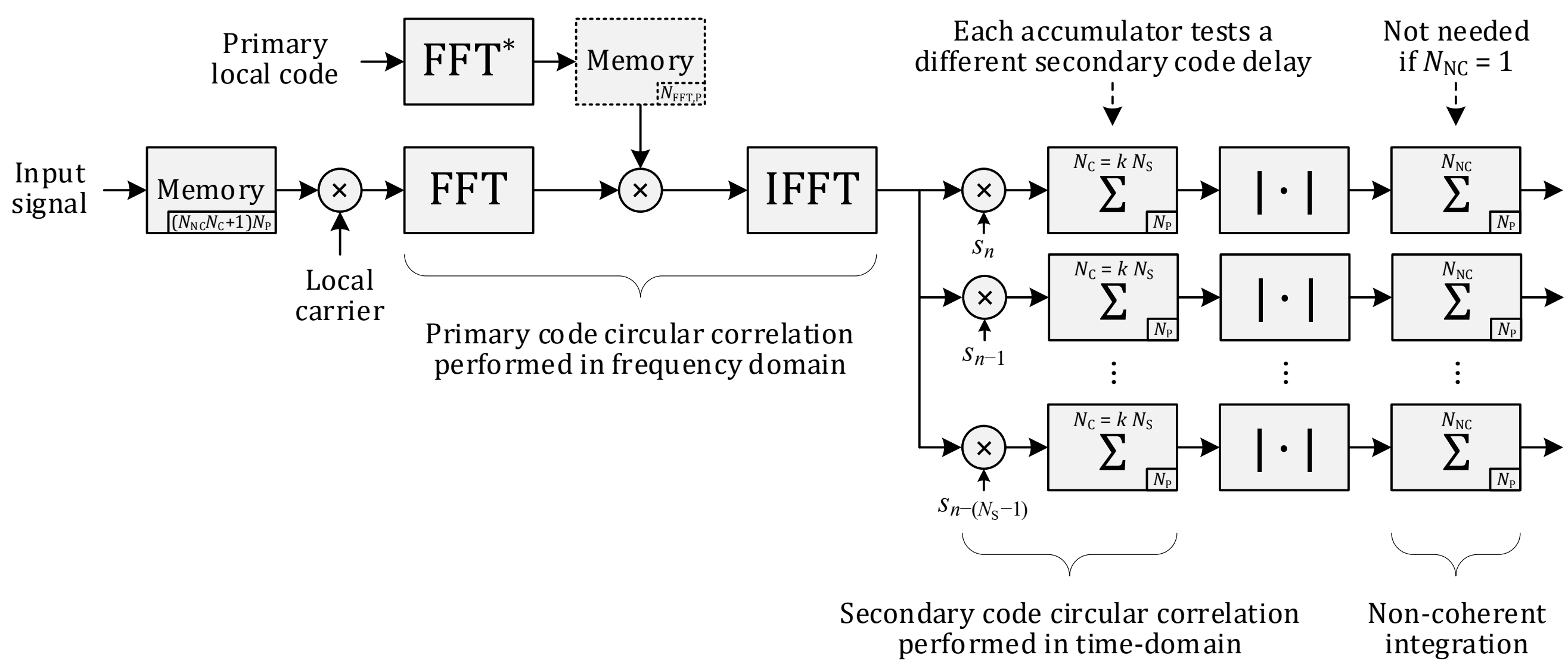

**Figure 4.19** Architecture for extended coherent integration for other GNSS signals via time-domain parallel full secondary code correlation (large $N_C$)

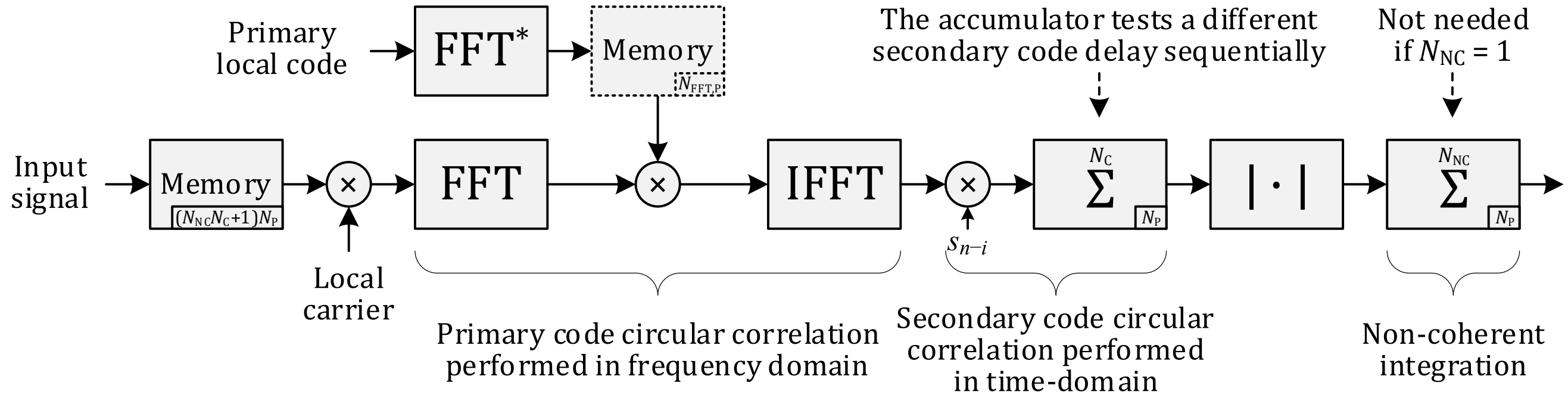

**Figure 4.20** Architecture for extended coherent integration for other GNSS signals via serial partial secondary code correlation (intermediate $N_C$)

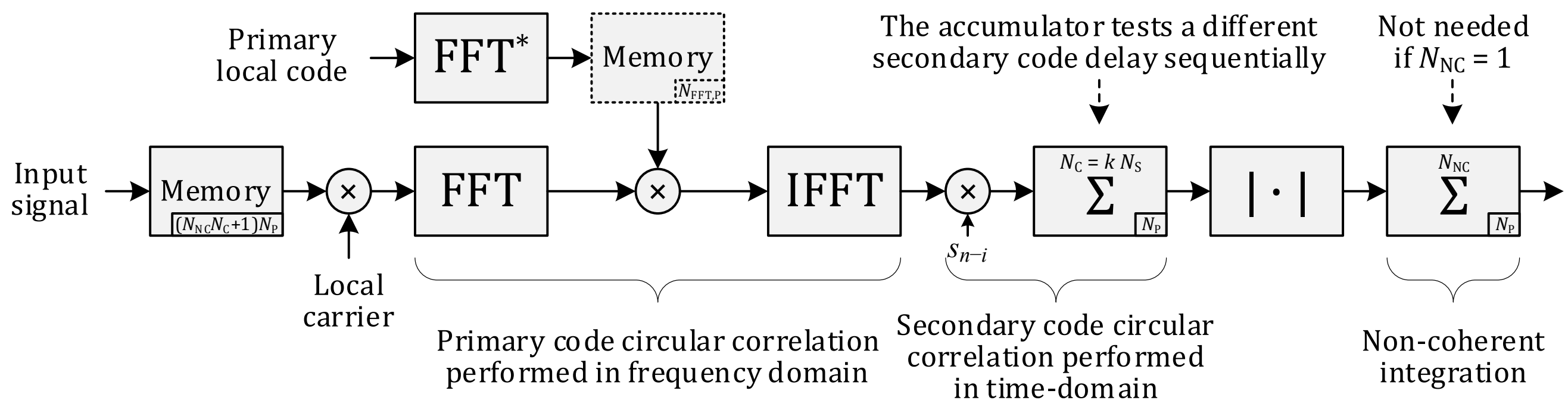

**Figure 4.21** Architecture for extended coherent integration for other GNSS signals via serial full secondary code correlation (large $N_C$)

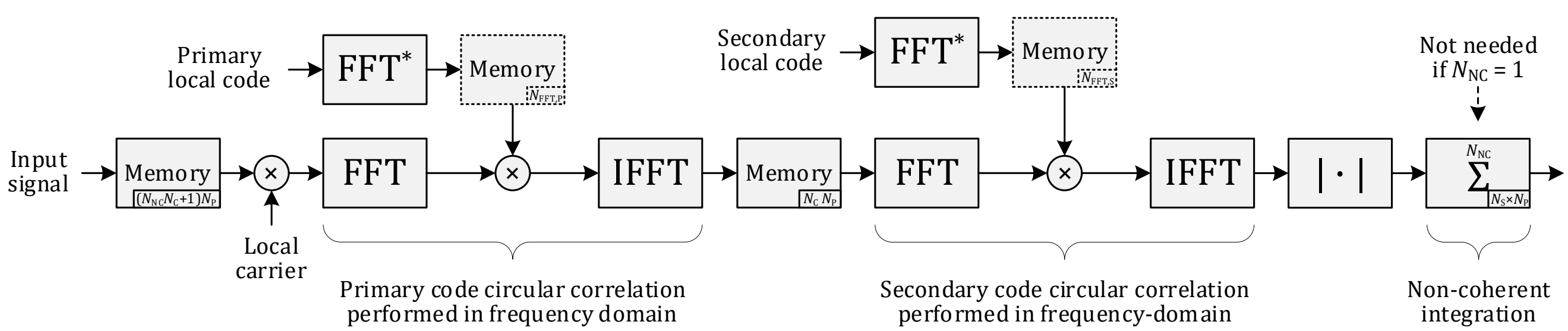

**Figure 4.22** Architecture for extended coherent integration for other GNSS signals via frequency-domain parallel partial secondary code correlation (intermediate $N_C$)

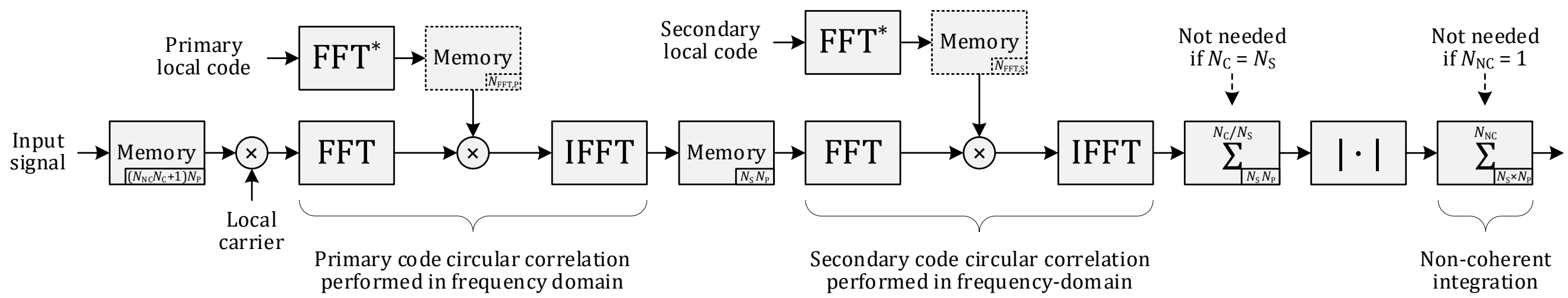

**Figure 4.23** Architecture for extended coherent integration for other GNSS signals via frequency-domain parallel full secondary code correlation (large $N_{\mathrm{C}}$)

### 4.2. Complexity metrics

Three metrics are considered in this study:

- Software operation count, to represent the computational effort required by a software receiver. The total operation count is defined as the sum of the numbers of real additions and real multiplications. These operations are intentionally assigned equal weight to provide a common measure of computational workload; their actual execution costs depend on the target processor and instruction set.
- Hardware processing time, expressed in clock cycles, to represent the processing effort required by a hardware receiver. A streaming pipelined implementation is assumed, and fixed FFT/IFFT pipeline latencies are neglected in the comparison, as detailed in Appendix B.
- Memory requirement, expressed in bits, since memory constitutes a significant part of the implementation cost of a receiver (van Diggelen, 2021).

For the software operation count, the following assumptions are used:

- A complex multiplication requires 4 real multiplications and 2 real additions.
- A complex addition requires 2 real additions.
- An FFT or IFFT of $N_{\mathrm{FFT}}$ points requires $N_{\mathrm{FFT}} \log_2(N_{\mathrm{FFT}})$ complex additions and $\frac{N_{\mathrm{FFT}}}{2} \log_2(N_{\mathrm{FFT}})$ complex multiplications (Lyons, 2011), corresponding to $3N_{\mathrm{FFT}} \log_2(N_{\mathrm{FFT}})$ real additions and $2\,N_{\mathrm{FFT}} \log_2(N_{\mathrm{FFT}})$ real multiplications.
- The square-root operation is not explicitly included in the magnitude computation because low-complexity approximations can be used. As a simplifying assumption, the magnitude operation is assigned the same arithmetic cost as $x^2 + y^2$, namely two real multiplications and one real addition.

For the memory requirement, the following is considered:

- The ADC resolution is 2-bit IQ, resulting in $B_{\mathrm{ADC}} = 4$bits per complex sample.
- The FFT/IFFT resolution is assumed to be 12 bits per real component, as in **TABLE 4.1**, giving $B_{\mathrm{FFT}} = B_{\mathrm{IFFT}} = 24$ bits per complex sample.
- The FFT memory required is taken from **TABLE 4.1**.
- The resolution after an accumulator is assumed to remain equal to its input resolution, implying that appropriate scaling is applied to prevent natural bit growth. Therefore, $B_{\mathrm{C}} = B_{\mathrm{IFFT}}$ for complex accumulators and $B_{\mathrm{NC}} = B_{\mathrm{C}}/2$ for real non-coherent accumulators.

To keep the main text focused, the detailed expressions used to derive these metrics for each architecture are provided in Appendix B.

### 4.3. Results

The results are evaluated using three implementation metrics: software operation count, hardware processing time, and memory requirement. These metrics cannot always be minimized simultaneously, since reducing computational workload or processing time may require additional memory. To capture this trade-off, a memory-processing product is introduced as a common comparison metric. For hardware receivers, it corresponds to the memory-time product, with processing time expressed in clock cycles, while for software receivers it corresponds to the memory-operation product. These products provide a compact measure of the trade-off between processing requirements and memory usage, analogous to the area-time product commonly used in hardware design. They are used as relative implementation-cost metrics rather than as direct estimates of receiver energy consumption.

The analysis first considers non-coherent-only integration. This case provides a useful baseline because all signals use the same receiver architecture, such that differences in implementation cost result directly from the signal characteristics, acquisition search space, and required number of non-coherent accumulations. Coherent-only and hybrid integration are then considered, for which signal-dependent integration architectures introduce additional processing-memory trade-offs.

For coherent and hybrid integration, several architectures may satisfy the same acquisition-performance requirement while providing different trade-offs between memory and processing cost. To provide a consistent signal-level comparison, the architecture minimizing the memory-time product is selected independently for the hardware comparison, while the architecture minimizing the memory-operation product is selected for the software comparison. The selected architecture may therefore vary with signal, attenuation, assistance condition, and implementation class. This selection is used only to obtain a balanced and consistent comparison between signals; it should not be interpreted as universally optimal, since a receiver subject to specific memory, latency, or processor constraints may favor a different architecture.

#### *4.3.1. Non-coherent integration*

This section evaluates the implementation complexity associated with the non-coherent integration lengths determined in Section 3. **Figure 4.24** presents the memory requirement, hardware processing time and software operation count for the assisted acquisition case, while **Figure 4.25** combines the memory and processing requirements through the memory-processing product introduced in Section 4.3. Since the same non-coherent architecture is used for every signal, these results provide a direct view of how the primary-code characteristics, acquisition search space, and required integration length translate into implementation cost.

The memory results exhibit two distinct regimes. At low attenuation, only a limited number of input primary-code periods must be stored, and the memory is therefore dominated by components whose size is determined by the number of samples per primary code and the corresponding FFT length. This explains the similar memory requirements of Pulsar X1 and GPS L1 C/A, which both use a 1 ms primary code and a 4096-point FFT, while GPS L1C and BeiDou B1C require substantially more memory because of their 10 ms primary codes and 65536-point FFTs. Galileo E1 and BeiDou B1I lie between these two cases.

As attenuation increases, the input-signal buffer progressively becomes the dominant memory contribution. Its size is proportional to $(N_{\mathrm{NC}}+1)N_{\mathrm{P}}$, so the memory requirement increasingly depends on the total number of input samples required to reach the target detection performance rather than only on the primary-code size. Consequently, signals requiring a large number of non-coherent accumulations experience a much faster increase in memory, whereas the shorter integration required by Pulsar X1 under attenuated conditions results in substantially slower memory growth (see **Figure 3.8** and **Figure 3.9** for the integration length and time).

The hardware processing time and software operation count are affected by the same mechanisms but show an even stronger dependence on the primary-code structure. The dominant repeated processing scales with both the number of non-coherent accumulations $N_{\mathrm{NC}}$ and the number of Doppler bins $N_{\mathrm{FB}}$, while the cost of each primary-code correlation increases with the FFT length. A longer primary-code period therefore has competing effects: it provides additional coherent processing gain before non-coherent integration and can reduce $N_{\mathrm{NC}}$, but it also requires a larger FFT and, because the Doppler-bin spacing decreases as coherent integration is extended, increases the number of frequency hypotheses that must be evaluated. The results show that the reduction in $N_{\mathrm{NC}}$ does not compensate for these additional costs for the longer-code signals.

GPS L1 C/A consequently provides the lowest processing cost among the analyzed MEO signals. Its 1 ms primary code requires only a 4096-point FFT and permits a comparatively coarse Doppler grid. GPS L1C and BeiDou B1C, despite requiring fewer non-coherent accumulations because each primary correlation already integrates over 10 ms, require 65536-point FFTs and a correspondingly finer Doppler search. BeiDou B1I retains a short 1 ms primary-code period, but its higher chip rate doubles the number of samples per code relative to L1 C/A, while its lower received signal power requires significantly more integration.

The comparison between Pulsar X1 and GPS L1 C/A shows a different behavior. Under low attenuation, GPS L1 C/A remains less computationally demanding even though Pulsar X1 has a higher nominal received power. Two effects explain this result. First, the aggregate interference generated by the LEO constellation significantly degrades the effective $C/N_0$ of X1 under strong-signal conditions; the degradation considered in this analysis decreases from 12.10 dB at 0 dB attenuation to 7.64 dB at 5 dB and 4.01 dB at 10 dB. Second, even in the assisted case, the larger residual Doppler uncertainty of X1 results in a larger frequency search space than for the MEO signals, increasing the number of frequency bins that must be processed.

As the attenuation increases, the aggregate-interference contribution becomes progressively smaller relative to the thermal-noise floor, and the higher nominal received power of X1 becomes the dominant factor. Because X1 simultaneously retains the same short 1 ms primary-code period and 4096-point FFT as GPS L1 C/A, the resulting reduction in $N_{\mathrm{NC}}$ directly translates into lower processing time, fewer software operations, and eventually lower memory requirements. This evolution explains the crossover between X1 and L1 C/A around 10 dB attenuation and the increasing X1 advantage at higher attenuation.

**Figure 4.25** combines the memory and processing requirements through the memory-processing product, using the memory-time product for hardware receivers and the memory-operation product for software receivers. Both implementation models exhibit the same overall evolution. GPS L1 C/A provides the lowest implementation cost at low attenuation, while Pulsar X1 and L1 C/A become comparable around 10 dB. As attenuation increases, X1 provides the lowest memory-processing product, with the separation from the MEO signals increasing rapidly as the required integration depth grows. At 20 dB attenuation, the memory-time product of X1 is approximately 60 times lower than that of GPS L1 C/A and 145 times lower than that of the best-performing modern MEO signal. At 30 dB, these ratios increase to approximately 100 and 160, respectively.

**Figure 4.26** provides a complementary view of the comparison in **Figure 4.25** by displaying memory requirement against processing time or operation count directly. This representation verifies whether a lower memory-processing product corresponds to simultaneous reductions in both constituent metrics or instead results from a trade-off between them. At 0 dB attenuation, GPS L1 C/A occupies the lowest-complexity region, with substantially lower processing cost than X1 for a similar memory requirement. At 10 dB, X1 and GPS L1 C/A converge to similar implementation points. At 20 dB and 30 dB, X1 moves clearly toward the lower-left region relative to all analyzed MEO signals, indicating that its advantage is not obtained by trading additional memory for reduced processing time: both metrics are reduced simultaneously. A similar evolution is observed for the hardware and software models, showing that the relative signal ranking is not specific to either implementation class.

The non-assisted case exhibits the same general evolution with attenuation as the assisted case, as shown in **Figure 4.27** and **Figure 4.28**. Removing assistance expands the frequency uncertainty for all signals, thereby increasing the number of Doppler hypotheses and consequently the hardware processing time and software operation count. The increase is larger for Pulsar X1 because of its higher orbital dynamics: the considered standalone frequency search space is 76.6 kHz for X1, compared with approximately 16–18 kHz for the analyzed MEO signals. However, this additional frequency-search cost should be considered together with the other dimensions of the acquisition search. For example, GPS L1C and BeiDou B1C have a primary code ten times longer than X1 and L1 C/A, resulting in a correspondingly larger code-delay search and substantially larger FFTs.

Consequently, Pulsar X1 is slightly less favorable relative to the other signals in the non-assisted case than in the assisted case, but the overall conclusions remain unchanged. GPS L1 C/A provides the lowest implementation cost under low attenuation, X1 and L1 C/A become comparable at intermediate attenuation, and X1 provides the lowest cost as attenuation increases. Assistance therefore improves the relative position of X1 by reducing the impact of its wider Doppler uncertainty, without fundamentally changing the observed signal ranking.

The non-coherent-only strategy is primarily used here as a baseline because it provides a common architecture applicable to all signals and therefore isolates the impact of signal characteristics on implementation complexity. At high attenuation, however, the required number of non-coherent accumulations and the resulting observation and processing times become very large (see **Figure 3.8** and **Figure 3.9**). Such an approach would generally not be attractive for a practical receiver operating under strongly attenuated conditions. Instead, extending the coherent integration interval, or combining coherent and non-coherent integration, can provide the required processing gain more efficiently. The following sections therefore investigate coherent-only and hybrid strategies to determine how the relative signal complexity changes when more practical integration approaches are considered.

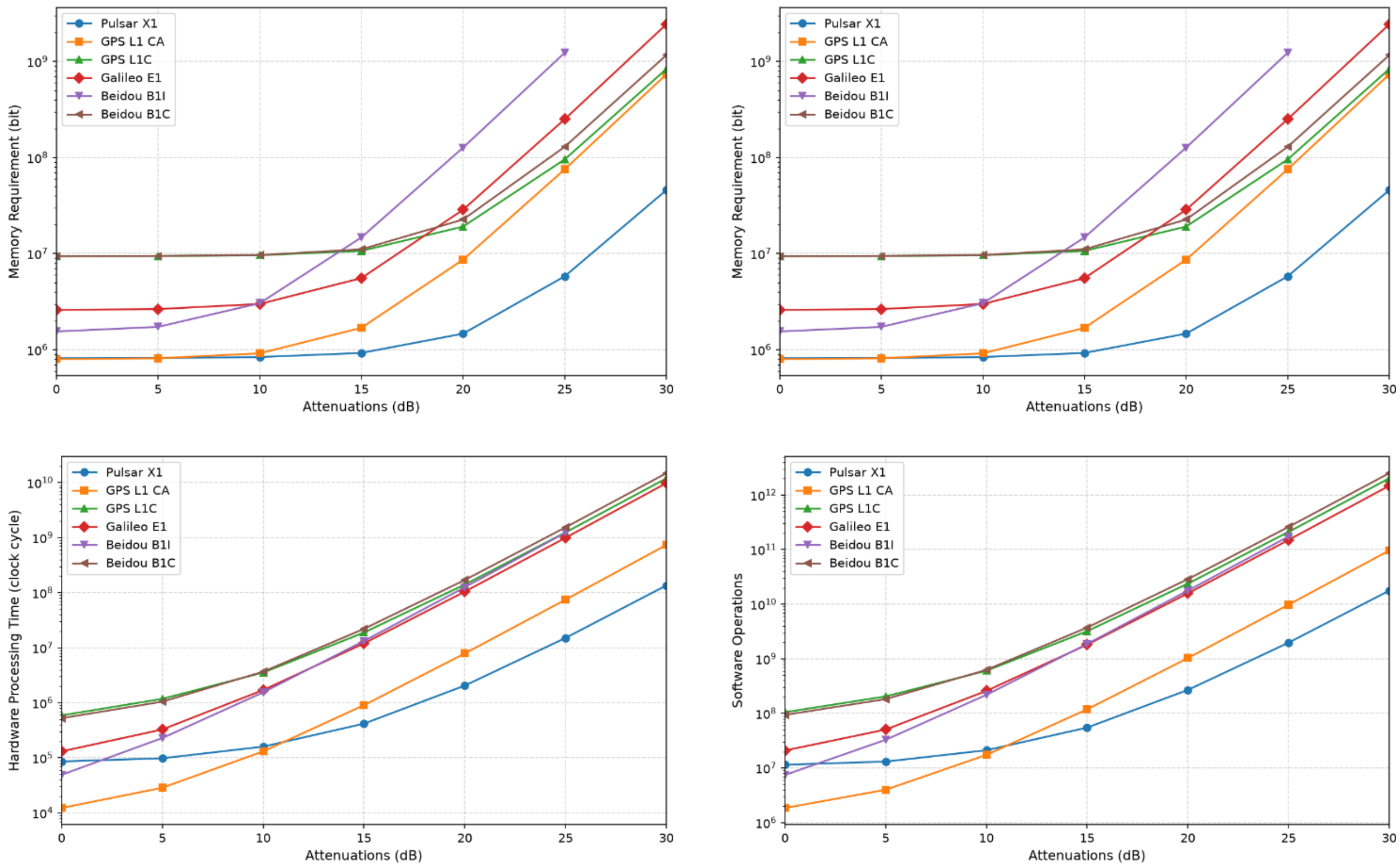

**Figure 4.24** Implementation metrics of non-coherent integration architecture for all signals, assisted case. Left: metrics for hardware receivers. Right: metrics for software receivers

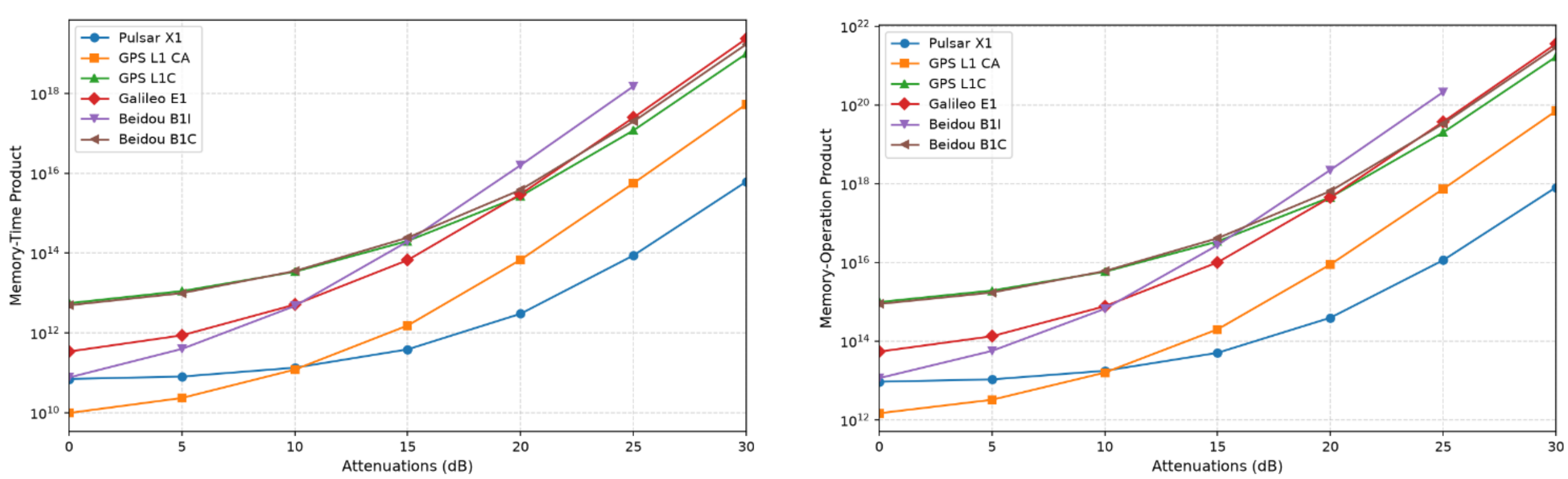

**Figure 4.25** Memory-processing products of non-coherent integration architecture for all signals, assisted case. Left: metrics for hardware receivers. Right: metrics for software receivers

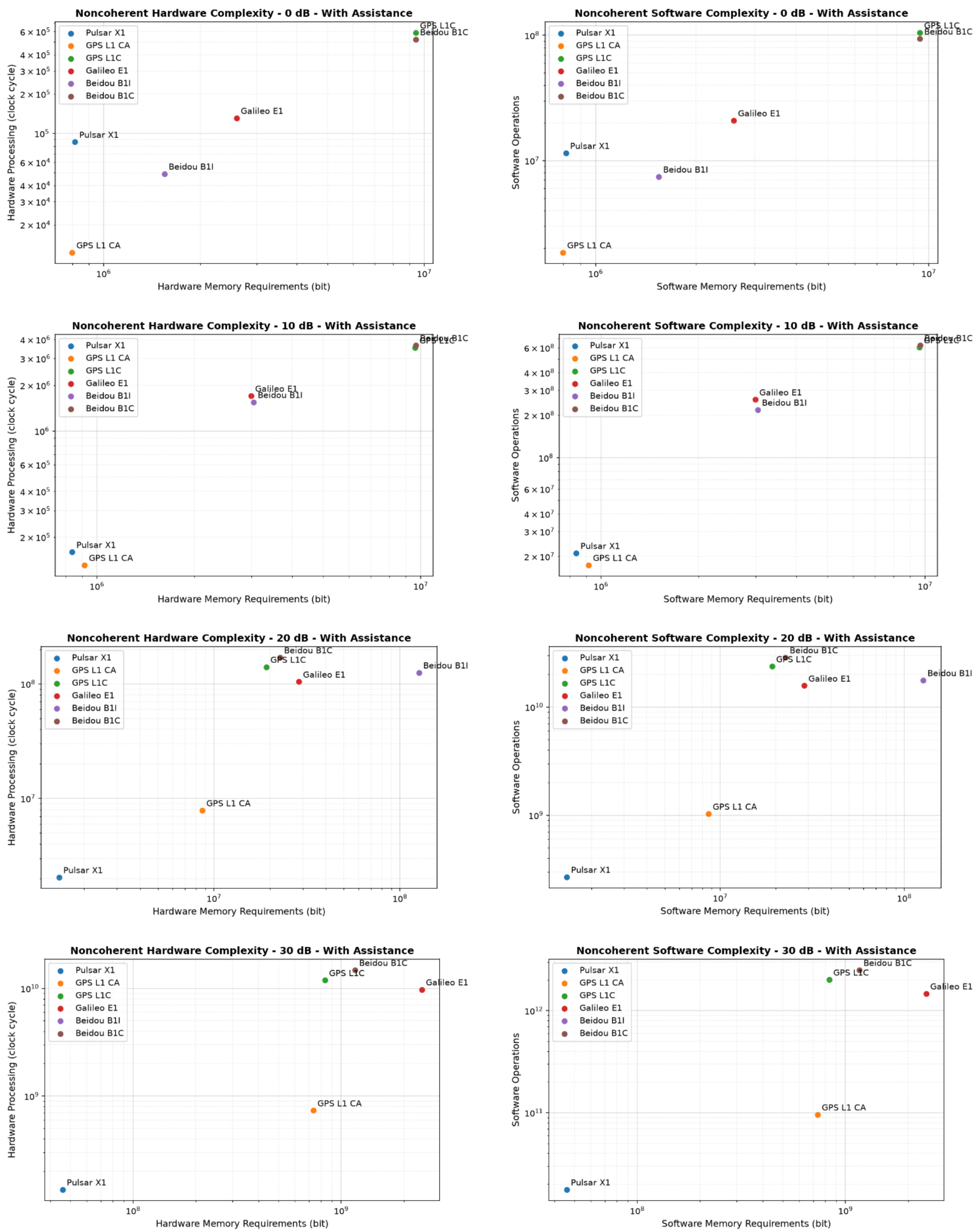

**Figure 4.26** Implementation metrics plane for non-coherent integration, assisted case, for 0, 10, 20 and 30 dB attenuations. Left: metrics for hardware receivers. Right: metrics for software receivers

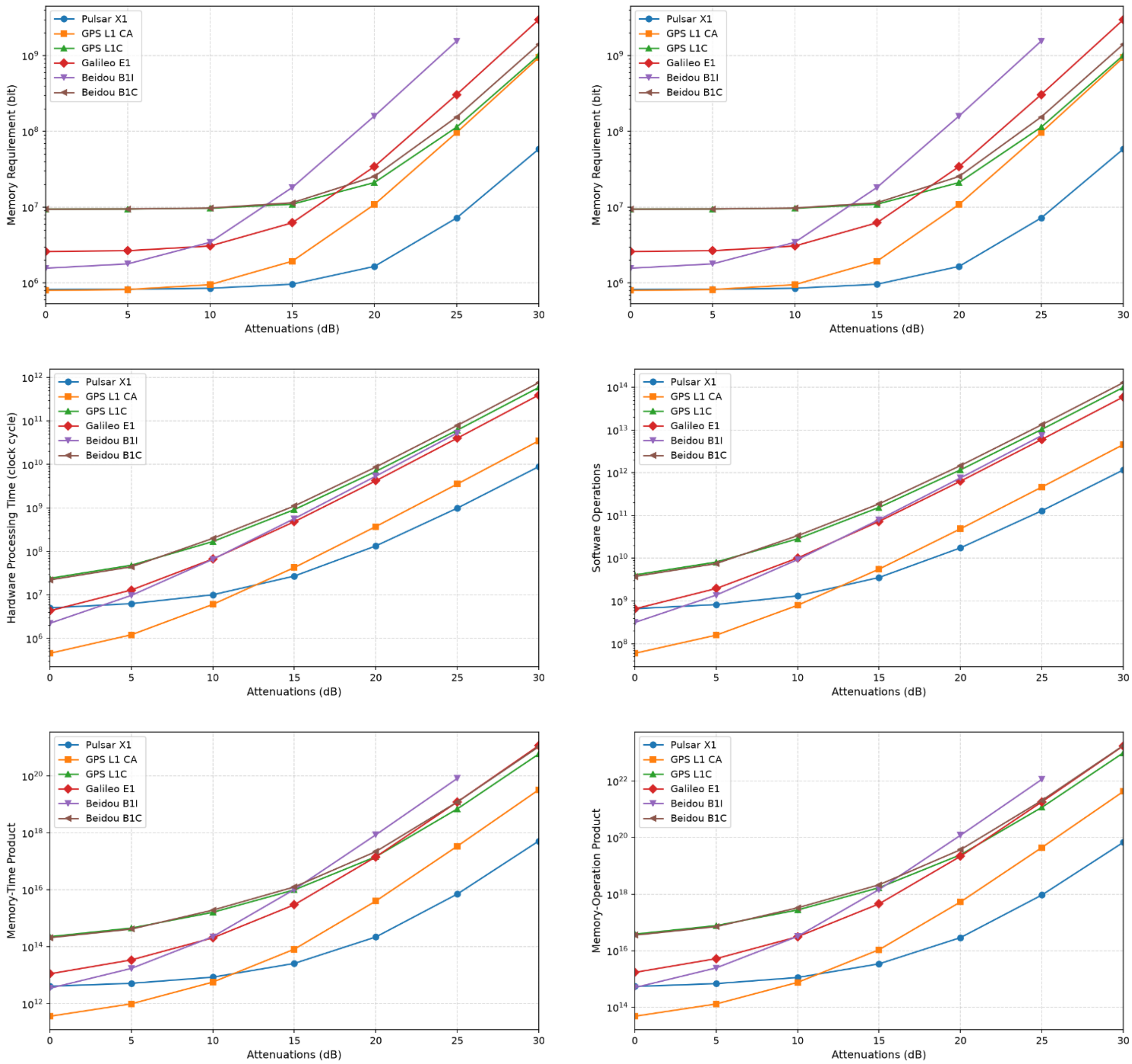

**Figure 4.27** Implementation metrics of non-coherent integration architecture for all signals, non-assisted case. Left: metrics for hardware receivers. Right: metrics for software receivers

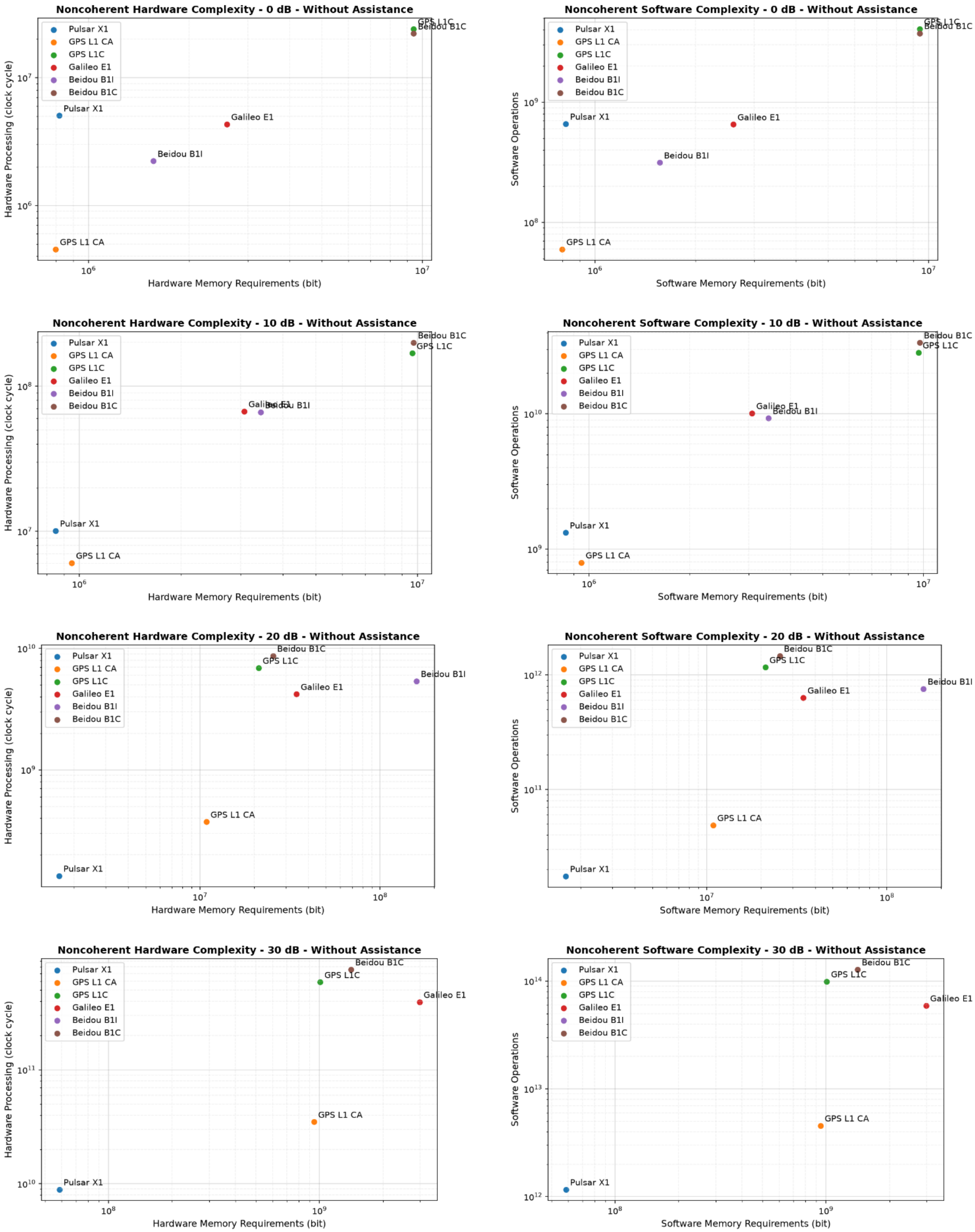

**Figure 4.28** Implementation-metrics plane for non-coherent integration, non-assisted case, for 0, 10, 20 and 30 dB attenuations. Left: metrics for hardware receivers. Right: metrics for software receivers

### *4.3.2. Coherent integration*

Unlike the non-coherent-only baseline, extending the coherent integration interval requires signal-dependent architectures to handle navigation-data or secondary-code transitions. Consequently, architectures providing the same coherent integration gain can exhibit substantially different trade-offs between memory requirement and processing cost. **Figure 4.29** illustrates these trade-offs for GPS L1 C/A and Pulsar X1 in the non-assisted case. These two signals are used as representative examples because they have the same 1 ms primary-code period and FFT size, while requiring fundamentally different approaches to extend coherent integration.

For GPS L1 C/A, the alternate block and full-bits methods remain relatively inexpensive while the required coherent integration interval is short. Once integration must extend beyond one navigation data bit, however, the unknown signs of successive 20 ms data bits introduce $2^{N_B-1}$ hypotheses. The effect is clearly visible at high attenuation in **Figure 4.29**. A parallel full-bits implementation minimizes processing time by evaluating the hypotheses simultaneously, but its memory requirement grows rapidly with the number of hypotheses. Serial implementations substantially reduce this memory requirement at the expense of additional processing time. Gray-code recursion limits this processing increase by allowing successive hypotheses to reuse the previous result, but it does not eliminate the exponential growth in the total number of hypotheses.

Pulsar X1 exhibits a similar trade-off when the multi-hypothesis method is used. For short coherent intervals, exhaustive evaluation of the $2^{N_C-1}$ possible secondary-code sign combinations remains practical, and parallel and serial implementations provide the expected memory-versus-processing trade-off. As attenuation increases and a larger $N_C$ is required, however, the exponential growth of the multi-hypothesis method rapidly makes it unattractive. Correlation with the known secondary code avoids this exponential scaling: partial secondary-code correlation is used while the coherent interval remains shorter than a complete secondary-code period, whereas full secondary-code correlation becomes applicable once one or more complete secondary-code periods are integrated.

To combine the processing-memory trade-offs into a common criterion, **Figure 4.30** shows the memory-time product for hardware implementations and the memory-operation product for software implementations for the candidate architectures of GPS L1 C/A, Pulsar X1, Galileo E1, and GPS L1C. BeiDou B1C is omitted from this architecture-level figure because its primary and secondary-code structure leads to behavior very similar to GPS L1C; it is retained in the subsequent signal-level comparison. The non-assisted case is shown here to illustrate the architecture transitions over the widest processing range. The same architecture minimization is performed independently for the assisted case in the final comparison.

For GPS L1 C/A, the different approaches have similar composite costs at low attenuation, with the alternate block method providing a small advantage while the required coherent interval remains below 10 ms. Once longer integration is required, the full-bits architectures become necessary. At large integration depths, the serial implementations provide a lower composite cost than the fully parallel implementation because the reduction in memory outweighs their increased processing time. Again, the Gray-code recursion is important in this regime because it reduces the cost of sequential hypothesis evaluation.

The modern pilot signals exhibit a different progression. At low attenuation, only a small number of primary-code periods must be combined, so multi-hypothesis processing remains competitive with partial secondary-code correlation. As the attenuation increases, the exponential dependence of the multi-hypothesis method on $N_C$ causes secondary-code correlation to become increasingly favorable. Partial secondary-code correlation is first preferred when the required coherent interval covers only part of the secondary code, followed by full secondary-code correlation when sufficiently long coherent integration is required. The exact transition depends on the primary-code duration, secondary-code length, required $N_C$, and the hardware or software cost model.

**Figure 4.31** compares the signals after independently selecting, at each attenuation, the architecture that minimizes the corresponding composite metric. At low attenuation, GPS L1 C/A and Pulsar X1 form the lowest-cost group, consistent with their short 1 ms primary-code periods and relatively modest integration requirements. As the attenuation increases, their behavior diverges. Pulsar X1 continues to benefit from the shorter integration enabled by its higher received signal power, while GPS L1 C/A becomes increasingly penalized by the need to resolve unknown navigation-data signs when coherent integration extends over multiple 20 ms data bits.

This result highlights an important limitation of GPS L1 C/A that was not visible in the non-coherent-only comparison. In Section 4.3.1, L1 C/A provided the lowest implementation cost among the analyzed MEO signals because of its short primary code and favorable link budget. Under long coherent-only integration, however, its data-only structure becomes a disadvantage: extending coherence over multiple navigation bits requires an exponentially increasing number of sign hypotheses. In contrast, the modern pilot signals contain known secondary codes, allowing long coherent integrations to be implemented through secondary-code correlation without exhaustive sign testing. Consequently, at high attenuation the coherent-only implementation cost of L1 C/A eventually exceeds that of the modern pilot signals.

Pulsar X1 avoids this high-attenuation growth because its secondary code can likewise be handled through correlation once multi-hypothesis processing becomes inefficient, while its higher received power limits the coherent depth required in the first place. As a result, X1 maintains a comparatively low composite implementation cost as attenuation increases. The assisted and non-assisted results exhibit the same general behavior, although assistance reduces the Doppler search and therefore lowers the processing component of the composite metric.

**Figure 4.32** provides the corresponding memory-versus-processing representation for the selected architectures in the non-assisted case. The figure confirms that the composite-product comparison does not merely favor one signal by accepting an extreme increase in the other metric. At low attenuation, the signals occupy relatively different memory-processing trade-off points, with GPS L1 C/A and X1 among the lower-cost implementations. As attenuation increases, X1 moves toward the lower-left region relative to the other signals, while the L1 C/A implementation moves strongly toward higher processing and memory requirements once long coherent integration across multiple data bits is required. The same trend is observed for both hardware processing time and software operation count.

The coherent-only case is primarily useful for exposing the implementation consequences of extending coherent integration. In a practical high-sensitivity receiver, very long coherent intervals may be constrained by oscillator stability, residual dynamics, and the cost of transition handling. For Pulsar X1 in particular, the longest coherent intervals should also be interpreted with the Doppler-rate limitation discussed in Section 3.1.1.3. The following section therefore considers hybrid integration, where the coherent interval can be limited to a practical value and the remaining processing gain is obtained through non-coherent accumulation.

Although the curves appear relatively close because of the logarithmic scale, the quantitative separation is substantial. At 10 dB attenuation, the minimum memory-time product of Pulsar X1 is approximately $2 \times 10^4$ times lower than that of the best-performing modern MEO signal, increasing to approximately $4 \times 10^4$ at 20 dB. Thus, the apparent proximity of the curves does not imply comparable implementation requirements.

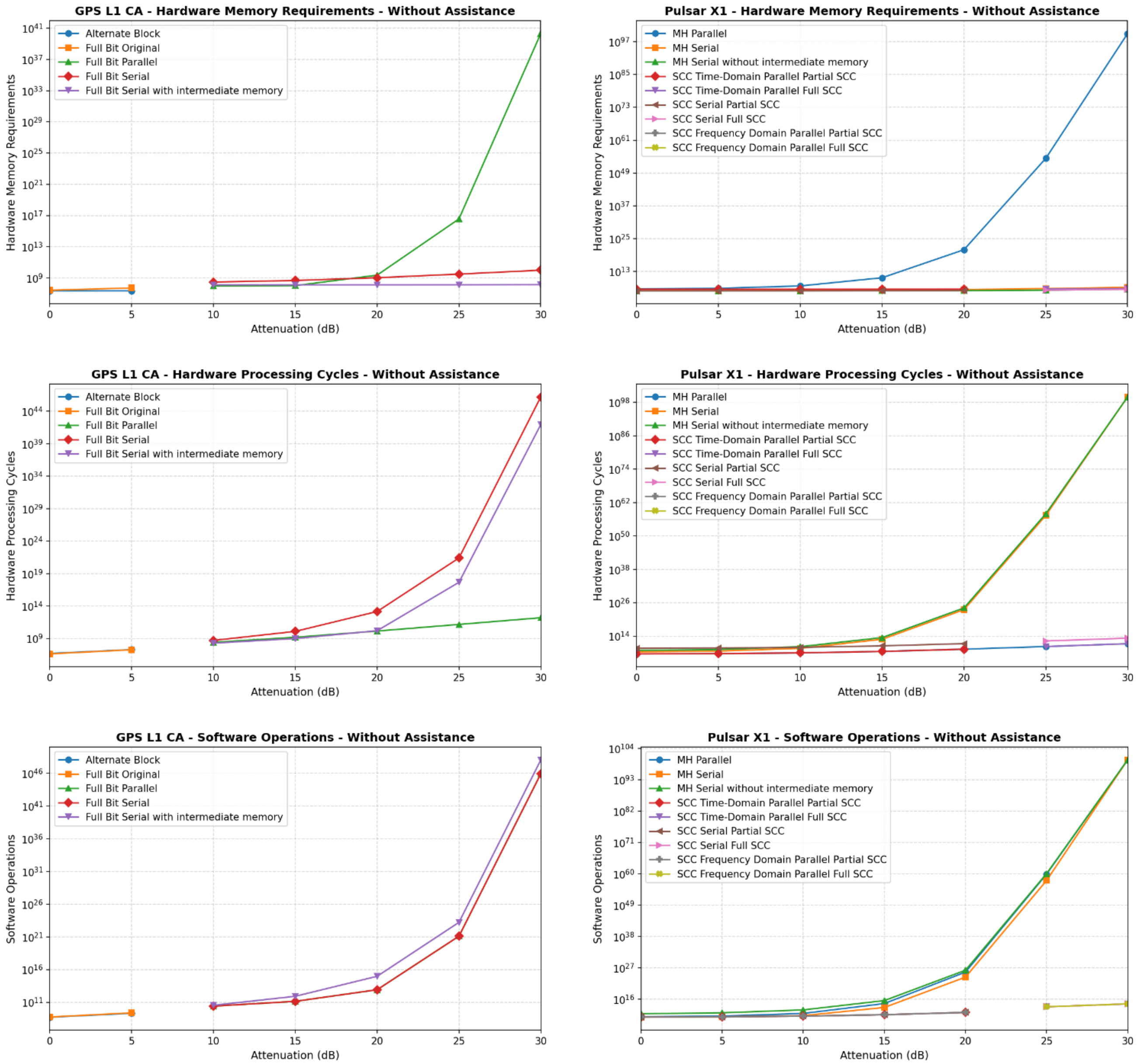

**Figure 4.29** Implementation metrics of candidate coherent-integration architectures for GPS L1 C/A and Pulsar X1, non-assisted case. Left: GPS L1 C/A; right: Pulsar X1.

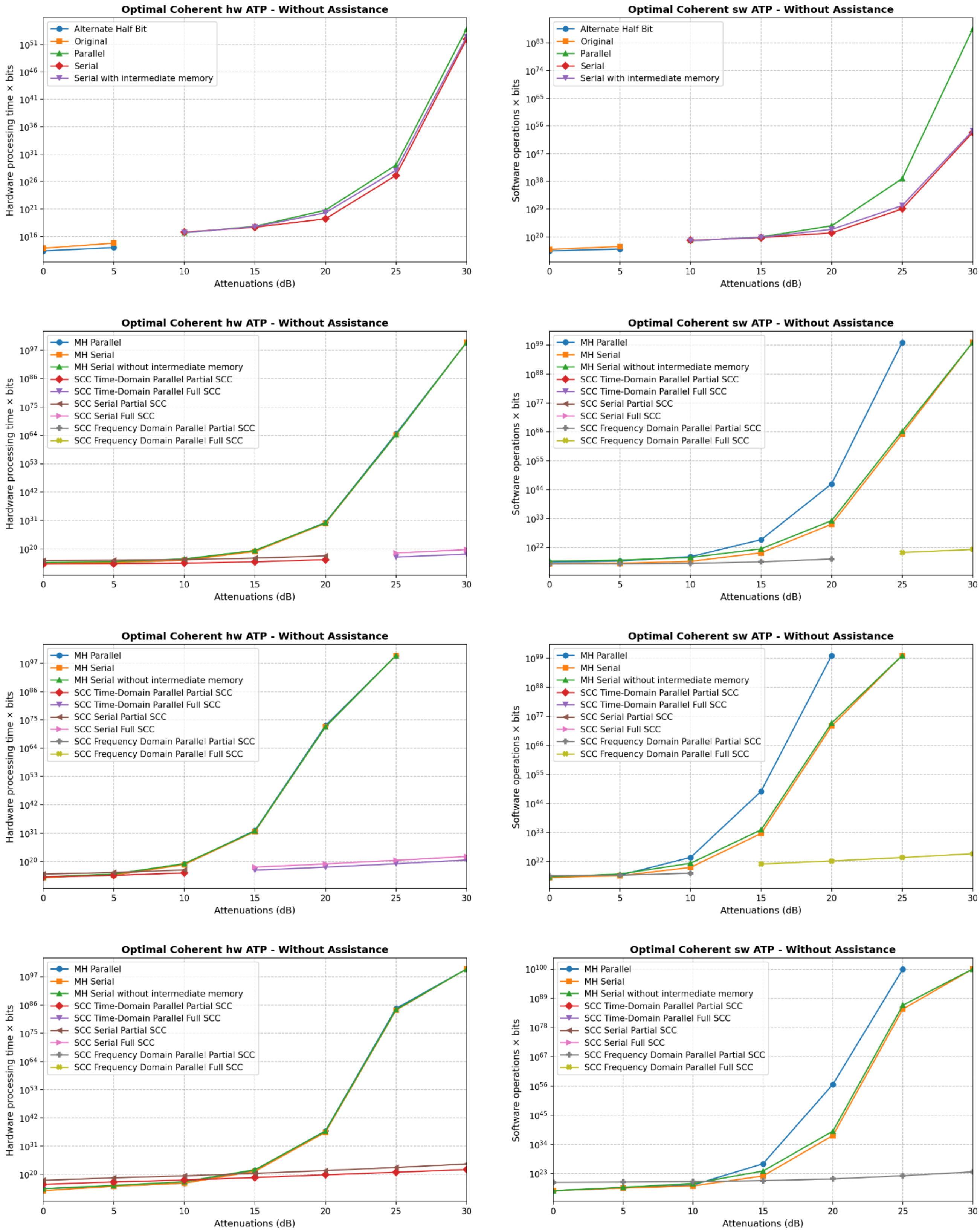

**Figure 4.30** Memory-processing products of candidate coherent-integration architectures, non-assisted case. Rows: GPS L1 C/A, Pulsar X1, Galileo E1, GPS L1C. Left: metrics for hardware receivers. Right: metrics for software receivers.

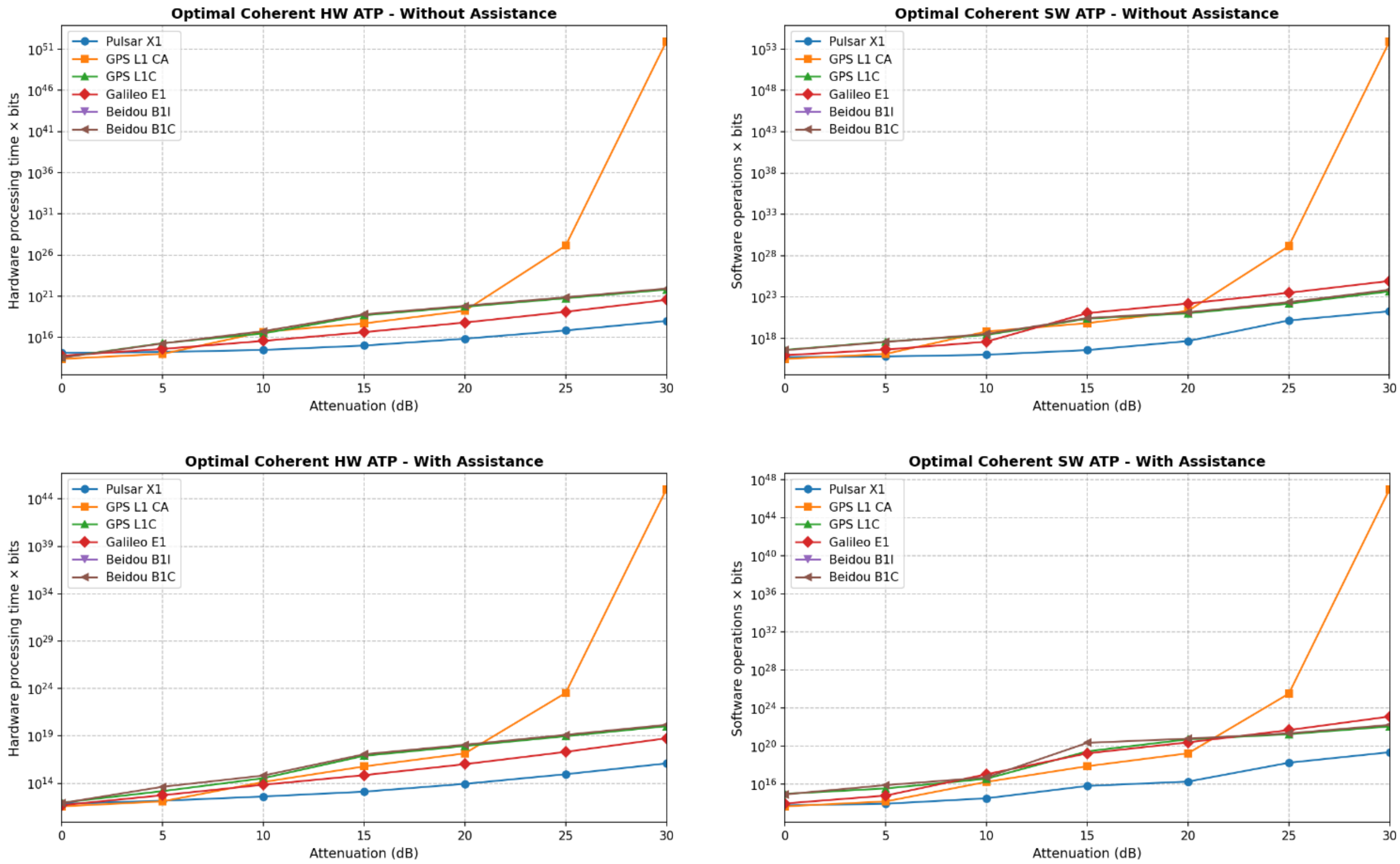

**Figure 4.31** Minimum memory-processing product across candidate coherent-integration architectures for each signal. Top: non-assisted case; bottom: assisted case. Left: memory-time product; right: memory-operation product.

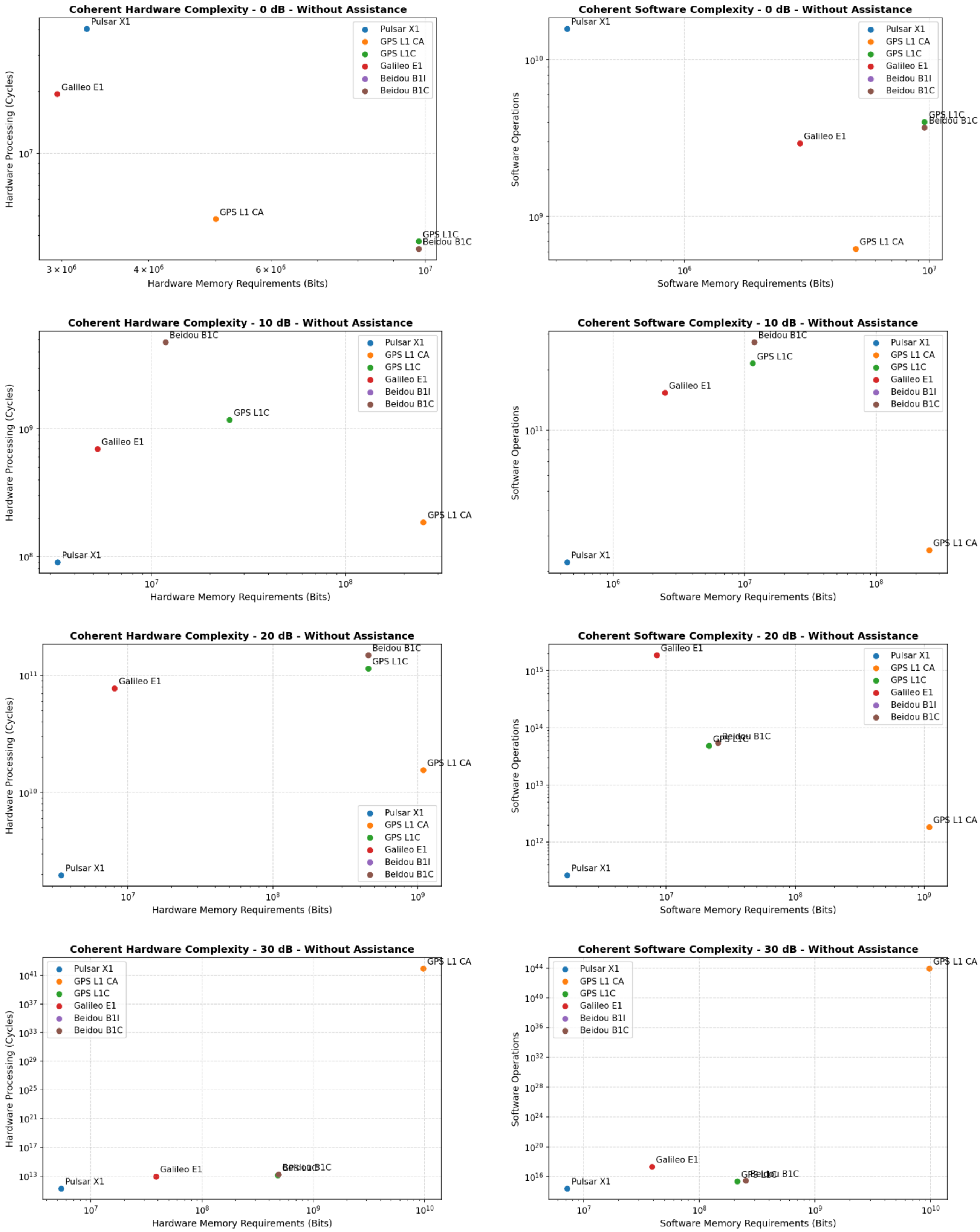

**Figure 4.32** Implementation-metric planes using the selected coherent-integration architecture, non-assisted case, for 0, 10, 20, and 30 dB attenuation. Left: hardware processing time versus memory; right: software operation count versus memory.

### *4.3.3. Hybrid integration*

The three hybrid integration cases introduced in Section 3.2 are now associated with the receiver architectures described in Section 4.1. For GPS L1 C/A, the $N_{\mathrm{C}}^{\max} = 10$, $N_{\mathrm{C}}^{\max} = 19$, and $N_{\mathrm{C}}^{\max} = 20$ cases correspond respectively to the alternate block, average bit-transition loss, and full-bits methods. For the modern pilot signals, the $N_{\mathrm{C}}^{\max} = 5$, $N_{\mathrm{C}}^{\max} = 10$, and $T_{\mathrm{C}}^{\max} = 100$ ms cases correspond respectively to the multi-hypothesis, partial secondary-code correlation, and full secondary-code correlation approaches.

Hybrid integration fundamentally changes the implementation behavior observed in the coherent-only case. Rather than increasing the coherent integration interval without bound as attenuation increases, the coherent depth is limited by the selected architecture and any additional processing gain is obtained through non-coherent accumulation. This prevents the very large combinatorial costs that occur in the coherent-only case when unknown navigation-data or secondary-code signs must be resolved over long coherent intervals.

**Figure 4.33** illustrates the resulting implementation trade-offs for GPS L1 C/A and Pulsar X1 in the assisted case. For GPS L1 C/A, the three methods remain within a relatively limited complexity range compared with the coherent-only full-bits architectures. At low attenuation, little or no non-coherent integration is required and the methods provide similar implementation requirements. As attenuation increases, the coherent-integration constraint becomes active and the architectures differ primarily in the amount of input data that must be processed and in the additional loss or parallelism associated with transition handling. The average bit-transition loss method becomes particularly competitive at large attenuation because it retains a simple architecture despite accepting an additional integration loss.

For Pulsar X1, the multi-hypothesis method remains attractive while only a small number of coherent primary-code periods are combined. In contrast with the coherent-only case, however, its exponential dependence on $N_{\mathrm{C}}$ is bounded by limiting the coherent integration to $N_{\mathrm{C}}^{\max} = 5$; further sensitivity is then obtained non-coherently. Partial secondary-code correlation similarly limits the coherent interval to ten primary-code periods, while full secondary-code correlation permits integration over the complete 100 ms secondary-code period. The three approaches therefore provide different balances between coherent processing gain and implementation cost rather than allowing the coherent complexity to grow indefinitely.

**Figure 4.34** compares the memory-processing products of the candidate hybrid architectures, still in the assisted case. The results show that the preferred architecture depends on the signal and attenuation. At low attenuation, short coherent intervals are sufficient and the simpler coherent-extension techniques remain attractive. As the required processing gain increases, architectures allowing a longer coherent interval can reduce the required number of non-coherent accumulations, but this benefit must be balanced against their additional memory and processing requirements. The architecture providing the lowest memory-processing product is therefore selected independently at each operating point, as in the coherent-only comparison.

**Figure 4.35** presents the resulting signal-level comparison after independently selecting the architecture that minimizes the corresponding memory-processing product. At low attenuation, GPS L1 C/A and Pulsar X1 form the lowest-cost group, but GPS L1 C/A retains an advantage: below 15 dB attenuation, its memory-processing product is approximately 2 to 6 times lower than that of X1, depending on the operating point. As attenuation increases, however, the relative behavior reverses. At 20 dB attenuation, the memory-processing product of X1 is already more than 6 times lower than that of GPS L1 C/A, and the separation continues to increase at higher attenuation. This crossover reflects the increasing influence of the higher received power of X1 as aggregate RNSS interference becomes less significant, while its short 1 ms primary code keeps the acquisition processing relatively inexpensive. The same general evolution is observed for both hardware and software implementations and in the assisted and non-assisted cases.

An important difference from the coherent-only results is that GPS L1 C/A no longer experiences the exponential complexity increase associated with coherent integration across an increasing number of unknown navigation bits. Once the selected coherent limit is reached, the remaining gain is instead obtained through non-coherent accumulation. Consequently, L1 C/A remains comparatively efficient among the MEO signals under high attenuation, while the modern pilot signals continue to incur the larger FFT and memory requirements associated with their longer primary codes. The hybrid strategy therefore removes one of the principal disadvantages of L1 C/A observed in the coherent-only comparison, while preserving the increasing advantage of X1 as attenuation increases.

**Figure 4.36** provides the corresponding memory-versus-processing representation for the selected hybrid architectures in the assisted case. At low attenuation, GPS L1 C/A occupies the lowest-processing region, while Pulsar X1 remains among the lowest-memory implementations. As attenuation increases, the X1 operating point progressively moves toward the lower-left region relative to the MEO signals. At 20 and 30 dB attenuation, its lower memory-processing product therefore reflects simultaneous reductions in both memory requirement and processing workload rather than a trade-off between the two metrics. The same behavior is observed for hardware processing time and software operation count.

Overall, the hybrid strategy provides a more practical high-sensitivity operating regime than either limiting coherence to one primary-code period or extending coherent integration without constraint. It retains part of the efficiency of coherent integration while preventing the transition-handling complexity from growing excessively. The resulting comparison confirms the same principal trend observed throughout this study: GPS L1 C/A is particularly efficient under favorable conditions, whereas the higher received power and short primary-code period of Pulsar X1 become increasingly advantageous as the signal is attenuated. Assistance further benefits X1 by reducing the processing penalty associated with its larger Doppler uncertainty.

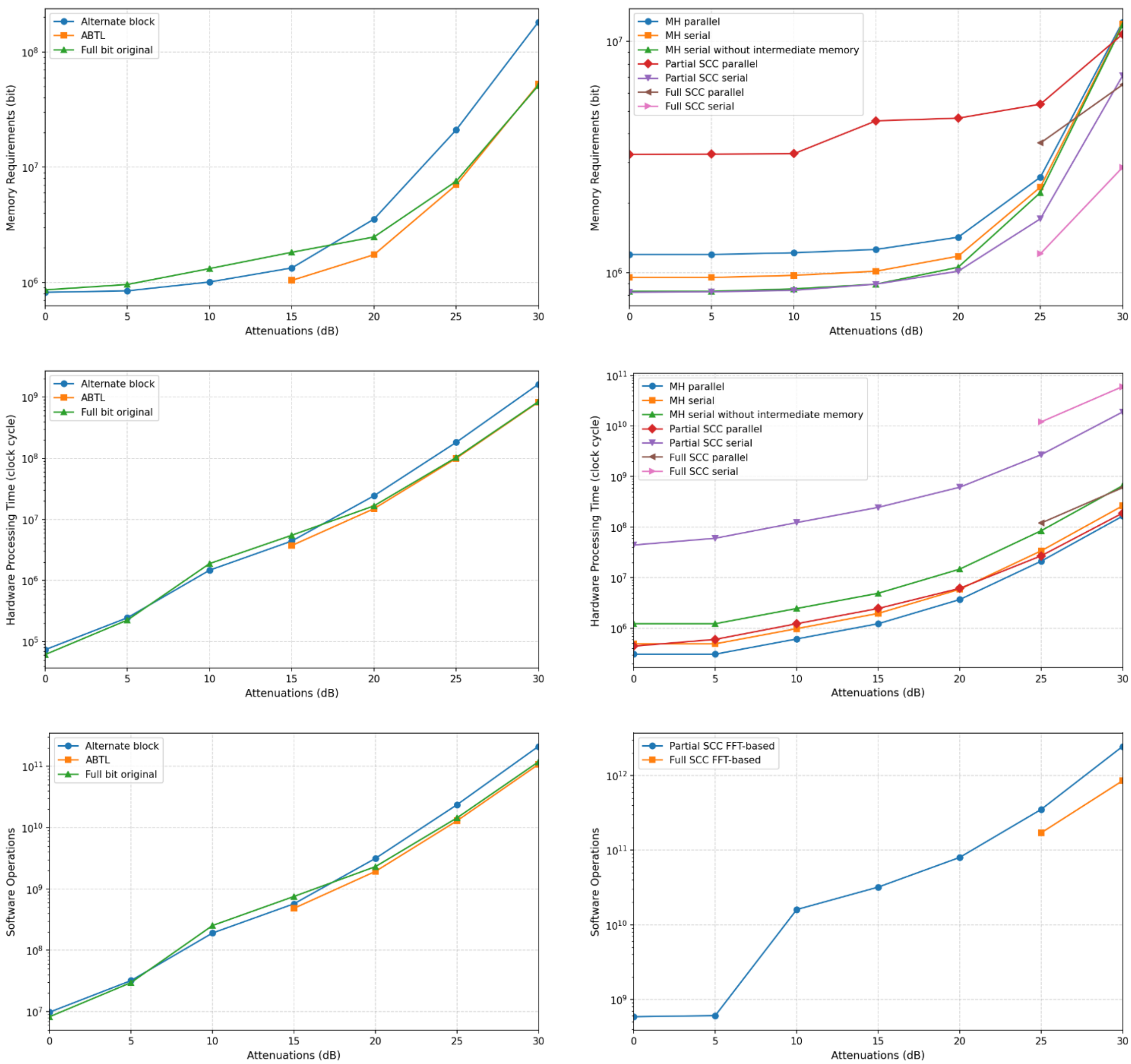

**Figure 4.33** Implementation metrics of candidate hybrid-integration architectures for GPS L1 C/A and Pulsar X1, assisted case. Left: GPS L1 C/A; right: Pulsar X1.

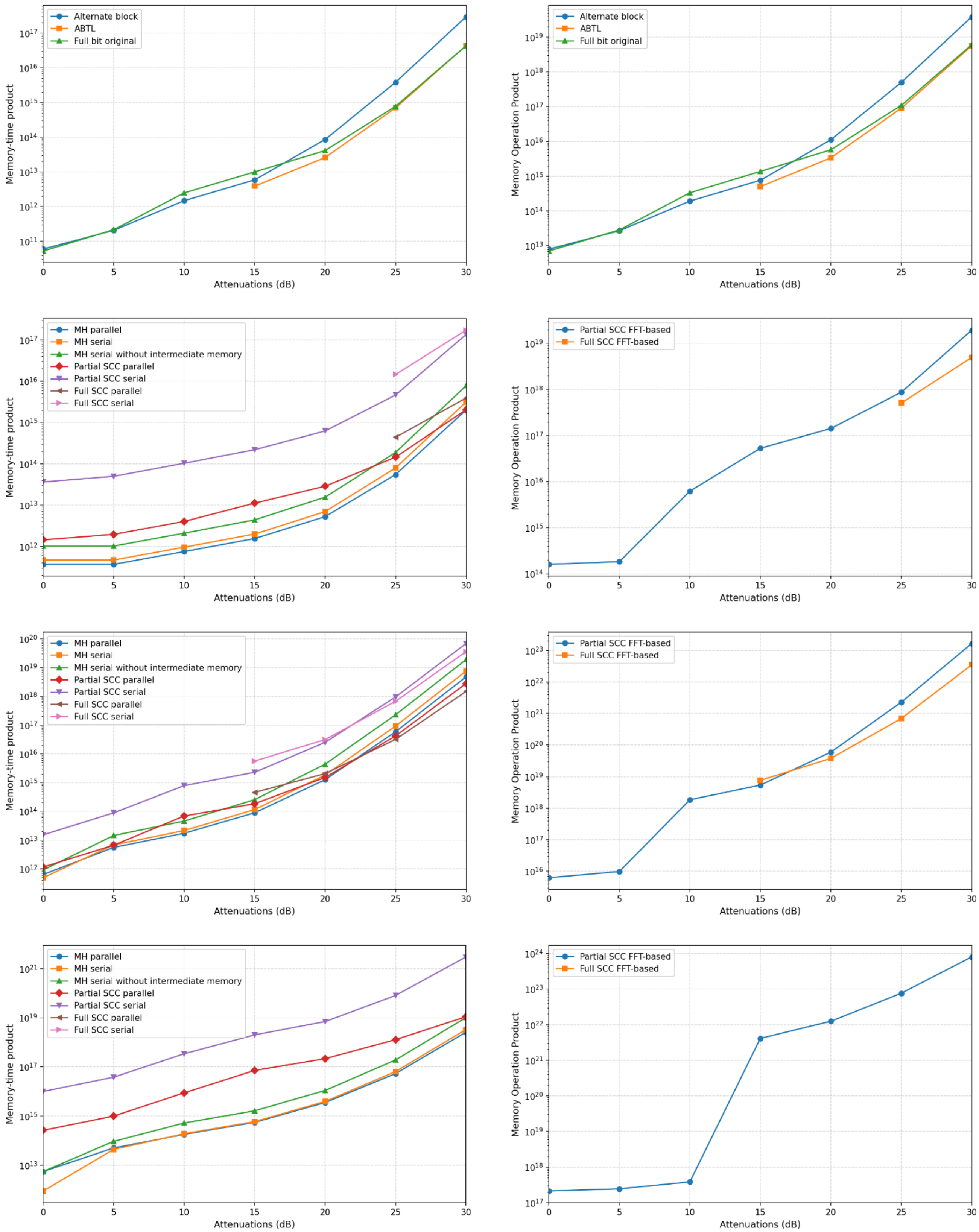

**Figure 4.34** Memory-processing products of candidate hybrid-integration architectures, assisted case. Rows: GPS L1 C/A, Pulsar X1, Galileo E1, GPS L1C. Left: metrics for hardware receivers. Right: metrics for software receivers.

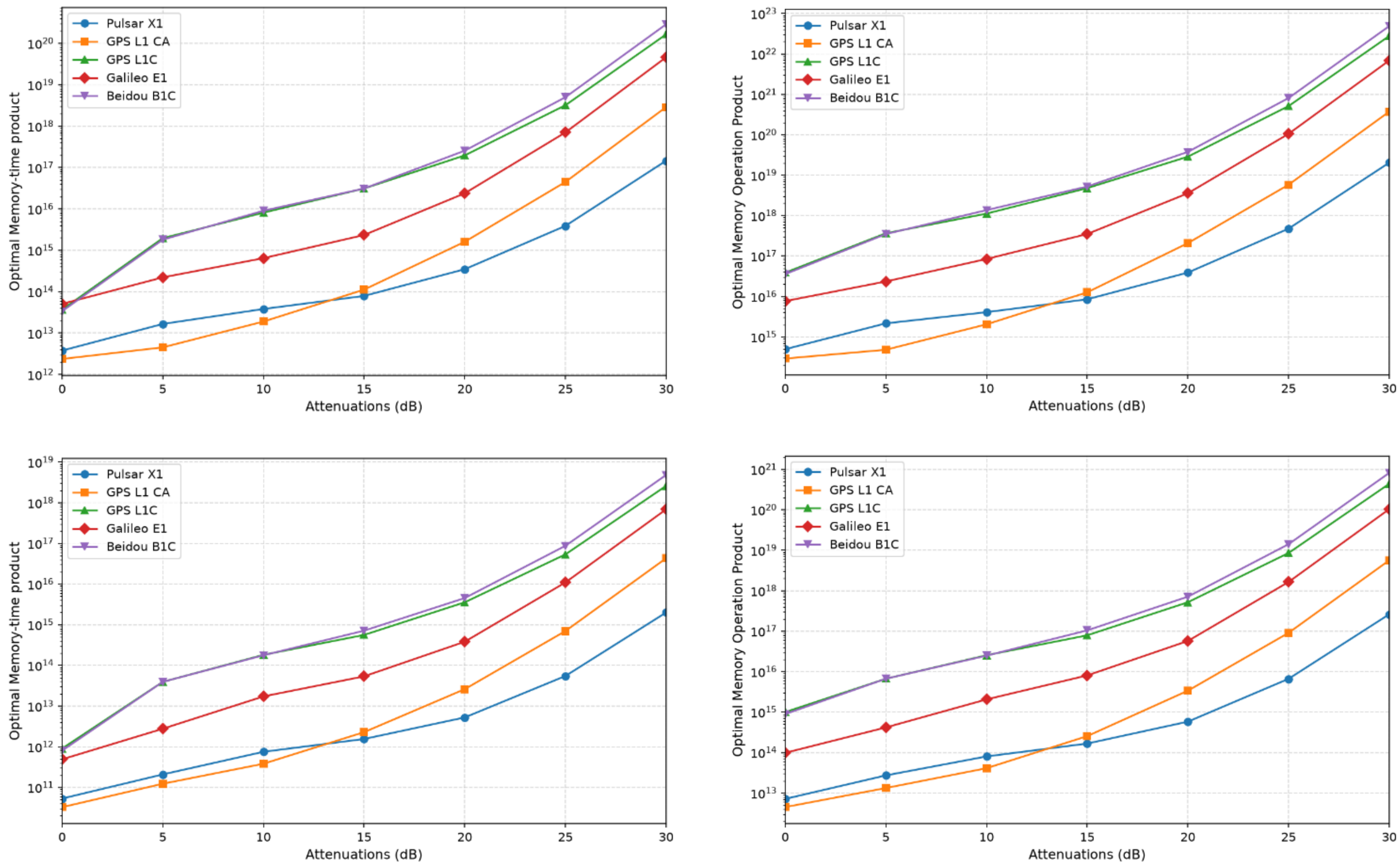

**Figure 4.35** Minimum memory-processing product across candidate hybrid-integration architectures for each signal. Top: non-assisted case; bottom: assisted case. Left: memory-time product; right: memory-operation product.

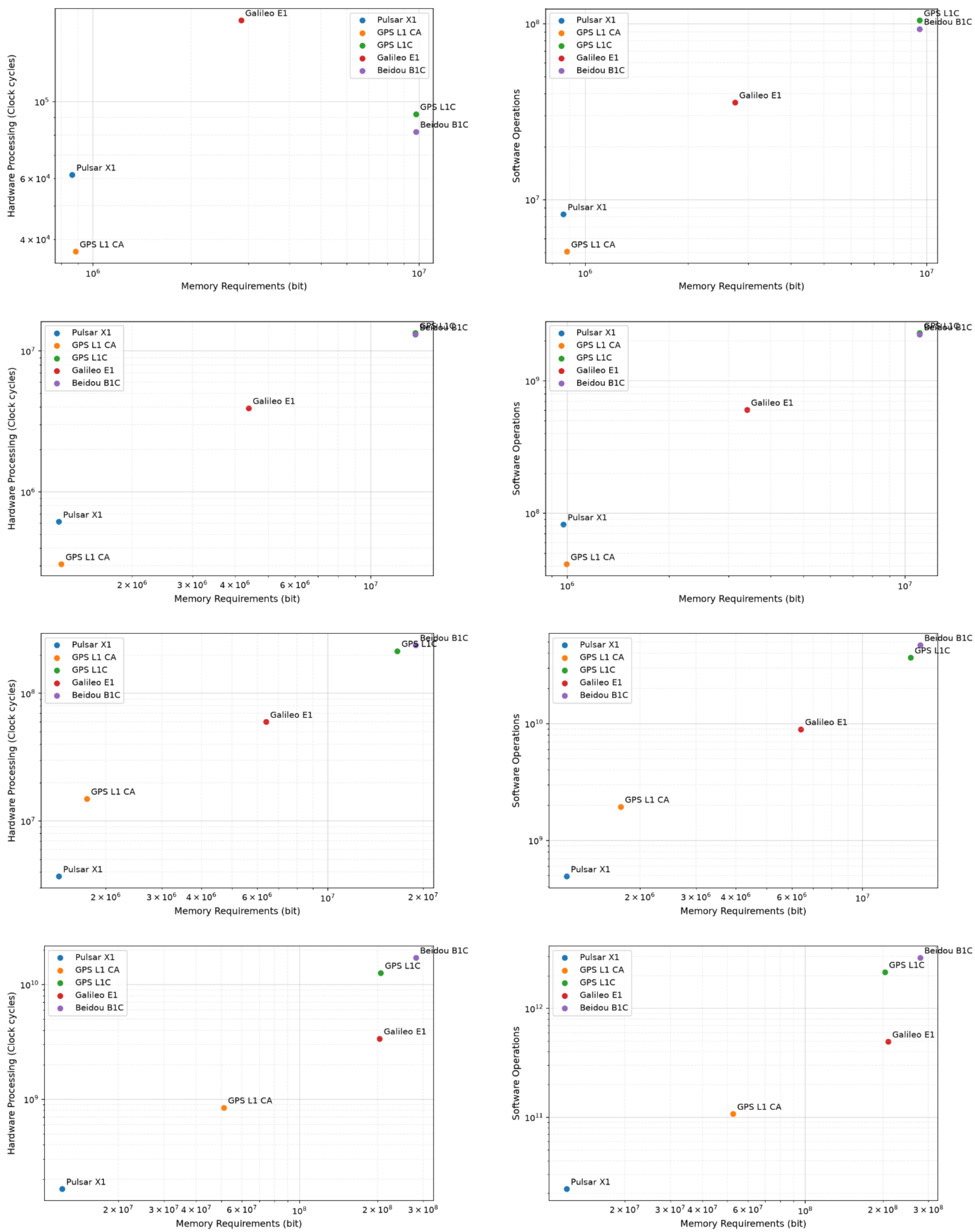

**Figure 4.36** Implementation-metric planes using the selected hybrid-integration architecture, assisted case, for 0, 10, 20, and 30 dB attenuation. Left: hardware processing time versus memory; right: software operation count versus memory.

## 5. CONCLUSION

This paper presented a unified methodology for comparing the acquisition complexity of representative L1-band navigation signals under identical acquisition-performance requirements. The analysis first determines the integration required to satisfy common probabilities of detection and false alarm while accounting for signal power, receiver losses, aggregate RNSS interference, and acquisition search space. These requirements are then translated into receiver implementation metrics including software operation count, hardware processing time, and memory requirement. Non-coherent-only, coherent-only, and hybrid integration strategies were evaluated using representative FFT-based acquisition architectures.

The results show that integration time alone is not sufficient to characterize acquisition complexity. Primary-code duration, code length, secondary-code or navigation-data structure, and acquisition search space strongly affect the implementation required to realize a given integration gain. In particular, long primary codes can reduce the number of required integrations while simultaneously increasing the FFT size, memory requirement, and number of Doppler hypotheses. Similarly, extending coherent integration can improve sensitivity efficiently but may introduce significant transition-handling complexity. For GPS L1 C/A, coherent integration across multiple unknown navigation bits eventually produces an exponential increase in the number of sign hypotheses, whereas modern pilot signals can instead exploit their known secondary codes through secondary-code correlation.

The comparison also shows a consistent dependence on signal attenuation. Under favorable reception conditions, GPS L1 C/A generally provides the lowest acquisition complexity because of its short 1 ms primary code, favorable link budget, and absence of pilot/data power sharing. For Pulsar X1, its higher nominal received power is partially offset at low attenuation by aggregate RNSS interference and, particularly without assistance, by its larger Doppler search space. As attenuation increases, however, the RNSS-interference contribution decreases relative to thermal noise and the higher received power of X1 becomes dominant. Combined with its short 1 ms primary code and 4096-point FFT, this results in substantially lower integration and implementation requirements than the analyzed MEO signals under attenuated signal conditions. In the hybrid comparison, GPS L1 C/A retains a memory-processing product approximately 2 to 6 times lower than X1 below 15 dB attenuation, whereas at 20 dB the trend reverses and X1 achieves a memory-processing product more than 6 times lower than GPS L1 C/A.

The coherent-only analysis emphasizes the implementation penalties that can arise when coherent integration is extended without constraint, while the hybrid analysis provides a more practical compromise. By limiting the coherent interval and obtaining the remaining processing gain through non-coherent accumulation, hybrid acquisition prevents the combinatorial transition-handling cost from growing indefinitely. This is particularly beneficial for GPS L1 C/A, which remains competitive among the MEO signals when long coherent integration across multiple data bits is avoided. Pulsar X1 retains its advantage at higher attenuation because its stronger received signal requires less additional processing gain, reducing both the coherent and non-coherent integration burden.

These results are particularly relevant to low-power and snapshot-based receivers, for which acquisition processing and memory directly contribute to receiver resource usage and potentially to energy consumption. The memory-processing products introduced in this work provide a compact means of comparing these implementation trade-offs, although they are not intended as direct energy estimates. More generally, the analysis demonstrates that the receiver implications of a navigation signal cannot be inferred from received power or integration time alone; signal structure, acquisition search space, and the architecture required to realize the processing gain must be considered jointly.

Future work will focus on validating the analytical complexity models through implementation and measurements on representative hardware and software receivers using real GNSS and Pulsar X1 signals. Such measurements would allow the predicted differences in processing time, memory usage, and energy consumption to be compared directly with practical acquisition performance under different attenuation and assistance conditions.

## ACKNOWLEDGMENTS

We acknowledge the support of the Canadian Space Agency (CSA) 24STDPU43.

## 6. APPENDIX A. EXAMPLE OF FULL WORKSHEET

| | Pulsar X1 | GPS L1 C/A | GPS L1C (SSB) | Galileo E1 (SSB) | Beidou B1I | Beidou B1C (SSB) | Units |
|---|---|---|---|---|---|---|---|
| **Signal Strength and Loss before FE** | | | | | | | |
| Nominal Signal Strength | -118.2 | -128.5 | -128.25 | -127.25 | -133 | -129 | dBm |
| Additional Attenuation | -10 | -10 | -10 | -10 | -10 | -10 | dB |
| Data/Pilot Channel Power Spreading Loss | -3.010299957 | 0 | -1.249387366 | -3.010299957 | 0 | -1.249387366 | dB |
| BOC/BPSK Loss | 0 | 0 | -3.571459376 | -3.424226808 | 0 | -3.571459376 | dB |
| Received Signal strength | -131.21 | -138.50 | -143.07 | -143.68 | -143.00 | -143.82 | dBm |
| | | | | | | | |
| **Front End** | | | | | | | |
| IF Bandwidth | 3 | 3 | 3 | 3 | 6 | 3 | MHz |
| $T_{eff}$ | 290 | 290 | 290 | 290 | 290 | 290 | K |
| Ideal C/N0 | 42.77 | 35.48 | 30.90 | 30.29 | 30.98 | 30.15 | dBHz |
| C/N0 Degradation | -4.01 | -0.35 | -0.29 | -0.19 | -0.13 | -0.17 | dB |
| Effective C/N0 | 38.76 | 35.13 | 30.61 | 30.10 | 30.85 | 29.98 | dBHz |
| Noise power | -105.19 | -108.85 | -108.91 | -109.01 | -106.06 | -109.03 | dBm |
| IF SNR | -26.02 | -29.65 | -34.16 | -34.67 | -36.94 | -34.79 | dB |
| | | | | | | | |
| **Coherent Integration** | | | | | | | |
| Sample rate | 2.046 | 2.046 | 2.046 | 2.046 | 4.092 | 2.046 | MHz |
| Minimal Coherent interval | 1 | 1 | 10 | 4 | 1 | 10 | ms |
| Number of coherent accumulation ($N_C$) | 1 | 1 | 1 | 1 | 1 | 1 | - |
| Coherent integration time ($T_C$) | 1 | 1 | 10 | 4 | 1 | 10 | ms |
| Number of samples accumulated ($M_C$) | 2046 | 2046 | 20460 | 8184 | 4092 | 20460 | |
| Ideal coherent gain | 33.11 | 33.11 | 43.11 | 39.13 | 36.12 | 43.11 | dB |
| Implementation Loss | -3.45 | -2.51 | -2.51 | -2.51 | -2.51 | -2.51 | dB |
| Actual coherent gain | 29.66 | 30.60 | 40.60 | 36.62 | 33.61 | 40.60 | dB |
| Coherent SNR | 3.64 | 0.95 | 6.44 | 1.95 | -3.33 | 5.81 | dB |
| sqrt(coherent SNR) ratio | **1.521** | **1.116** | **2.100** | **1.252** | **0.682** | **1.953** | ratio |
| | | | | | | | |
| **Magnitude computation** | | | | | | | |
| Squaring loss | 1.23 | -0.63 | 2.76 | 0.09 | -4.20 | 2.45 | dB |
| Post RSS SNR | 4.87 | 0.32 | 9.20 | 2.04 | -7.52 | 8.27 | dB |
| Post RSS SNRratio | **1.753** | **1.038** | **2.885** | **1.265** | **0.421** | **2.590** | ratio |
| | | | | | | | |
| **Non-coherent Integration without assistance** | | | | | | | |
| # nc sums, $M_{NC}$ | 16 | 40 | 7 | 31 | 237 | 9 | - |
| Total integration time $T_{NC}$ | 16 | 40 | 70 | 124 | 237 | 90 | ms |
| Non-coherent gain | 12.04 | 16.02 | 8.45 | 14.91 | 23.75 | 9.54 | dB |
| Final SNR | 16.92 | 16.34 | 17.65 | 16.96 | 16.22 | 17.81 | dB |
| sqrt(final SNR) ratio | **7.011** | **6.563** | **7.633** | **7.045** | **6.474** | **7.770** | ratio |
| | | | | | | | |
| **Detection Performance without assistance** | | | | | | | |
| FA Threshold | 5.407825 | 5.146407 | 5.950274 | 5.62448 | 5.259464 | 5.936787 | |
| PFA (cell) | 3.18974E-08 | 1.32762E-07 | 1.33847E-09 | 9.30338E-09 | 7.2238E-08 | 1.45331E-09 | |
| Doppler Search Step | 500 | 500 | 50 | 125 | 500 | 50 | Hz |
| Doppler Search Space | 154 | 37 | 367 | 132 | 34 | 338 | bin |
| Code Search Space | 2046 | 2046 | 20460 | 8184 | 4092 | 20460 | sample |
| Total Number of Cells | 315084 | 75702 | 7508820 | 1080288 | 139128 | 6915480 | cell |
| PFA (global) | 1.0E-02 | 1.0E-02 | 1.0E-02 | 1.0E-02 | 1.0E-02 | 1.0E-02 | |
| q | 1.415 | 1.325 | 1.474 | 1.360 | 1.174 | 1.464 | |
| PD (cell) | 0.87 | 0.86 | 0.87 | 0.85 | 0.85 | 0.89 | |
| | | | | | | | |
| **Non-coherent Integration with assistance** | | | | | | | |
| # nc sums, $M_{NC}$ | 13 | 32 | 6 | 26 | 189 | 7 | - |
| Total integration time $T_{NC}$ | 13 | 32 | 60 | 104 | 189 | 70 | ms |
| Non-coherent gain | 11.14 | 15.05 | 7.78 | 14.15 | 22.76 | 8.45 | dB |
| Final SNR | 16.01 | 15.37 | 16.98 | 16.19 | 15.24 | 16.72 | dB |
| sqrt(final SNR) ratio | **6.320** | **5.870** | **7.067** | **6.452** | **5.781** | **6.853** | ratio |
| | | | | | | | |
| **Detection Performance with assistance** | | | | | | | |
| FA Threshold | 4.652787 | 4.421004 | 5.310776 | 4.986753 | 4.568504 | 5.289271 | |
| PFA (cell) | 1.63739E-06 | 4.91217E-06 | 5.45797E-08 | 3.07012E-07 | 2.45609E-06 | 6.14024E-08 | |
| Doppler Search Step | 500 | 500 | 50 | 125 | 500 | 50 | Hz |
| Doppler Search Space | 3 | 1 | 9 | 4 | 1 | 8 | bin |
| Code Search Space | 2046 | 2046 | 20460 | 8184 | 4092 | 20460 | sample |
| Total Number of Cells | 6138 | 2046 | 184140 | 32736 | 4092 | 163680 | cell |
| PFA (global) | 1.0E-02 | 1.0E-02 | 1.0E-02 | 1.0E-02 | 1.0E-02 | 1.0E-02 | |
| q | 1.415 | 1.325 | 1.474 | 1.360 | 1.174 | 1.464 | |
| PD (cell) | 0.88 | 0.86 | 0.88 | 0.86 | 0.85 | 0.86 | |

**Figure 6.1** Example worksheet for non-coherent integration for 10 dB attenuation

| | Pulsar X1 | GPS L1 C/A | GPS L1C (SSB) | Galileo E1 (SSB) | Beidou B1I | Beidou B1C (SSB) | Units |
|---|---|---|---|---|---|---|---|
| **Signal Strength and Loss before FE** | | | | | | | |
| Nominal Signal Strength | -118.2 | -128.5 | -128.25 | -127.25 | -133 | -129 | dBm |
| Additional Attenuation | -10 | -10 | -10 | -10 | -10 | -10 | dB |
| Data/Pilot Channel Power Spreading Loss | -3.010299957 | 0 | -1.249387366 | -3.010299957 | 0 | -1.249387366 | dB |
| BOC/BPSK Loss | 0 | 0 | -3.571459376 | -3.424226808 | 0 | -3.571459376 | dB |
| Received Signal strength | -131.21 | -138.50 | -143.07 | -143.68 | -143.00 | -143.82 | dBm |
| | | | | | | | |
| **Front End** | | | | | | | |
| IF Bandwidth | 3 | 3 | 3 | 3 | 6 | 3 | MHz |
| $T_{eff}$ | 290 | 290 | 290 | 290 | 290 | 290 | K |
| Ideal C/N0 | 42.77 | 35.48 | 30.90 | 30.29 | 30.98 | 30.15 | dBHz |
| C/N0 Degradation | -4.01 | -0.35 | -0.29 | -0.19 | -0.13 | -0.17 | dB |
| Effective C/N0 | 38.76 | 35.13 | 30.61 | 30.10 | 30.85 | 29.98 | dBHz |
| Noise power | -105.19 | -108.85 | -108.91 | -109.01 | -106.06 | -109.03 | dBm |
| IF SNR | -26.02 | -29.65 | -34.16 | -34.67 | -36.94 | -34.79 | dB |
| | | | | | | | |
| **Coherent Integration without assistance** | | | | | | | |
| Sample rate | 2.046 | 2.046 | 2.046 | 2.046 | 4.092 | 2.046 | MHz |
| Minimal Coherent interval | 1 | 1 | 10 | 4 | 1 | 10 | ms |
| Number of coherent accumulation ($N_C$) | 12 | 21 | 7 | 18 | 59 | 8 | - |
| Coherent integration time ($T_C$) | 12 | 21 | 70 | 72 | 59 | 80 | ms |
| Number of samples accumulated ($M_C$) | 24552 | 42966 | 143220 | 147312 | 241428 | 163680 | |
| Ideal coherent gain | 43.90 | 46.33 | 51.56 | 51.68 | 53.83 | 52.14 | dB |
| Implementation Loss | -3.45 | -2.51 | -2.51 | -2.51 | -2.51 | -2.51 | dB |
| Actual coherent gain | 40.45 | 43.82 | 49.05 | 49.17 | 51.32 | 49.63 | dB |
| Coherent SNR | 14.44 | 14.18 | 14.89 | 14.50 | 14.38 | 14.84 | dB |
| Linear coherent SNR - square rooted | **5.270** | **5.114** | **5.555** | **5.310** | **5.237** | **5.523** | ratio |
| Linear coherent SNR | **27.772** | **26.154** | **30.855** | **28.197** | **27.427** | **30.502** | |
| | | | | | | | |
| **Magnitude computation without** | | | | | | | |
| Squaring loss | 5.18 | 5.13 | 5.26 | 5.19 | 5.17 | 5.25 | dB |
| Post RSS SNR | 19.61 | 19.31 | 20.15 | 19.69 | 19.55 | 20.09 | dB |
| Post RSS SNRratio | **9.565** | **9.232** | **10.175** | **9.651** | **9.495** | **10.106** | ratio |
| | | | | | | | |
| **Detection Performance without assistance** | | | | | | | |
| FA Threshold | 6.283433 | 6.143529 | 6.689882 | 6.539096 | 6.405519 | 6.697591 | |
| PFA (cell) | 2.671E-09 | 6.371E-09 | 1.913E-10 | 5.186E-10 | 1.231E-09 | 1.81664E-10 | |
| Doppler Search Step | 41.66666667 | 23.80952381 | 7.142857143 | 6.944444444 | 8.474576271 | 6.25 | Hz |
| Doppler Search Space | 1839 | 771 | 2568 | 2368 | 1995 | 2704 | bins |
| Code Search Space | 2046 | 2046 | 20460 | 8184 | 4092 | 20460 | samples |
| Total Number Of Cell | 3762594 | 1577466 | 52541280 | 19379712 | 8163540 | 55323840 | cells |
| PFA (global) | **1.0E-02** | **1.0E-02** | **1.0E-02** | **1.0E-02** | **1.0E-02** | **1.0E-02** | |
| PD (cell) | **0.89** | **0.88** | **0.89** | **0.85** | **0.86** | **0.88** | |
| | | | | | | | |
| **Coherent Integration with assistance** | | | | | | | |
| Sample rate | 2.046 | 2.046 | 2.046 | 2.046 | 4.092 | 2.046 | MHz |
| Minimal Coherent interval | 1 | 1 | 10 | 4 | 1 | 10 | ms |
| Number of coherent accumulation ($N_C$) | 10 | 17 | 6 | 16 | 49 | 7 | - |
| Coherent integration time ($T_C$) | 10 | 17 | 60 | 64 | 49 | 70 | ms |
| Number of samples accumulated ($M_C$) | 20460 | 34782 | 122760 | 130944 | 200508 | 143220 | |
| Ideal coherent gain | 43.11 | 45.41 | 50.89 | 51.17 | 53.02 | 51.56 | dB |
| Implementation Loss | -3.45 | -2.51 | -2.51 | -2.51 | -2.51 | -2.51 | dB |
| Actual coherent gain | 39.66 | 42.90 | 48.38 | 48.66 | 50.51 | 49.05 | dB |
| Coherent SNR | 13.64 | 13.26 | 14.22 | 13.99 | 13.58 | 14.26 | dB |
| Linear coherent SNR - square rooted | **4.811** | **4.601** | **5.143** | **5.006** | **4.773** | **5.166** | ratio |
| Linear coherent SNR | **23.143** | **21.172** | **26.448** | **25.064** | **22.778** | **26.689** | |
| | | | | | | | |
| **Magnitude computation with assistance** | | | | | | | |
| Squaring loss | 5.03 | 4.95 | 5.14 | 5.10 | 5.02 | 5.15 | dB |
| Post RSS SNR | 18.67 | 18.21 | 19.36 | 19.09 | 18.59 | 19.41 | dB |
| Post RSS SNRratio | **8.584** | **8.136** | **9.293** | **9.002** | **8.502** | **9.343** | ratio |
| | | | | | | | |
| **Detection Performance with assistance** | | | | | | | |
| FA Threshold | 5.590156 | 5.452102 | 6.072626 | 5.923158 | 5.762949 | 6.09126 | |
| PFA (cell) | 1.638E-07 | 3.509E-07 | 9.824E-09 | 2.408E-08 | 6.140E-08 | 8.7718E-09 | |
| Doppler Search Step | 50 | 29.41176471 | 8.333333333 | 7.8125 | 10.20408163 | 7.142857143 | Hz |
| Doppler Search Space | 30 | 14 | 50 | 51 | 40 | 56 | bin |
| Code Search Space | 2046 | 2046 | 20460 | 8184 | 4092 | 20460 | samples |
| Total Number Of Cell | 61380 | 28644 | 1023000 | 417384 | 163680 | 1145760 | cell |
| PFA (global) | **1.0E-02** | **1.0E-02** | **1.0E-02** | **1.0E-02** | **1.0E-02** | **1.0E-02** | |
| PD (cell) | **0.90** | **0.87** | **0.90** | **0.89** | **0.86** | **0.90** | |

**Figure 6.2** Example worksheet for coherent integration for 10 dB attenuation

## 7. APPENDIX B. DETERMINATION OF COMPLEXITY METRICS

This appendix derives the complexity metrics used in Section 4.2 for each acquisition architecture introduced in Section 4.1. For software receivers, each table decomposes the architecture into elementary processing stages, and the corresponding equation gives the total number of real additions and multiplications. For hardware receivers, processing-time expressions are derived from the timing diagrams shown in this appendix, assuming streaming pipelined processing and neglecting fixed FFT/IFFT latency. Memory expressions are obtained directly from the buffers and accumulators shown in the corresponding architecture figures. Each derivation therefore explicitly references the architecture from which it is obtained.

### 7.1. Complexity for FFT-based primary code correlation

The FFT-based primary-code correlation of **Figure 4.1** is common to all acquisition architectures considered in this work. Its software complexity is therefore derived first and reused throughout the remainder of this appendix.

#### *7.1.1. Software operation count*

**TABLE 7.1** details the elementary operations required by the FFT-based primary-code correlation of **Figure 4.1**. The FFT of the local code is computed only once per acquisition attempt, whereas the remaining operations are repeated for every frequency hypothesis. The conjugation of the local-code spectrum is assumed to be included in the frequency-domain multiplication and therefore does not require additional operations.

**TABLE 7.1** Software operation count for FFT-based primary-code correlation (**Figure 4.1**)

| Computation stage | Vector size | # real additions | # real multiplications | # repetitions |
|---|---|---|---|---|
| Product with local carrier | $2N_\mathrm{P}$ | $4N_\mathrm{P}$ | $8N_\mathrm{P}$ | $N_\mathrm{FB}$ |
| Pre-FFT zero-padding | - | 0 | 0 | $N_\mathrm{FB}$ |
| Input FFT | $N_\mathrm{FFT,P}$ | $3N_\mathrm{FFT,P}\log_2(N_\mathrm{FFT,P})$ | $2N_\mathrm{FFT,P}\log_2(N_\mathrm{FFT,P})$ | $N_\mathrm{FB}$ |
| Local code FFT | $N_\mathrm{FFT,P}$ | $3N_\mathrm{FFT,P}\log_2(N_\mathrm{FFT,P})$ | $2N_\mathrm{FFT,P}\log_2(N_\mathrm{FFT,P})$ | 1 |
| Product of FFT results | $N_\mathrm{FFT,P}$ | $2N_\mathrm{FFT,P}$ | $4N_\mathrm{FFT,P}$ | $N_\mathrm{FB}$ |
| IFFT of FFTs product | $N_\mathrm{FFT,P}$ | $3N_\mathrm{FFT,P}\log_2(N_\mathrm{FFT,P})$ | $2N_\mathrm{FFT,P}\log_2(N_\mathrm{FFT,P})$ | $N_\mathrm{FB}$ |
| Splice signal length to $N_\mathrm{P}$ | in: $N_\mathrm{FFT,P}$<br>out: $N_\mathrm{P}$ | 0 | 0 | $N_\mathrm{FB}$ |
| Total | - | $3N_\mathrm{FFT,P}\log_2(N_\mathrm{FFT,P})$<br>$+N_\mathrm{FB}\left[4N_\mathrm{P}+2N_\mathrm{FFT,P}\right.$<br>$\left.+6N_\mathrm{FFT,P}\log_2(N_\mathrm{FFT,P})\right]$ | $2N_\mathrm{FFT,P}\log_2(N_\mathrm{FFT,P})$<br>$+N_\mathrm{FB}\left[8N_\mathrm{P}+4N_\mathrm{FFT,P}\right.$<br>$\left.+4N_\mathrm{FFT,P}\log_2(N_\mathrm{FFT,P})\right]$ | - |

#### *7.1.2. Hardware processing time*

A standalone processing-time expression is not given for **Figure 4.1** because the primary-code correlation is always embedded within one of the complete acquisition architectures considered in the following sections.

#### *7.1.3. Memory requirements*

The internal FFT/IFFT memory requirements used for the hardware model are those provided in **TABLE 4.1**. The additional memories required for the input signal, local-code spectrum, and coherent or non-coherent accumulators are included separately for each architecture.

### 7.2. Complexity for non-coherent integration

For the non-coherent architecture of Figure 4.2, $N_\mathrm{NC}+1$ primary-code periods of input samples must be stored because of the DBZP processing.

### *7.2.1. Software operation count*

**TABLE 7.2** gives the software operation count for the non-coherent architecture of **Figure 4.2**. For each frequency hypothesis, $N_{\mathrm{NC}}$ primary-code correlations are computed, followed by magnitude computation and non-coherent accumulation.

**TABLE 7.2** Software operation count for non-coherent integration (**Figure 4.2** )

| Computation stage | Vector size | # real addition | # real multiplications | # repetitions |
|---|---|---|---|---|
| Local code FFT | $N_{\mathrm{FFT,P}}$ | $3N_{\mathrm{FFT,P}}\log_2(N_{\mathrm{FFT,P}})$ | $2N_{\mathrm{FFT,P}}\log_2(N_{\mathrm{FFT,P}})$ | 1 |
| Primary code correlation excluding local code FFT | in: $2N_{\mathrm{P}}$<br>out: $N_{\mathrm{P}}$ | $4N_{\mathrm{P}}+2N_{\mathrm{FFT,P}}$ $+6N_{\mathrm{FFT,P}}\log_2(N_{\mathrm{FFT,P}})$ | $8N_{\mathrm{P}}+4N_{\mathrm{FFT,P}}$ $+4N_{\mathrm{FFT,P}}\log_2(N_{\mathrm{FFT,P}})$ | $N_{\mathrm{FB}}N_{\mathrm{NC}}$ |
| Magnitude computation | $N_{\mathrm{P}}$ | $N_{\mathrm{P}}$ | $2N_{\mathrm{P}}$ | $N_{\mathrm{FB}}N_{\mathrm{NC}}$ |
| Non-coherent integration | $N_{\mathrm{P}}$ | $N_{\mathrm{P}}$ | 0 | $N_{\mathrm{FB}}N_{\mathrm{NC}}$ |
| Total | - | $3N_{\mathrm{FFT,P}}\log_2(N_{\mathrm{FFT,P}})$ $+N_{\mathrm{FB}}N_{\mathrm{NC}}[6N_{\mathrm{P}}+2N_{\mathrm{FFT,P}}$ $+6N_{\mathrm{FFT,P}}\log_2(N_{\mathrm{FFT,P}})]$ | $2N_{\mathrm{FFT,P}}\log_2(N_{\mathrm{FFT,P}})$ $+N_{\mathrm{FB}}N_{\mathrm{NC}}[10N_{\mathrm{P}}+4N_{\mathrm{FFT,P}}$ $+4N_{\mathrm{FFT,P}}\log_2(N_{\mathrm{FFT,P}})]$ | - |

Summing the real additions and multiplications gives

$$N_{\mathrm{FB}}N_{\mathrm{NC}}\big(16N_{\mathrm{P}}+6N_{\mathrm{FFT,P}}+10N_{\mathrm{FFT,P}}\log_2(N_{\mathrm{FFT,P}})\big)+5N_{\mathrm{FFT,P}}\log_2(N_{\mathrm{FFT,P}}). \tag{6}$$

### *7.2.2. Hardware processing time*

**Figure 7.1** illustrates the processing sequence for $N_{\mathrm{NC}}=4$. Fixed FFT/IFFT pipeline latency is neglected throughout the analysis. Once the pipeline is filled, one primary-code correlation is processed every $N_{\mathrm{FFT,P}}$ clock cycles. The total number of clock cycles required to evaluate all $N_{\mathrm{FB}}$ frequency hypotheses is therefore

$$N_{\mathrm{FB}}N_{\mathrm{NC}}N_{\mathrm{FFT,P}}. \tag{7}$$

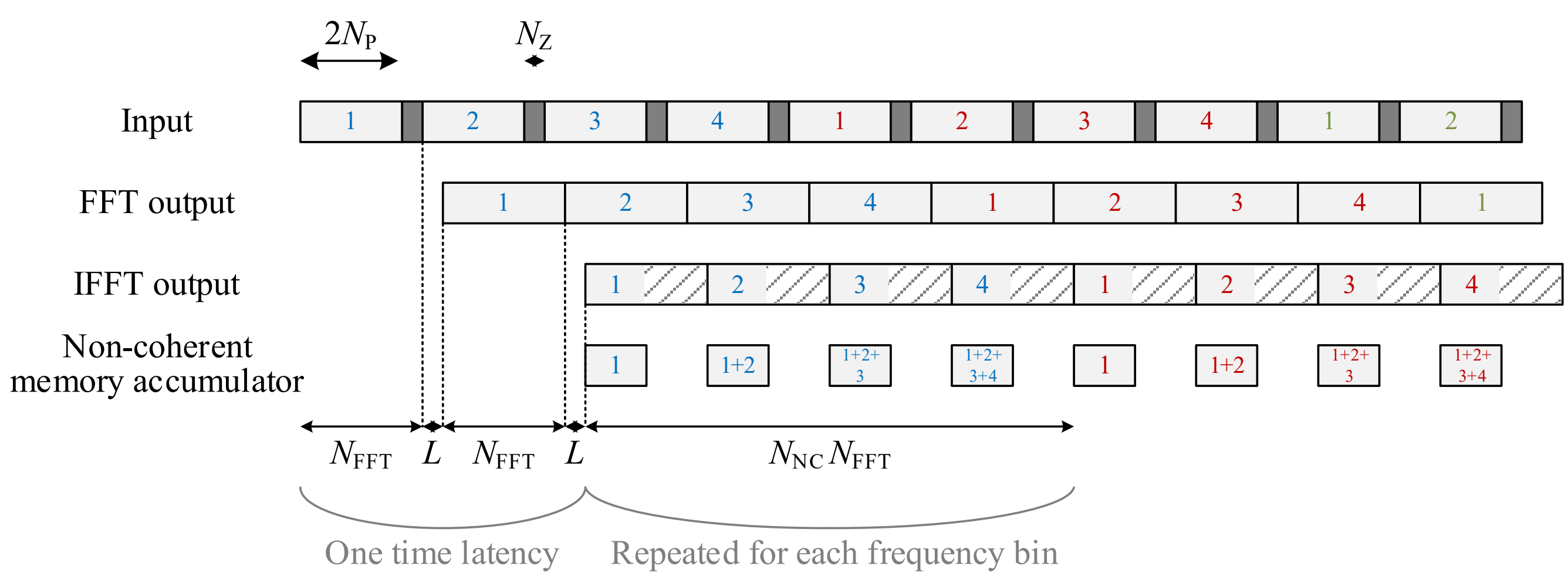

**Figure 7.1** Hardware timing for non-coherent integration, $N_{\mathrm{NC}}=4$ (**Figure 4.2**)

### *7.2.3. Memory requirements*

The memory requirement includes the input-signal buffer, the stored FFT of the local primary code, the internal FFT/IFFT memories, and one real non-coherent accumulator. Since the primary code is real, only half of its symmetric complex FFT is stored. The resulting total memory requirement is

$$(N_{\mathrm{NC}}+1)N_{\mathrm{P}}B_{\mathrm{ADC}}+N_{\mathrm{FFT,P}}B_{\mathrm{FFT,P}}/2+2M_{\mathrm{FFT,P}}+N_{\mathrm{P}}B_{\mathrm{NC}}. \tag{8}$$

### 7.3. Complexity for GPS L1 C/A with alternate block method

For the alternate block method, the number of stored input periods is $2N_{\mathrm{C}}+1$ for coherent-only integration and $2N_{\mathrm{NC}}N_{\mathrm{C}}+1$ when non-coherent integration is subsequently applied.

#### *7.3.1. Software operation count*

**TABLE 7.3** gives the operation count for the coherent-only architecture of **Figure 4.4**.

**TABLE 7.3** Software operation count for GPS L1 C/A alternate block method, coherent only (**Figure 4.4**)

| Computation stage | Vector size | # real additions | # real multiplications | # repetitions |
|---|---|---|---|---|
| Local code FFT | $N_{\mathrm{FFT,P}}$ | $3N_{\mathrm{FFT,P}}\log_2(N_{\mathrm{FFT,P}})$ | $2N_{\mathrm{FFT,P}}\log_2(N_{\mathrm{FFT,P}})$ | 1 |
| Primary code correlation excluding local code FFT | in: $2N_{\mathrm{P}}$<br>out: $N_{\mathrm{P}}$ | $4N_{\mathrm{P}}+2N_{\mathrm{FFT,P}}$ $+6N_{\mathrm{FFT,P}}\log_2(N_{\mathrm{FFT,P}})$ | $8N_{\mathrm{P}}+4N_{\mathrm{FFT,P}}$ $+4N_{\mathrm{FFT,P}}\log_2(N_{\mathrm{FFT,P}})$ | $2N_{\mathrm{FB}}N_{\mathrm{C}}$ |
| Extra coherent integration | $N_{\mathrm{P}}$ | $2N_{\mathrm{P}}$ | 0 | $2N_{\mathrm{FB}}N_{\mathrm{C}}$ |
| Magnitude computation | $N_{\mathrm{P}}$ | $N_{\mathrm{P}}$ | $2N_{\mathrm{P}}$ | $2N_{\mathrm{FB}}$ |
| Total | - | $3N_{\mathrm{FFT,P}}\log_2(N_{\mathrm{FFT,P}})$ $+2N_{\mathrm{FB}}[N_{\mathrm{P}}$ $+N_{\mathrm{C}}(6N_{\mathrm{P}}+2N_{\mathrm{FFT,P}}$ $+6N_{\mathrm{FFT,P}}\log_2(N_{\mathrm{FFT,P}}))]$ | $2N_{\mathrm{FFT,P}}\log_2(N_{\mathrm{FFT,P}})$ $+2N_{\mathrm{FB}}[2N_{\mathrm{P}}$ $+N_{\mathrm{C}}(8N_{\mathrm{P}}+4N_{\mathrm{FFT,P}}$ $+4N_{\mathrm{FFT,P}}\log_2(N_{\mathrm{FFT,P}}))]$ | - |

The resulting total number of operations is

$$2N_{\mathrm{FB}}\left(3N_{\mathrm{P}}+N_{\mathrm{C}}\left(14N_{\mathrm{P}}+6N_{\mathrm{FFT,P}}+10N_{\mathrm{FFT,P}}\log_2(N_{\mathrm{FFT,P}})\right)\right)+5N_{\mathrm{FFT,P}}\log_2(N_{\mathrm{FFT,P}}). \tag{9}$$

**TABLE 7.4** gives the corresponding operation count for the hybrid architecture of **Figure 4.5**.

**TABLE 7.4** Software operation count for GPS L1 C/A alternate block method, hybrid (**Figure 4.5**)

| Computation stage | Vector size | # real additions | # real multiplications | # repetitions |
|---|---|---|---|---|
| Local code FFT | $N_{\mathrm{FFT,P}}$ | $3N_{\mathrm{FFT,P}}\log_2(N_{\mathrm{FFT,P}})$ | $2N_{\mathrm{FFT,P}}\log_2(N_{\mathrm{FFT,P}})$ | 1 |
| Primary code correlation excluding local code FFT | in: $2N_{\mathrm{P}}$<br>out: $N_{\mathrm{P}}$ | $4N_{\mathrm{P}}+2N_{\mathrm{FFT,P}}$ $+6N_{\mathrm{FFT,P}}\log_2(N_{\mathrm{FFT,P}})$ | $8N_{\mathrm{P}}+4N_{\mathrm{FFT,P}}$ $+4N_{\mathrm{FFT,P}}\log_2(N_{\mathrm{FFT,P}})$ | $2N_{\mathrm{FB}}N_{\mathrm{NC}}N_{\mathrm{C}}$ |
| Extra coherent integration | $N_{\mathrm{P}}$ | $2N_{\mathrm{P}}$ | 0 | $2N_{\mathrm{FB}}N_{\mathrm{NC}}N_{\mathrm{C}}$ |
| Magnitude computation | $N_{\mathrm{P}}$ | $N_{\mathrm{P}}$ | $2N_{\mathrm{P}}$ | $2N_{\mathrm{FB}}N_{\mathrm{NC}}$ |
| Non-coherent integration | $N_{\mathrm{P}}$ | $N_{\mathrm{P}}$ | 0 | $2N_{\mathrm{FB}}N_{\mathrm{NC}}$ |
| Total | - | $3N_{\mathrm{FFT,P}}\log_2(N_{\mathrm{FFT,P}})$ $+2N_{\mathrm{FB}}N_{\mathrm{NC}}[2N_{\mathrm{P}}$ $+N_{\mathrm{C}}(6N_{\mathrm{P}}+2N_{\mathrm{FFT,P}}$ $+6N_{\mathrm{FFT,P}}\log_2(N_{\mathrm{FFT,P}}))]$ | $2N_{\mathrm{FFT,P}}\log_2(N_{\mathrm{FFT,P}})$ $+2N_{\mathrm{FB}}N_{\mathrm{NC}}[2N_{\mathrm{P}}$ $+N_{\mathrm{C}}(8N_{\mathrm{P}}+4N_{\mathrm{FFT,P}}$ $+4N_{\mathrm{FFT,P}}\log_2(N_{\mathrm{FFT,P}}))]$ | - |

The resulting total number of operations is

$$2N_{\mathrm{FB}}N_{\mathrm{NC}}\left(4N_{\mathrm{P}}+N_{\mathrm{C}}\left(14N_{\mathrm{P}}+6N_{\mathrm{FFT,P}}+10N_{\mathrm{FFT,P}}\log_2(N_{\mathrm{FFT,P}})\right)\right)+5N_{\mathrm{FFT,P}}\log_2(N_{\mathrm{FFT,P}}). \tag{10}$$

#### *7.3.2. Hardware processing time*

**Figure 7.2** illustrates the processing sequence for $N_{\mathrm{C}}=2$ and $N_{\mathrm{NC}}=2$. Since the two coherent paths are processed independently, the total number of clock cycles required to evaluate the complete frequency search is

$$2N_{\mathrm{FB}}N_{\mathrm{NC}}N_{\mathrm{C}}N_{\mathrm{FFT,P}}. \tag{11}$$

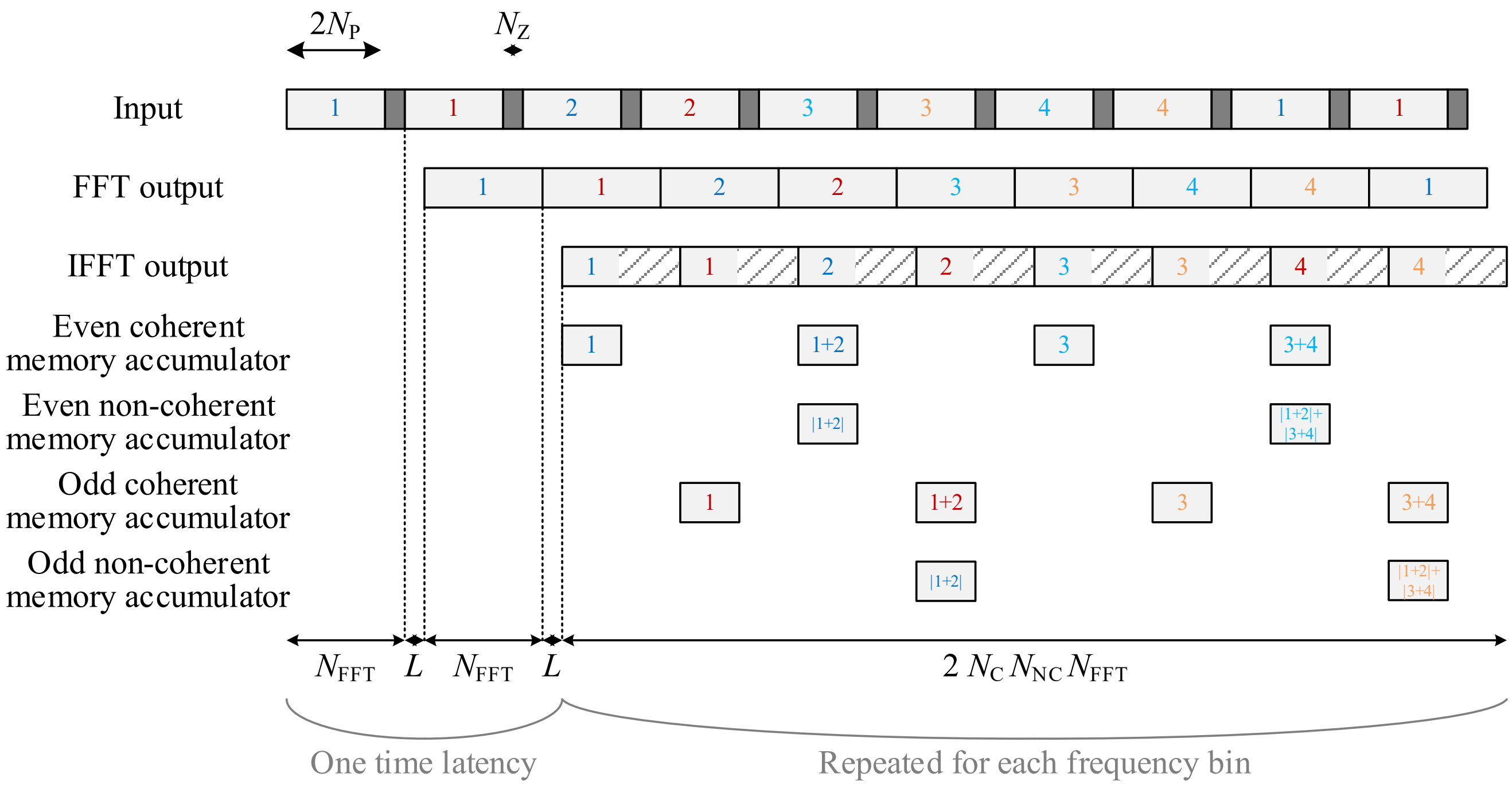

**Figure 7.2** Hardware timing for GPS L1 C/A alternate block method, $N_\mathrm{C} = 2$, $N_\mathrm{NC} = 2$ (**Figure 4.5**)

### *7.3.3. Memory requirements*

For the coherent-only architecture of **Figure 4.4**, the memory requirement is

$$(2N_\mathrm{C} + 1)N_\mathrm{P}B_\mathrm{ADC} + N_\mathrm{FFT,P}B_\mathrm{FFT,P}/2 + 2M_\mathrm{FFT,P} + N_\mathrm{P}B_\mathrm{C}. \tag{12}$$

When non-coherent integration is included as in **Figure 4.5**, two coherent and two non-coherent accumulator memories are required, giving

$$(2N_\mathrm{NC}N_\mathrm{C} + 1)N_\mathrm{P}B_\mathrm{ADC} + N_\mathrm{FFT,P}B_\mathrm{FFT,P}/2 + 2M_\mathrm{FFT,P} + 2N_\mathrm{P}B_\mathrm{C} + 2N_\mathrm{P}B_\mathrm{NC}. \tag{13}$$

## 7.4. Complexity for GPS L1 C/A with full-bits method

The full-bits architecture depends on the required coherent integration length. For $N_\mathrm{C} \leq 20$, the architecture of **Figure 4.6** is used. For $N_\mathrm{C} = 20N_\mathrm{B} > 20$, the 20 ms coherent outputs must additionally be combined across $N_\mathrm{B}$ data bits using either a parallel implementation, a serial implementation with intermediate memory, or a serial implementation without intermediate memory, as shown in **Figure 4.7** to **Figure 4.9**. The serial implementations exploit Gray-code ordering to reduce the amount of processing required between successive hypotheses.

The number of input primary-code periods used and stored is $N_\mathrm{C}(N_\mathrm{NC} + 1) - 1$\) for $N_\mathrm{C} \leq 20$, and $20(N_\mathrm{B} + 1) - 1$ for $N_\mathrm{C} = 20N_\mathrm{B} > 20$ with $N_\mathrm{NC} = 1$.

### *7.4.1. Software operation count*

**TABLE 7.5** gives the operation count for the full-bits architecture of **Figure 4.6** with $N_\mathrm{C} \leq 20$.

**TABLE 7.5** Software operation count for GPS L1 C/A full-bits method, $N_\mathrm{C} \leq 20$ (**Figure 4.6**)

| Computation stage | Vector size | # real additions | # real multiplications | # repetitions |
|---|---|---|---|---|
| Local code FFT | $N_\mathrm{FFT,P}$ | $3N_\mathrm{FFT,P}\log_2(N_\mathrm{FFT,P})$ | $2N_\mathrm{FFT,P}\log_2(N_\mathrm{FFT,P})$ | 1 |
| Primary code correlation excluding local code FFT | in: $2N_\mathrm{P}$<br>out: $N_\mathrm{P}$ | $4N_\mathrm{P} + 2N_\mathrm{FFT,P}$<br>$+ 6N_\mathrm{FFT,P}\log_2(N_\mathrm{FFT,P})$ | $8N_\mathrm{P} + 4N_\mathrm{FFT,P}$<br>$+ 4N_\mathrm{FFT,P}\log_2(N_\mathrm{FFT,P})$ | $[N_\mathrm{C}(N_\mathrm{NC} + 1)$<br>$- 1]N_\mathrm{FB}$ |
| Extra coherent integration | $N_\mathrm{P}$ | $2N_\mathrm{P}N_\mathrm{C}$ | 0 | $N_\mathrm{FB}N_\mathrm{NC}N_\mathrm{C}$ |
| Magnitude computation | $N_{\mathrm{P}}$ | $N_{\mathrm{P}} N_{\mathrm{C}}$ | $2N_{\mathrm{P}} N_{\mathrm{C}}$ | $N_{\mathrm{FB}} N_{\mathrm{NC}}$ |
| Non-coherent integration | $N_{\mathrm{P}}$ | $N_{\mathrm{P}} N_{\mathrm{C}}$ | 0 | $N_{\mathrm{FB}} N_{\mathrm{NC}}$ |
| Total | - | $3N_{\mathrm{FFT,P}} \log_2(N_{\mathrm{FFT,P}})$ $+[N_{\mathrm{C}}(N_{\mathrm{NC}}+1)-1]N_{\mathrm{FB}}[4N_{\mathrm{P}}$ $+2N_{\mathrm{FFT,P}}$ $+6N_{\mathrm{FFT,P}}\log_2(N_{\mathrm{FFT,P}})]$ $+2N_{\mathrm{P}}N_{\mathrm{C}}N_{\mathrm{NC}}N_{\mathrm{FB}}(N_{\mathrm{C}}+1)$ | $2N_{\mathrm{FFT,P}} \log_2(N_{\mathrm{FFT,P}})$ $+[N_{\mathrm{C}}(N_{\mathrm{NC}}+1)-1]N_{\mathrm{FB}}[8N_{\mathrm{P}}$ $+4N_{\mathrm{FFT,P}}$ $+4N_{\mathrm{FFT,P}}\log_2(N_{\mathrm{FFT,P}})]$ $+2N_{\mathrm{P}}N_{\mathrm{C}}N_{\mathrm{NC}}N_{\mathrm{FB}}$ | - |

The resulting total number of operations is

$$[N_{\mathrm{C}}(N_{\mathrm{NC}}+1)-1]N_{\mathrm{FB}}[12N_{\mathrm{P}}+6N_{\mathrm{FFT,P}}+10N_{\mathrm{FFT,P}}\log_2(N_{\mathrm{FFT,P}})]+2N_{\mathrm{P}}N_{\mathrm{C}}N_{\mathrm{NC}}N_{\mathrm{FB}}(N_{\mathrm{C}}+2) + 5N_{\mathrm{FFT,P}}\log_2(N_{\mathrm{FFT,P}}). \quad (14)$$

For $N_{\mathrm{C}} = 20N_{\mathrm{B}} > 20$, **TABLE 7.6** gives the operation count for the parallel architecture of **Figure 4.7**, where all $2^{N_{\mathrm{B}}-1}$ sign hypotheses are evaluated simultaneously.

**TABLE 7.6** Software operation count for GPS L1 C/A full-bits method, parallel, $N_{\mathrm{C}} > 20$ (**Figure 4.7**)

| Computation stage | Vector size | # real additions | # real multiplications | # repetitions |
|---|---|---|---|---|
| Local code FFT | $N_{\mathrm{FFT,P}}$ | $3N_{\mathrm{FFT,P}}\log_2(N_{\mathrm{FFT,P}})$ | $2N_{\mathrm{FFT,P}}\log_2(N_{\mathrm{FFT,P}})$ | 1 |
| Primary code correlation excluding local code FFT | in: $2N_{\mathrm{P}}$ out: $N_{\mathrm{P}}$ | $4N_{\mathrm{P}}+2N_{\mathrm{FFT,P}}$ $+6N_{\mathrm{FFT,P}}\log_2(N_{\mathrm{FFT,P}})$ | $8N_{\mathrm{P}}+4N_{\mathrm{FFT,P}}$ $+4N_{\mathrm{FFT,P}}\log_2(N_{\mathrm{FFT,P}})$ | $[20(N_{\mathrm{B}}+1)$ $-1]N_{\mathrm{FB}}$ |
| Extra coherent integration over 1 data bit | $N_{\mathrm{P}}$ | $20\times 2N_{\mathrm{P}}$ | 0 | $20N_{\mathrm{FB}}N_{\mathrm{B}}$ |
| Extra coherent integration over $N_{\mathrm{B}}$ data bit | $N_{\mathrm{P}}$ | $20\times 2^{N_{\mathrm{B}}-1}2N_{\mathrm{P}}$ | 0 | $N_{\mathrm{FB}}N_{\mathrm{B}}$ |
| Magnitude computation | $N_{\mathrm{P}}$ | $20\times 2^{N_{\mathrm{B}}-1}N_{\mathrm{P}}$ | $20\times 2^{N_{\mathrm{B}}-1}2N_{\mathrm{P}}$ | $N_{\mathrm{FB}}$ |
| Total | - | $3N_{\mathrm{FFT,P}}\log_2(N_{\mathrm{FFT,P}})$ $+[20(N_{\mathrm{B}}+1)-1]N_{\mathrm{FB}}[4N_{\mathrm{P}}$ $+2N_{\mathrm{FFT,P}}$ $+6N_{\mathrm{FFT,P}}\log_2(N_{\mathrm{FFT,P}})]$ $+N_{\mathrm{P}}N_{\mathrm{FB}}(800N_{\mathrm{B}}$ $+(20N_{\mathrm{B}}+10)2^{N_{\mathrm{B}}})$ | $2N_{\mathrm{FFT,P}}\log_2(N_{\mathrm{FFT,P}})$ $+[20(N_{\mathrm{B}}+1)-1]N_{\mathrm{FB}}[8N_{\mathrm{P}}$ $+4N_{\mathrm{FFT,P}}$ $+4N_{\mathrm{FFT,P}}\log_2(N_{\mathrm{FFT,P}})]$ $+N_{\mathrm{FB}}N_{\mathrm{P}}(20\times 2^{N_{\mathrm{B}}})$ | - |

The resulting total number of operations is

$$[20(N_{\mathrm{B}}+1)-1]N_{\mathrm{FB}}[12N_{\mathrm{P}}+6N_{\mathrm{FFT,P}}+10N_{\mathrm{FFT,P}}\log_2(N_{\mathrm{FFT,P}})]+N_{\mathrm{FB}}N_{\mathrm{P}}(800N_{\mathrm{B}}+(20N_{\mathrm{B}}+30)\times 2^{N_{\mathrm{B}}}) + 5N_{\mathrm{FFT,P}}\log_2(N_{\mathrm{FFT,P}}). \quad (15)$$

**TABLE 7.7** considers the serial architecture with intermediate memory shown in **Figure 4.8**. The first hypothesis is formed from the $N_{\mathrm{B}}$ stored 20 ms correlation vectors, after which Gray-code ordering allows each additional hypothesis to be obtained by updating only one contribution.

**TABLE 7.7** Software operation count for GPS L1 C/A full-bits method, serial with intermediate memory, $N_{\mathrm{C}} > 20$ (**Figure 4.8**)

| Computation stage | Vector size | # real additions | # real multiplications | # repetitions |
|---|---|---|---|---|
| Local code FFT | $N_{\mathrm{FFT,P}}$ | $3N_{\mathrm{FFT,P}}\log_2(N_{\mathrm{FFT,P}})$ | $2N_{\mathrm{FFT,P}}\log_2(N_{\mathrm{FFT,P}})$ | 1 |
| Primary code correlation excluding local code FFT | in: $2N_{\mathrm{P}}$ out: $N_{\mathrm{P}}$ | $4N_{\mathrm{P}}+2N_{\mathrm{FFT,P}}$ $+6N_{\mathrm{FFT,P}}\log_2(N_{\mathrm{FFT,P}})$ | $8N_{\mathrm{P}}+4N_{\mathrm{FFT,P}}$ $+4N_{\mathrm{FFT,P}}\log_2(N_{\mathrm{FFT,P}})$ | $[20(N_{\mathrm{B}}+1)$ $-1]N_{\mathrm{FB}}$ |
| Extra coherent integration over 1 data bit | $N_\mathrm{P}$ | $20 \times 2N_\mathrm{P}$ | 0 | $20N_\mathrm{FB}N_\mathrm{B}$ |
| First output computation over $N_\mathrm{B}$ data bits | $N_\mathrm{P}$ | $20 \times 2(N_\mathrm{B} - 1)N_\mathrm{P}$ | 0 | $N_\mathrm{FB}$ |
| Other outputs computation using Gray recursion | $N_\mathrm{P}$ | $20 \times 2(2^{N_\mathrm{B}-1} - 1)N_\mathrm{P}$ | 0 | $N_\mathrm{FB}$ |
| Magnitude computation | $N_\mathrm{P}$ | $20 \times 2^{N_\mathrm{B}-1}N_\mathrm{P}$ | $20 \times 2^{N_\mathrm{B}-1}2N_\mathrm{P}$ | $N_\mathrm{FB}$ |
| Total | - | $3N_\mathrm{FFT,P}\log_2(N_\mathrm{FFT,P}) + [20(N_\mathrm{B}+1) - 1]N_\mathrm{FB}[4N_\mathrm{P} + 2N_\mathrm{FFT,P} + 6N_\mathrm{FFT,P}\log_2(N_\mathrm{FFT,P})] + 20N_\mathrm{P}N_\mathrm{FB}(42N_\mathrm{B} - 4 + 3 \times 2^{N_\mathrm{B}-1})$ | $2N_\mathrm{FFT,P}\log_2(N_\mathrm{FFT,P}) + [20(N_\mathrm{B}+1) - 1]N_\mathrm{FB}[8N_\mathrm{P} + 4N_\mathrm{FFT,P} + 4N_\mathrm{FFT,P}\log_2(N_\mathrm{FFT,P})] + 20N_\mathrm{P}N_\mathrm{FB}2^{N_\mathrm{B}}$ | - |

The resulting total number of operations is

$$[20(N_\mathrm{B}+1) - 1]N_\mathrm{FB}[12N_\mathrm{P} + 6N_\mathrm{FFT,P} + 10N_\mathrm{FFT,P}\log_2(N_\mathrm{FFT,P})] + N_\mathrm{P}N_\mathrm{FB}(840N_\mathrm{B} + 50 \times 2^{N_\mathrm{B}} - 80) + 5N_\mathrm{FFT,P}\log_2(N_\mathrm{FFT,P}). \tag{16}$$

**TABLE 7.8** considers the serial architecture without intermediate memory shown in **Figure 4.9**. In this case, the modified 20 ms contribution must be recomputed at each Gray-code transition. Across the 20 shifted branches, this requires 39 additional primary-code correlations for every new hypothesis.

**TABLE 7.8** Software operation count for GPS L1 C/A full-bits method, serial without intermediate memory, $N_\mathrm{C} > 20$ (**Figure 4.9**)

| Computation stage | Vector size | # real additions | # real multiplications | # repetitions |
|---|---|---|---|---|
| Local code FFT | $N_\mathrm{FFT,P}$ | $3N_\mathrm{FFT,P}\log_2(N_\mathrm{FFT,P})$ | $2N_\mathrm{FFT,P}\log_2(N_\mathrm{FFT,P})$ | 1 |
| Primary code correlation excluding local code FFT | in: $2N_\mathrm{P}$ out: $N_\mathrm{P}$ | $4N_\mathrm{P} + 2N_\mathrm{FFT,P} + 6N_\mathrm{FFT,P}\log_2(N_\mathrm{FFT,P})$ | $8N_\mathrm{P} + 4N_\mathrm{FFT,P} + 4N_\mathrm{FFT,P}\log_2(N_\mathrm{FFT,P})$ | $N_\mathrm{FB}[20(N_\mathrm{B} + 1) - 1 + 39(2^{N_\mathrm{B}-1} - 1)]$ |
| Extra coherent integration over 1 data bit | $N_\mathrm{P}$ | $20 \times 2N_\mathrm{P}$ | 0 | $20N_\mathrm{FB}(N_\mathrm{B} + 2^{N_\mathrm{B}-1} - 1)$ |
| First output computation over $N_\mathrm{B}$ data bits | $N_\mathrm{P}$ | $20 \times (N_\mathrm{B} - 1)2N_\mathrm{P}$ | 0 | $N_\mathrm{FB}$ |
| Other outputs computation using Gray recursion | $N_\mathrm{P}$ | $20 \times (2^{N_\mathrm{B}-1} - 1)2N_\mathrm{P}$ | 0 | $N_\mathrm{FB}$ |
| Magnitude computation | $N_\mathrm{P}$ | $20 \times 2^{N_\mathrm{B}-1}N_\mathrm{P}$ | $20 \times 2^{N_\mathrm{B}-1}2N_\mathrm{P}$ | $N_\mathrm{FB}$ |
| Total | - | $3N_\mathrm{FFT,P}\log_2(N_\mathrm{FFT,P}) + N_\mathrm{FB}[20(N_\mathrm{B}+1) - 1 + 39(2^{N_\mathrm{B}-1} - 1)][4N_\mathrm{P} + 2N_\mathrm{FFT,P} + 6N_\mathrm{FFT,P}\log_2(N_\mathrm{FFT,P})] + 20N_\mathrm{P}N_\mathrm{FB}(42N_\mathrm{B} + 43 \times 2^{N_\mathrm{B}-1} - 44)$ | $2N_\mathrm{FFT,P}\log_2(N_\mathrm{FFT,P}) + N_\mathrm{FB}[20(N_\mathrm{B}+1) - 1 + 39(2^{N_\mathrm{B}-1} - 1)][8N_\mathrm{P} + 4N_\mathrm{FFT,P} + 4N_\mathrm{FFT,P}\log_2(N_\mathrm{FFT,P})] + 20N_\mathrm{P}N_\mathrm{FB}2^{N_\mathrm{B}}$ | - |

The resulting total number of operations is

$$N_{\mathrm{FB}}[20(N_{\mathrm{B}}+1)-1+39(2^{N_{\mathrm{B}}-1}-1)]\left[12N_{\mathrm{P}}+6N_{\mathrm{FFT,P}}+10N_{\mathrm{FFT,P}}\log_2(N_{\mathrm{FFT,P}})\right] \\ +20N_{\mathrm{P}}N_{\mathrm{FB}}(42N_{\mathrm{B}}+45\times 2^{N_{\mathrm{B}}-1}-44)+5N_{\mathrm{FFT,P}}\log_2(N_{\mathrm{FFT,P}}). \tag{17}$$

### *7.4.2. Hardware processing time*

For $N_{\mathrm{C}} \leq 20$, **Figure 7.3** illustrates the timing of the full-bits architecture of **Figure 4.6**. The corresponding number of clock cycles is

$$N_{\mathrm{FB}}\left(N_{\mathrm{C}}N_{\mathrm{NC}}N_{\mathrm{FFT,P}}+(N_{\mathrm{C}}-1)N_{\mathrm{FFT,P}}\right)=N_{\mathrm{FB}}N_{\mathrm{FFT,P}}\left((N_{\mathrm{NC}}+1)N_{\mathrm{C}}-1\right). \tag{18}$$

For $N_{\mathrm{C}} = 20N_{\mathrm{B}} > 20$, all sign hypotheses are evaluated concurrently in the parallel architecture of **Figure 4.7**. Its processing time is therefore determined only by the number of required input primary-code periods:

$$N_{\mathrm{FB}}N_{\mathrm{FFT,P}}(20(N_{\mathrm{B}}+1)-1). \tag{19}$$

In the serial architecture with intermediate memory of **Figure 4.8**, the input signal is processed once and the intermediate 20 ms correlation vectors are stored. **Figure 7.4** illustrates the corresponding timing, giving

$$N_{\mathrm{FB}}\left((20(N_{\mathrm{B}}+1)-1)N_{\mathrm{FFT,P}}+2^{N_{\mathrm{B}}-1}N_{\mathrm{P}}\right) \tag{20}$$

In the serial architecture without intermediate memory of **Figure 4.9**, each Gray-code transition requires recomputation of the modified 20 ms contribution. The resulting processing time is

$$N_{\mathrm{FB}}N_{\mathrm{FFT,P}}\left(20(N_{\mathrm{B}}+1)-1+39(2^{N_{\mathrm{B}}-1}-1)\right) \tag{21}$$

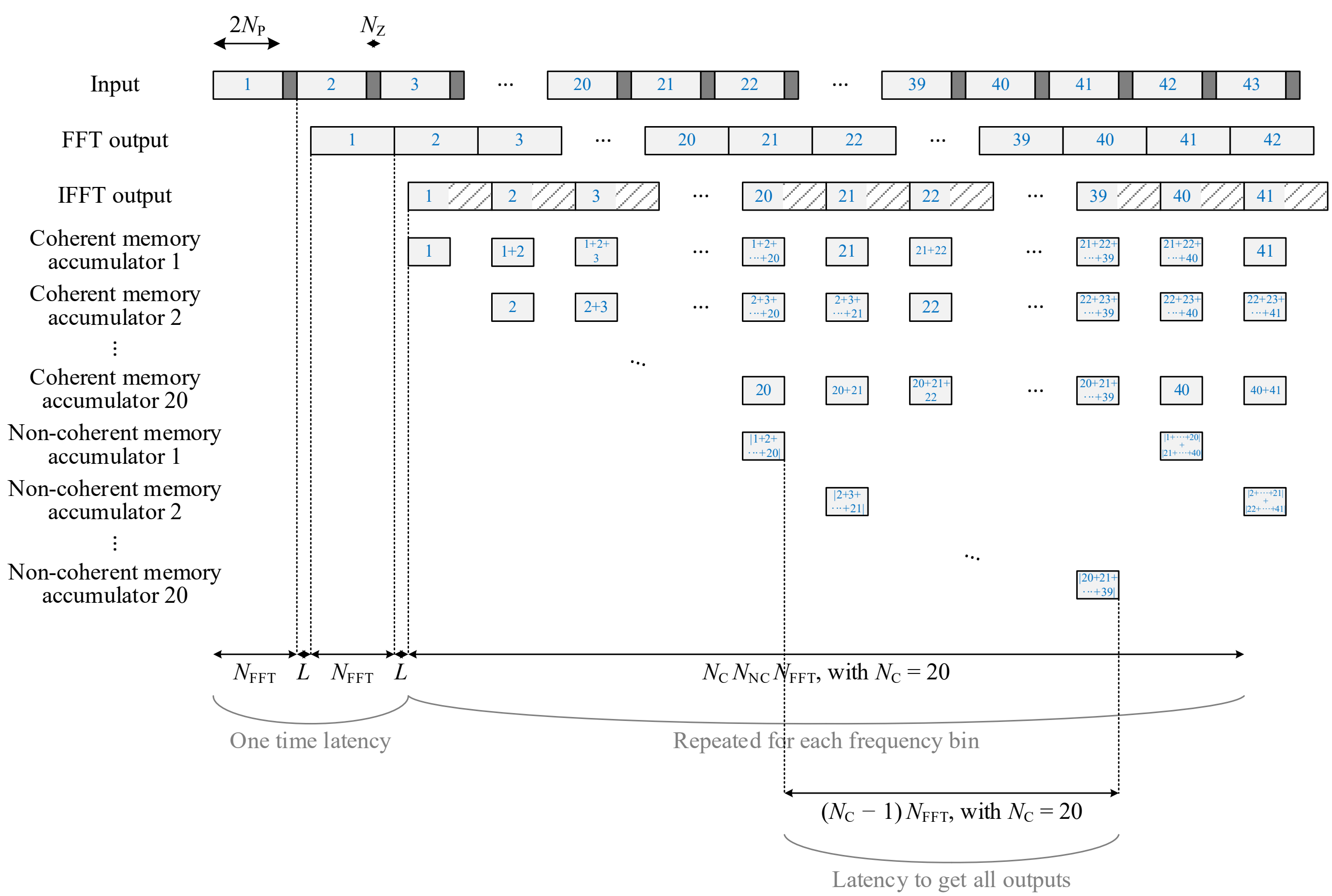

**Figure 7.3** Hardware timing for GPS L1 C/A full-bits method, $N_{\mathrm{C}} = 20$, $N_{\mathrm{NC}} = 2$ (**Figure 4.6**)

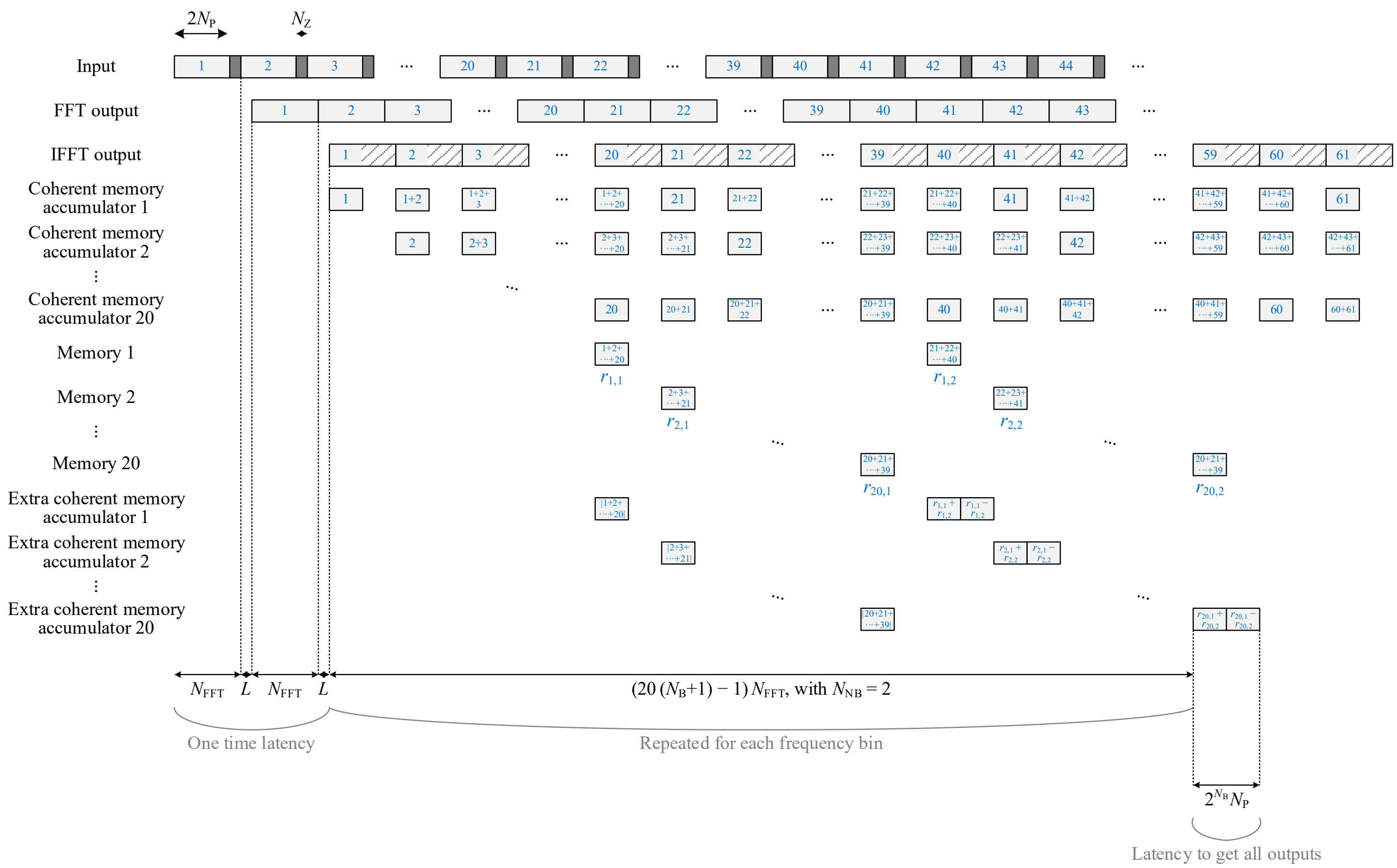

**Figure 7.4** Hardware timing for GPS L1 C/A full-bits method, serial with intermediate memory, $N_C = 40$ (**Figure 4.8**)

### *7.4.3. Memory requirements*

For the architecture of Figure 4.6 with $N_C \le 20$, the required memory is

$$\begin{cases} (2N_C - 1)N_P B_{ADC} + N_{FFT,P}B_{FFT,P}/2 + 2M_{FFT,P} + N_C N_P B_C, & \text{if } N_{NC} = 1 \\ (N_C(N_{NC} + 1) - 1)N_P B_{ADC} + N_{FFT,P}B_{FFT,P}/2 + 2M_{FFT,P} + N_C N_P B_C + N_C N_P B_{NC}, & \text{if } N_{NC} > 1 \end{cases}. \tag{22}$$

For $N_C = 20N_B > 20$, the parallel architecture of **Figure 4.7** requires $20 \times 2^{N_B - 1}$ second-stage coherent accumulators in addition to the 20 first-stage accumulators, giving

$$(20(N_B + 1) - 1)N_P B_{ADC} + N_{FFT,P}B_{FFT,P}/2 + 2M_{FFT,P} + (1 + 2^{N_B - 1})20N_P B_C. \tag{23}$$

For the serial architecture with intermediate memory of **Figure 4.8**, the $N_B$ intermediate 20 ms vectors are stored for each of the 20 branches, giving

$$(20(N_B + 1) - 1)N_P B_{ADC} + N_{FFT,P}B_{FFT,P}/2 + 2M_{FFT,P} + (N_B + 2)20N_P B_C. \tag{24}$$

Finally, removing these intermediate memories as in **Figure 4.9** gives

$$(20(N_B + 1) - 1)N_P B_{ADC} + N_{FFT,P}B_{FFT,P}/2 + 2M_{FFT,P} + 40N_P B_C. \tag{25}$$

## 7.5. Complexity for GPS L1 C/A with average bit-transition loss method

For the average bit-transition loss architecture of **Figure 4.10**, $N_{NC}N_C + 1$ primary-code periods of input samples are stored.

### *7.5.1. Software operation count*

**TABLE 7.9** gives the software operation count for the architecture of **Figure 4.10**.

**TABLE 7.9** Software operation count for GPS L1 C/A average bit-transition loss method (**Figure 4.10**)

| Computation stage | Vector size | # real additions | # real multiplications | # repetitions |
|---|---|---|---|---|
| Local code FFT | $N_{\text{FFT,P}}$ | $3N_{\text{FFT,P}}\log_2(N_{\text{FFT,P}})$ | $2N_{\text{FFT,P}}\log_2(N_{\text{FFT,P}})$ | 1 |
| Primary code correlation excluding local code FFT | in: $2N_{\text{P}}$ out: $N_{\text{P}}$ | $4N_{\text{P}} + 2N_{\text{FFT,P}} + 6N_{\text{FFT,P}}\log_2(N_{\text{FFT,P}})$ | $8N_{\text{P}} + 4N_{\text{FFT,P}} + 4N_{\text{FFT,P}}\log_2(N_{\text{FFT,P}})$ | $N_{\text{C}}N_{\text{NC}}N_{\text{FB}}$ |
| Extra coherent integration | $N_{\text{P}}$ | $2N_{\text{P}}$ | 0 | $N_{\text{C}}N_{\text{NC}}N_{\text{FB}}$ |
| Magnitude computation | $N_{\text{P}}$ | $N_{\text{P}}$ | $2N_{\text{P}}$ | $N_{\text{NC}}N_{\text{FB}}$ |
| Non-coherent integration | $N_{\text{P}}$ | $N_{\text{P}}$ | 0 | $N_{\text{NC}}N_{\text{FB}}$ |
| Total | - | $3N_{\text{FFT,P}}\log_2(N_{\text{FFT,P}}) + N_{\text{NC}}N_{\text{FB}}[2N_{\text{P}} + N_{\text{C}}(10N_{\text{P}} + 2N_{\text{FFT,P}} + 6N_{\text{FFT,P}}\log_2(N_{\text{FFT,P}}))]$ | $2N_{\text{FFT,P}}\log_2(N_{\text{FFT,P}}) + N_{\text{NC}}N_{\text{FB}}[2N_{\text{P}} + N_{\text{C}}(8N_{\text{P}} + 4N_{\text{FFT,P}} + 4N_{\text{FFT,P}}\log_2(N_{\text{FFT,P}}))]$ | - |

Summing the real additions and multiplications gives

$$N_{\text{NC}}N_{\text{FB}}[4N_{\text{P}} + N_{\text{C}}(14N_{\text{P}} + 6N_{\text{FFT,P}} + 10N_{\text{FFT,P}}\log_2(N_{\text{FFT,P}}))] + 5N_{\text{FFT,P}}\log_2(N_{\text{FFT,P}}). \quad (26)$$

### *7.5.2. Hardware processing time*

**Figure 7.5** illustrates the processing sequence for $N_{\text{C}} = 3$ and $N_{\text{NC}} = 2$. Since $N_{\text{C}}N_{\text{NC}}$ primary-code periods are processed for each frequency hypothesis, the total number of clock cycles is

$$N_{\text{FB}}N_{\text{C}}N_{\text{NC}}N_{\text{FFT,P}}. \quad (27)$$

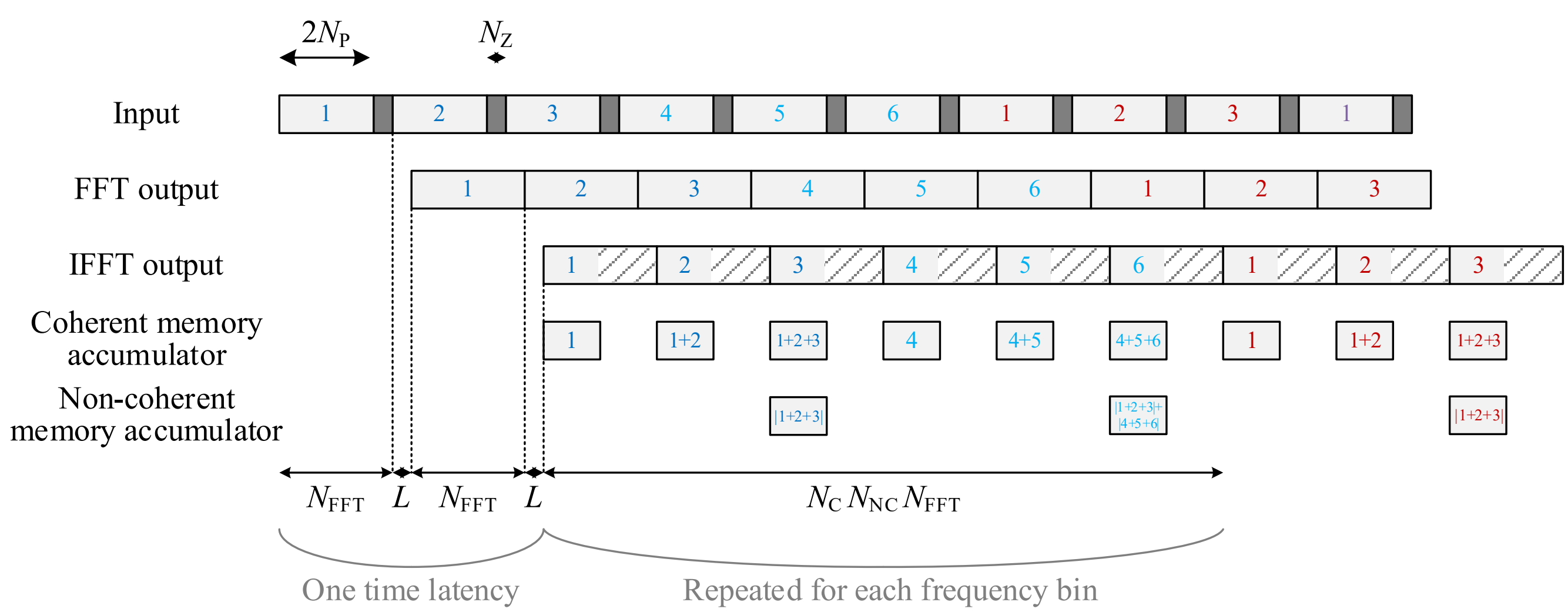

**Figure 7.5** Hardware timing for GPS L1 C/A average bit-transition loss method, $N_{\text{C}} = 3$, $N_{\text{NC}} = 2$ (**Figure 4.10**)

### *7.5.3. Memory requirements*

For the architecture of **Figure 4.10**, the required memory is

$$(N_{\text{NC}}N_{\text{C}} + 1)N_{\text{P}}B_{\text{ADC}} + N_{\text{FFT,P}}B_{\text{FFT,P}}/2 + 2M_{\text{FFT,P}} + N_{\text{P}}B_{\text{C}} + N_{\text{P}}B_{\text{NC}} \quad (28)$$

### 7.6. Complexity for multi-hypothesis method

The multi-hypothesis architectures of **Figure 4.12** to **Figure 4.17** are used when a small number $N_{\text{C}}$ of successive primary-code periods must be coherently integrated while their relative signs are unknown. The $2^{N_{\text{C}}-1}$ unique sign hypotheses can be evaluated either in parallel or serially using Gray-code recursion. The number of stored input periods is $N_{\text{C}} + 1$ for coherent-only acquisition and $N_{\text{NC}}N_{\text{C}} + 1$ for hybrid acquisition.

### *7.6.1. Software operation count*

**TABLE 7.10** gives the operation count for the parallel coherent-only architecture of **Figure 4.12**.

**TABLE 7.10** Software operation count for multi-hypothesis method, parallel coherent only (**Figure 4.12**)

| Computation stage | Vector size | # real additions | # real multiplications | # repetitions |
|---|---|---|---|---|
| Local code FFT | $N_{\mathrm{FFT,P}}$ | $3N_{\mathrm{FFT,P}}\log_2(N_{\mathrm{FFT,P}})$ | $2N_{\mathrm{FFT,P}}\log_2(N_{\mathrm{FFT,P}})$ | 1 |
| Primary code correlation excluding local code FFT | in: $2N_{\mathrm{P}}$<br>out: $N_{\mathrm{P}}$ | $4N_{\mathrm{P}}+2N_{\mathrm{FFT,P}}$<br>$+6N_{\mathrm{FFT,P}}\log_2(N_{\mathrm{FFT,P}})$ | $8N_{\mathrm{P}}+4N_{\mathrm{FFT,P}}$<br>$+4N_{\mathrm{FFT,P}}\log_2(N_{\mathrm{FFT,P}})$ | $N_{\mathrm{FB}}N_{\mathrm{C}}$ |
| Combinations | $N_{\mathrm{P}}$ | $2N_{\mathrm{C}}2^{N_{\mathrm{C}}-1}N_{\mathrm{P}}$ | 0 | $N_{\mathrm{FB}}$ |
| Magnitude computation | $N_{\mathrm{P}}$ | $2^{N_{\mathrm{C}}-1}N_{\mathrm{P}}$ | $2^{N_{\mathrm{C}}-1}2N_{\mathrm{P}}$ | $N_{\mathrm{FB}}$ |
| Total | - | $3N_{\mathrm{FFT,P}}\log_2(N_{\mathrm{FFT,P}})$<br>$+N_{\mathrm{FB}}[N_{\mathrm{C}}(4N_{\mathrm{P}}+2N_{\mathrm{FFT,P}}$<br>$+6N_{\mathrm{FFT,P}}\log_2(N_{\mathrm{FFT,P}})$<br>$+2^{N_{\mathrm{C}}}N_{\mathrm{P}})+2^{N_{\mathrm{C}}-1}N_{\mathrm{P}}]$ | $2N_{\mathrm{FFT,P}}\log_2(N_{\mathrm{FFT,P}})$<br>$+N_{\mathrm{FB}}[N_{\mathrm{C}}(8N_{\mathrm{P}}+4N_{\mathrm{FFT,P}}$<br>$+4N_{\mathrm{FFT,P}}\log_2(N_{\mathrm{FFT,P}}))$<br>$+2^{N_{\mathrm{C}}}N_{\mathrm{P}}]$ | - |

The resulting total number of operations is

$$N_{\mathrm{FB}}\left[N_{\mathrm{C}}\left(12N_{\mathrm{P}}+6N_{\mathrm{FFT,P}}+10N_{\mathrm{FFT,P}}\log_2(N_{\mathrm{FFT,P}})+2^{N_{\mathrm{C}}}N_{\mathrm{P}}\right)+3\times 2^{N_{\mathrm{C}}-1}N_{\mathrm{P}}\right]+5N_{\mathrm{FFT,P}}\log_2(N_{\mathrm{FFT,P}}). \quad (29)$$

**TABLE 7.11** gives the operation count for the serial architecture with intermediate memory of **Figure 4.13**. Gray-code recursion allows each hypothesis after the first one to be obtained by modifying only one contribution.

**TABLE 7.11** Software operation count for multi-hypothesis method, serial with intermediate memory, coherent only (**Figure 4.13**)

| Computation stage | Vector size | # real additions | # real multiplications | # repetitions |
|---|---|---|---|---|
| Local code FFT | $N_{\mathrm{FFT,P}}$ | $3N_{\mathrm{FFT,P}}\log_2(N_{\mathrm{FFT,P}})$ | $2N_{\mathrm{FFT,P}}\log_2(N_{\mathrm{FFT,P}})$ | 1 |
| Primary code correlation excluding local code FFT | in: $2N_{\mathrm{P}}$<br>out: $N_{\mathrm{P}}$ | $4N_{\mathrm{P}}+2N_{\mathrm{FFT,P}}$<br>$+6N_{\mathrm{FFT,P}}\log_2(N_{\mathrm{FFT,P}})$ | $8N_{\mathrm{P}}+4N_{\mathrm{FFT,P}}$<br>$+4N_{\mathrm{FFT,P}}\log_2(N_{\mathrm{FFT,P}})$ | $N_{\mathrm{C}}N_{\mathrm{FB}}$ |
| First output computation | $N_{\mathrm{P}}$ | $2(N_{\mathrm{C}}-1)N_{\mathrm{P}}$ | 0 | $N_{\mathrm{FB}}$ |
| Other outputs computation | $N_{\mathrm{P}}$ | $2(2^{N_{\mathrm{C}}-1}-1)N_{\mathrm{P}}$ | 0 | $N_{\mathrm{FB}}$ |
| Magnitude computation | $N_{\mathrm{P}}$ | $2^{N_{\mathrm{C}}-1}N_{\mathrm{P}}$ | $2^{N_{\mathrm{C}}-1}2N_{\mathrm{P}}$ | $N_{\mathrm{FB}}$ |
| Total | - | $3N_{\mathrm{FFT,P}}\log_2(N_{\mathrm{FFT,P}})$<br>$+N_{\mathrm{FB}}[N_{\mathrm{C}}(6N_{\mathrm{P}}+2N_{\mathrm{FFT,P}}$<br>$+6N_{\mathrm{FFT,P}}\log_2(N_{\mathrm{FFT,P}}))$<br>$+(3\times 2^{N_{\mathrm{C}}-1}-4)N_{\mathrm{P}}]$ | $2N_{\mathrm{FFT,P}}\log_2(N_{\mathrm{FFT,P}})$<br>$+N_{\mathrm{FB}}[N_{\mathrm{C}}(8N_{\mathrm{P}}+4N_{\mathrm{FFT,P}}$<br>$+4N_{\mathrm{FFT,P}}\log_2(N_{\mathrm{FFT,P}}))$<br>$+2^{N_{\mathrm{C}}}N_{\mathrm{P}}]$ | - |

The resulting total number of operations is

$$N_{\mathrm{FB}}\left[N_{\mathrm{C}}\left(14N_{\mathrm{P}}+6N_{\mathrm{FFT,P}}+10N_{\mathrm{FFT,P}}\log_2(N_{\mathrm{FFT,P}})\right)+(5\times 2^{N_{\mathrm{C}}-1}-4)N_{\mathrm{P}}\right]+5N_{\mathrm{FFT,P}}\log_2(N_{\mathrm{FFT,P}}). \quad (30)$$

**TABLE 7.12** gives the corresponding operation count when the intermediate correlation vectors are recomputed rather than stored, as in **Figure 4.14**.

**TABLE 7.12** Software operation count for multi-hypothesis method, serial without intermediate memory, coherent only (**Figure 4.14**)

| Computation stage | Vector size | # real additions | # real multiplications | # repetitions |
|---|---|---|---|---|
| Local code FFT | $N_{\mathrm{FFT,P}}$ | $3N_{\mathrm{FFT,P}}\log_2(N_{\mathrm{FFT,P}})$ | $2N_{\mathrm{FFT,P}}\log_2(N_{\mathrm{FFT,P}})$ | 1 |
| Primary code correlation excluding local code FFT | in: $2N_{\mathrm{P}}$<br>out: $N_{\mathrm{P}}$ | $4N_{\mathrm{P}}+2N_{\mathrm{FFT,P}}+6N_{\mathrm{FFT,P}}\log_2(N_{\mathrm{FFT,P}})$ | $8N_{\mathrm{P}}+4N_{\mathrm{FFT,P}}+4N_{\mathrm{FFT,P}}\log_2(N_{\mathrm{FFT,P}})$ | $N_{\mathrm{FB}}(N_{\mathrm{C}}+2^{N_{\mathrm{C}}-1}-1)$ |
| First output computation | $N_{\mathrm{P}}$ | $2(N_{\mathrm{C}}-1)N_{\mathrm{P}}$ | 0 | $N_{\mathrm{FB}}$ |
| Other outputs computation | $N_{\mathrm{P}}$ | $2(2^{N_{\mathrm{C}}-1}-1)N_{\mathrm{P}}$ | 0 | $N_{\mathrm{FB}}$ |
| Magnitude computation | $N_{\mathrm{P}}$ | $2^{N_{\mathrm{C}}-1}N_{\mathrm{P}}$ | $2^{N_{\mathrm{C}}-1}2N_{\mathrm{P}}$ | $N_{\mathrm{FB}}$ |
| Total | - | $3N_{\mathrm{FFT,P}}\log_2(N_{\mathrm{FFT,P}})+N_{\mathrm{FB}}\left[(N_{\mathrm{C}}+2^{N_{\mathrm{C}}-1}-1)\left(4N_{\mathrm{P}}+2N_{\mathrm{FFT,P}}+6N_{\mathrm{FFT,P}}\log_2(N_{\mathrm{FFT,P}})\right)+(3\times 2^{N_{\mathrm{C}}-1}+2N_{\mathrm{C}}-4)N_{\mathrm{P}}\right]$ | $2N_{\mathrm{FFT,P}}\log_2(N_{\mathrm{FFT,P}})+N_{\mathrm{FB}}\left[(N_{\mathrm{C}}+2^{N_{\mathrm{C}}-1}-1)\left(8N_{\mathrm{P}}+4N_{\mathrm{FFT,P}}+4N_{\mathrm{FFT,P}}\log_2(N_{\mathrm{FFT,P}})\right)+2^{N_{\mathrm{C}}}N_{\mathrm{P}}\right]$ | - |

The resulting total number of operations is

$$N_{\mathrm{FB}}\left[(N_{\mathrm{C}}+2^{N_{\mathrm{C}}-1}-1)\left(12N_{\mathrm{P}}+6N_{\mathrm{FFT,P}}+10N_{\mathrm{FFT,P}}\log_2(N_{\mathrm{FFT,P}})\right)+(5\times 2^{N_{\mathrm{C}}-1}+2N_{\mathrm{C}}-4)N_{\mathrm{P}}\right]+5N_{\mathrm{FFT,P}}\log_2(N_{\mathrm{FFT,P}}). \quad (31)$$

**TABLE 7.13** provides the operation count for the parallel hybrid architecture shown in **Figure 4.15**.

**TABLE 7.13** Software operation count for multi-hypothesis method, parallel hybrid (**Figure 4.15**)

| Computation stage | Vector size | # real additions | # real multiplications | # repetitions |
|---|---|---|---|---|
| Local code FFT | $N_{\mathrm{FFT,P}}$ | $3N_{\mathrm{FFT,P}}\log_2(N_{\mathrm{FFT,P}})$ | $2N_{\mathrm{FFT,P}}\log_2(N_{\mathrm{FFT,P}})$ | 1 |
| Primary code correlation excluding local code FFT | in: $2N_{\mathrm{P}}$<br>out: $N_{\mathrm{P}}$ | $4N_{\mathrm{P}}+2N_{\mathrm{FFT,P}}+6N_{\mathrm{FFT,P}}\log_2(N_{\mathrm{FFT,P}})$ | $8N_{\mathrm{P}}+4N_{\mathrm{FFT,P}}+4N_{\mathrm{FFT,P}}\log_2(N_{\mathrm{FFT,P}})$ | $N_{\mathrm{FB}}N_{\mathrm{NC}}N_{\mathrm{C}}$ |
| Combinations | $N_{\mathrm{P}}$ | $2N_{\mathrm{C}}2^{N_{\mathrm{C}}-1}N_{\mathrm{P}}$ | 0 | $N_{\mathrm{FB}}N_{\mathrm{NC}}$ |
| Magnitude computation | $N_{\mathrm{P}}$ | $2^{N_{\mathrm{C}}-1}N_{\mathrm{P}}$ | $2^{N_{\mathrm{C}}-1}2N_{\mathrm{P}}$ | $N_{\mathrm{FB}}N_{\mathrm{NC}}$ |
| Non-coherent integration | $N_{\mathrm{P}}$ | $N_{\mathrm{P}}$ | 0 | $N_{\mathrm{FB}}N_{\mathrm{NC}}$ |
| Total | - | $3N_{\mathrm{FFT,P}}\log_2(N_{\mathrm{FFT,P}})+N_{\mathrm{FB}}N_{\mathrm{NC}}\left[N_{\mathrm{C}}\left(4N_{\mathrm{P}}+2N_{\mathrm{FFT,P}}+6N_{\mathrm{FFT,P}}\log_2(N_{\mathrm{FFT,P}})+2^{N_{\mathrm{C}}}N_{\mathrm{P}}\right)+(2^{N_{\mathrm{C}}-1}+1)N_{\mathrm{P}}\right]$ | $2N_{\mathrm{FFT,P}}\log_2(N_{\mathrm{FFT,P}})+N_{\mathrm{FB}}N_{\mathrm{NC}}\left[N_{\mathrm{C}}\left(8N_{\mathrm{P}}+4N_{\mathrm{FFT,P}}+4N_{\mathrm{FFT,P}}\log_2(N_{\mathrm{FFT,P}})\right)+2^{N_{\mathrm{C}}}N_{\mathrm{P}}\right]$ | - |

The resulting total number of operations is

$$N_{\mathrm{FB}}N_{\mathrm{NC}}\left[N_{\mathrm{C}}\left(12N_{\mathrm{P}}+6N_{\mathrm{FFT,P}}+10N_{\mathrm{FFT,P}}\log_2(N_{\mathrm{FFT,P}})+2^{N_{\mathrm{C}}}N_{\mathrm{P}}\right)+(3\times 2^{N_{\mathrm{C}}-1}+1)N_{\mathrm{P}}\right]+5N_{\mathrm{FFT,P}}\log_2(N_{\mathrm{FFT,P}}). \quad (32)$$

**TABLE 7.14** provides the operation count for the serial hybrid architecture with intermediate memory shown in **Figure 4.16**.

**TABLE 7.14** Software operation count for multi-hypothesis method, serial with intermediate memory, hybrid (**Figure 4.16**)

| Computation stage | Vector size | # real additions | # real multiplications | # repetitions |
|---|---|---|---|---|
| Local code FFT | $N_{\mathrm{FFT,P}}$ | $3N_{\mathrm{FFT,P}}\log_2(N_{\mathrm{FFT,P}})$ | $2N_{\mathrm{FFT,P}}\log_2(N_{\mathrm{FFT,P}})$ | 1 |
| Primary code correlation excluding local code FFT | in: $2N_{\mathrm{P}}$<br>out: $N_{\mathrm{P}}$ | $4N_{\mathrm{P}}+2N_{\mathrm{FFT,P}}+6N_{\mathrm{FFT,P}}\log_2(N_{\mathrm{FFT,P}})$ | $8N_{\mathrm{P}}+4N_{\mathrm{FFT,P}}+4N_{\mathrm{FFT,P}}\log_2(N_{\mathrm{FFT,P}})$ | $N_{\mathrm{FB}}N_{\mathrm{NC}}N_{\mathrm{C}}$ |
| First output computation | $N_{\mathrm{P}}$ | $2(N_{\mathrm{C}}-1)N_{\mathrm{P}}$ | 0 | $N_{\mathrm{FB}}N_{\mathrm{NC}}$ |
| Other outputs computation | $N_{\mathrm{P}}$ | $2(2^{N_{\mathrm{C}}-1}-1)N_{\mathrm{P}}$ | 0 | $N_{\mathrm{FB}}N_{\mathrm{NC}}$ |
| Magnitude computation | $N_{\mathrm{P}}$ | $2^{N_{\mathrm{C}}-1}N_{\mathrm{P}}$ | $2^{N_{\mathrm{C}}-1}2N_{\mathrm{P}}$ | $N_{\mathrm{FB}}N_{\mathrm{NC}}$ |
| Non-coherent integration | $N_{\mathrm{P}}$ | $N_{\mathrm{P}}$ | 0 | $N_{\mathrm{FB}}N_{\mathrm{NC}}$ |
| Total | - | $3N_{\mathrm{FFT,P}}\log_2(N_{\mathrm{FFT,P}}) + N_{\mathrm{FB}}N_{\mathrm{NC}}\left[N_{\mathrm{C}}\left(6N_{\mathrm{P}}+2N_{\mathrm{FFT,P}} + 6N_{\mathrm{FFT,P}}\log_2(N_{\mathrm{FFT,P}})\right) + (3\times 2^{N_{\mathrm{C}}-1}-3)N_{\mathrm{P}}\right]$ | $2N_{\mathrm{FFT,P}}\log_2(N_{\mathrm{FFT,P}}) + N_{\mathrm{FB}}N_{\mathrm{NC}}\left[N_{\mathrm{C}}\left(8N_{\mathrm{P}} + 4N_{\mathrm{FFT,P}} + 4N_{\mathrm{FFT,P}}\log_2(N_{\mathrm{FFT,P}})\right) + 2^{N_{\mathrm{C}}}N_{\mathrm{P}}\right]$ | - |

The resulting total number of operations is

$$N_{\mathrm{FB}}N_{\mathrm{NC}}\left[N_{\mathrm{C}}\left(14N_{\mathrm{P}}+6N_{\mathrm{FFT,P}}+10N_{\mathrm{FFT,P}}\log_2(N_{\mathrm{FFT,P}})\right)+(5^{N_{\mathrm{C}}-1}-3)N_{\mathrm{P}}\right]+5N_{\mathrm{FFT,P}}\log_2(N_{\mathrm{FFT,P}}). \quad (33)$$

**TABLE 7.15** provides the operation count for the serial hybrid architecture without intermediate memory shown in **Figure 4.17**.

**TABLE 7.15** Software operation count for multi-hypothesis method, serial without intermediate memory, hybrid (**Figure 4.17**)

| Computation stage | Vector size | # real additions | # real multiplications | # repetitions |
|---|---|---|---|---|
| Local code FFT | $N_{\mathrm{FFT,P}}$ | $3N_{\mathrm{FFT,P}}\log_2(N_{\mathrm{FFT,P}})$ | $2N_{\mathrm{FFT,P}}\log_2(N_{\mathrm{FFT,P}})$ | 1 |
| Primary code correlation excluding local code FFT | in: $2N_{\mathrm{P}}$ out: $N_{\mathrm{P}}$ | $4N_{\mathrm{P}}+2N_{\mathrm{FFT,P}} + 6N_{\mathrm{FFT,P}}\log_2(N_{\mathrm{FFT,P}})$ | $8N_{\mathrm{P}}+4N_{\mathrm{FFT,P}} + 4N_{\mathrm{FFT,P}}\log_2(N_{\mathrm{FFT,P}})$ | $N_{\mathrm{FB}}N_{\mathrm{NC}}(N_{\mathrm{C}} + 2^{N_{\mathrm{C}}-1}-1)$ |
| First output computation | $N_{\mathrm{P}}$ | $2(N_{\mathrm{C}}-1)N_{\mathrm{P}}$ | 0 | $N_{\mathrm{FB}}N_{\mathrm{NC}}$ |
| Other outputs computation | $N_{\mathrm{P}}$ | $2(2^{N_{\mathrm{C}}-1}-1)N_{\mathrm{P}}$ | 0 | $N_{\mathrm{FB}}N_{\mathrm{NC}}$ |
| Magnitude computation | $N_{\mathrm{P}}$ | $2^{N_{\mathrm{C}}-1}N_{\mathrm{P}}$ | $2^{N_{\mathrm{C}}-1}2N_{\mathrm{P}}$ | $N_{\mathrm{FB}}N_{\mathrm{NC}}$ |
| Non-coherent integration | $N_{\mathrm{P}}$ | $N_{\mathrm{P}}$ | 0 | $N_{\mathrm{FB}}N_{\mathrm{NC}}$ |
| Total | - | $3N_{\mathrm{FFT,P}}\log_2(N_{\mathrm{FFT,P}}) + N_{\mathrm{FB}}N_{\mathrm{NC}}\left[(N_{\mathrm{C}}+2^{N_{\mathrm{C}}-1} - 1)\left(4N_{\mathrm{P}}+2N_{\mathrm{FFT,P}} + 6N_{\mathrm{FFT,P}}\log_2(N_{\mathrm{FFT,P}})\right) + (3\times 2^{N_{\mathrm{C}}-1}+2N_{\mathrm{C}}-3)N_{\mathrm{P}}\right]$ | $2N_{\mathrm{FFT,P}}\log_2(N_{\mathrm{FFT,P}}) + N_{\mathrm{FB}}N_{\mathrm{NC}}\left[(N_{\mathrm{C}}+2^{N_{\mathrm{C}}-1} - 1)\left(8N_{\mathrm{P}}+4N_{\mathrm{FFT,P}} + 4N_{\mathrm{FFT,P}}\log_2(N_{\mathrm{FFT,P}})\right) + 2^{N_{\mathrm{C}}}N_{\mathrm{P}}\right]$ | - |

The resulting total number of operations is

$$N_{\mathrm{FB}}N_{\mathrm{NC}}\left[(N_{\mathrm{C}}+2^{N_{\mathrm{C}}-1}-1)\left(12N_{\mathrm{P}}+6N_{\mathrm{FFT,P}}+10N_{\mathrm{FFT,P}}\log_2(N_{\mathrm{FFT,P}})\right)+(5\times 2^{N_{\mathrm{C}}-1}+2N_{\mathrm{C}}-3)N_{\mathrm{P}}\right] + 5N_{\mathrm{FFT,P}}\log_2(N_{\mathrm{FFT,P}}). \quad (34)$$

### *7.6.2. Hardware processing time*

**Figure 7.6** illustrates the timing of the parallel coherent-only architecture of **Figure 4.12**, giving

$$N_{\mathrm{FB}}N_{\mathrm{C}}N_{\mathrm{FFT,P}}. \quad (35)$$

**Figure 7.7** illustrates the serial architecture with intermediate memory of **Figure 4.13**, giving

$$N_{\mathrm{FB}}\left[(N_{\mathrm{C}}-1)N_{\mathrm{FFT,P}}+2^{N_{\mathrm{C}}-1}N_{\mathrm{P}}\right]. \quad (36)$$

**Figure 7.8** illustrates the serial architecture without intermediate memory of **Figure 4.14**, giving

$$N_{\mathrm{FB}}(N_{\mathrm{C}}+2^{N_{\mathrm{C}}-1}-1)N_{\mathrm{FFT,P}}. \quad (37)$$

For the hybrid architectures of **Figure 4.15** to **Figure 4.17**, the same coherent processing is repeated independently for each of the $N_{\mathrm{NC}}$ non-coherent blocks. The resulting processing times are

$$N_{\mathrm{FB}}N_{\mathrm{NC}}N_{\mathrm{C}}N_{\mathrm{FFT,P}}, \tag{38}$$

for the parallel architecture,

$$N_{\mathrm{FB}}N_{\mathrm{NC}}\left[(N_{\mathrm{C}}-1)N_{\mathrm{FFT,P}}+2^{N_{\mathrm{C}}-1}N_{\mathrm{P}}\right], \tag{39}$$

for the serial architecture with intermediate memory, and

$$N_{\mathrm{FB}}N_{\mathrm{NC}}(N_{\mathrm{C}}+2^{N_{\mathrm{C}}-1}-1)N_{\mathrm{FFT,P}}. \tag{40}$$

for the serial architecture without intermediate memory.

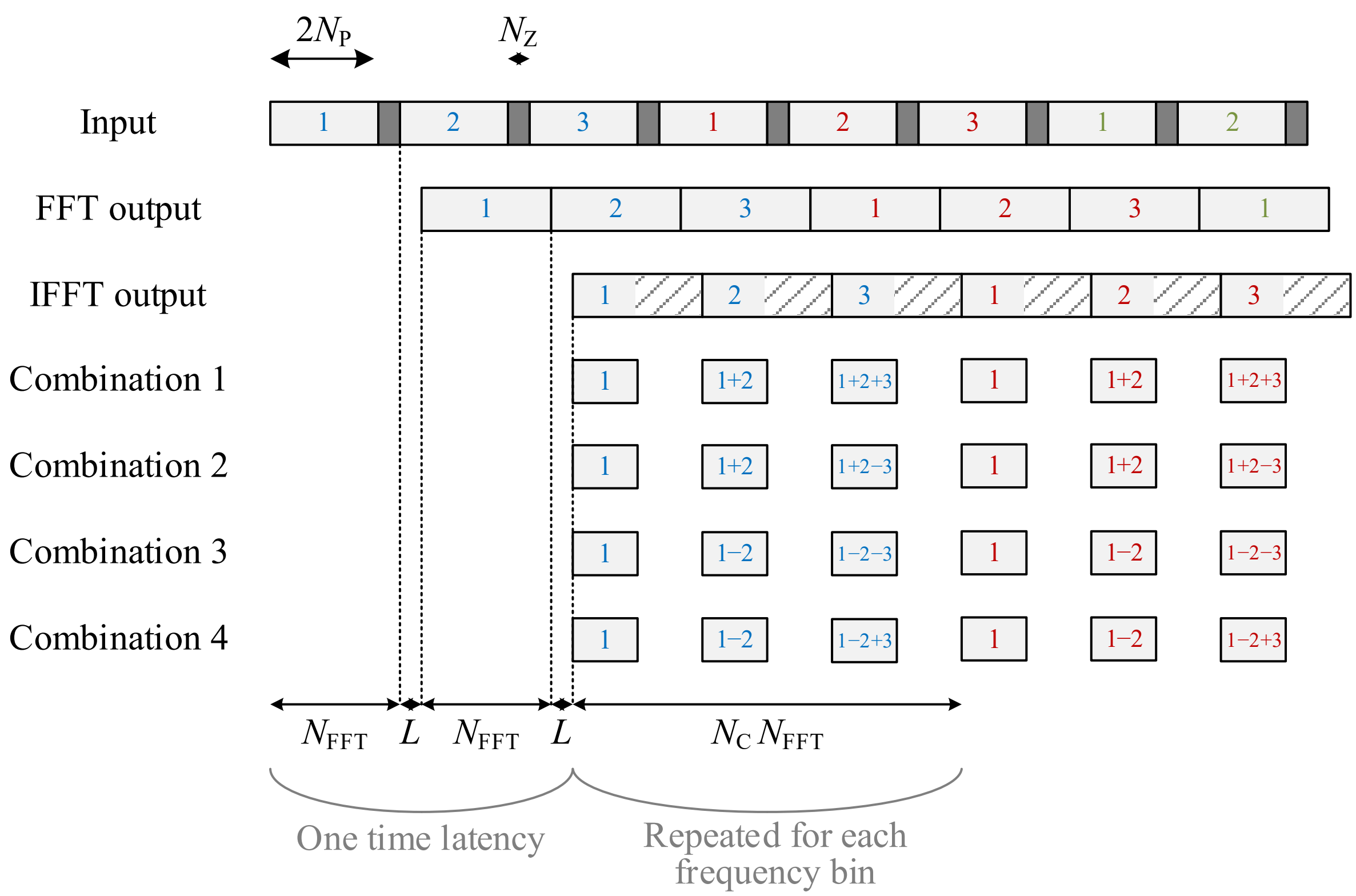

**Figure 7.6** Hardware timing for multi-hypothesis method, parallel coherent only, $N_{\mathrm{C}}=3$ (**Figure 4.12**)

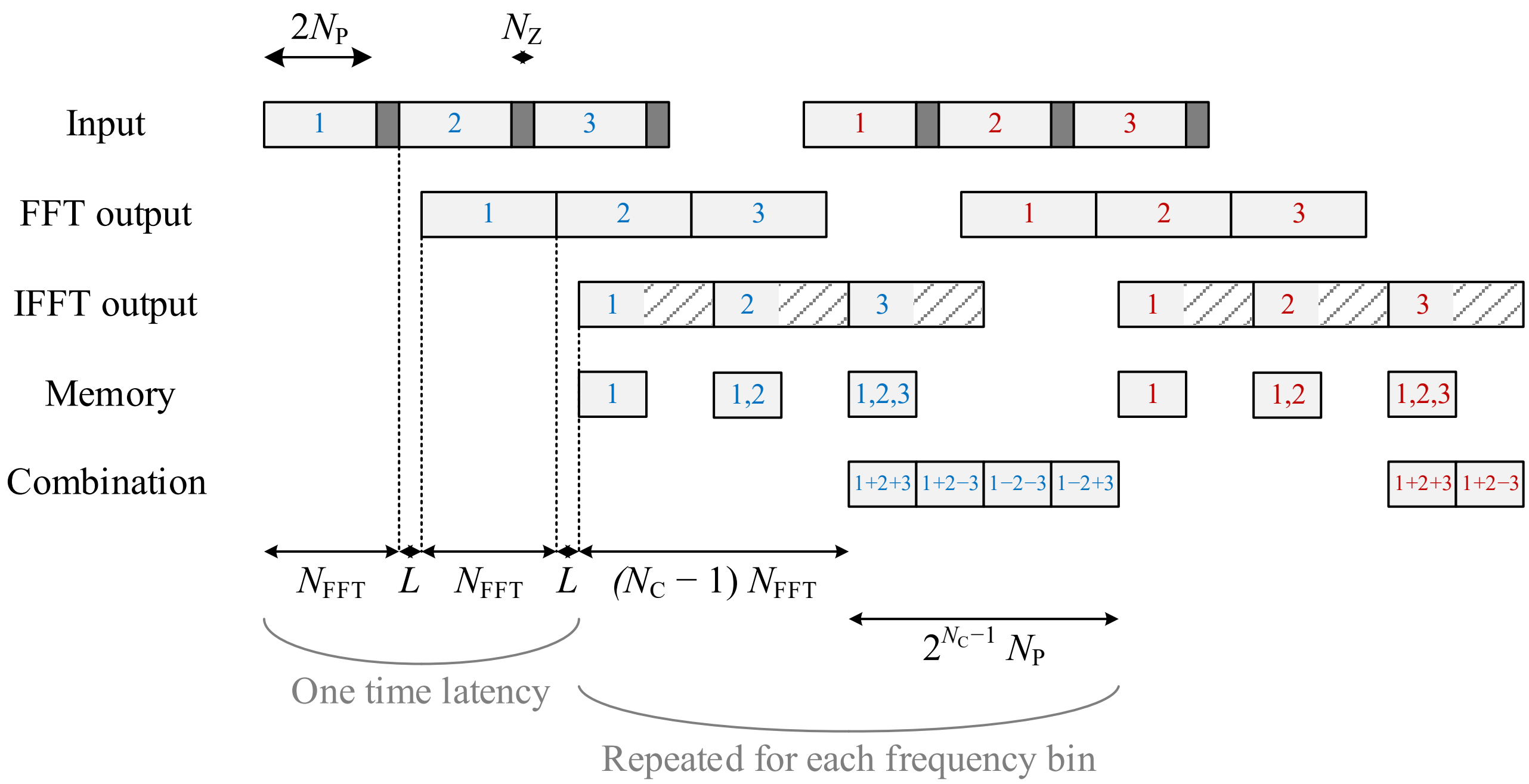

**Figure 7.7** Hardware timing for multi-hypothesis method, serial with intermediate memory, coherent only, $N_C = 3$ (**Figure 4.13**)

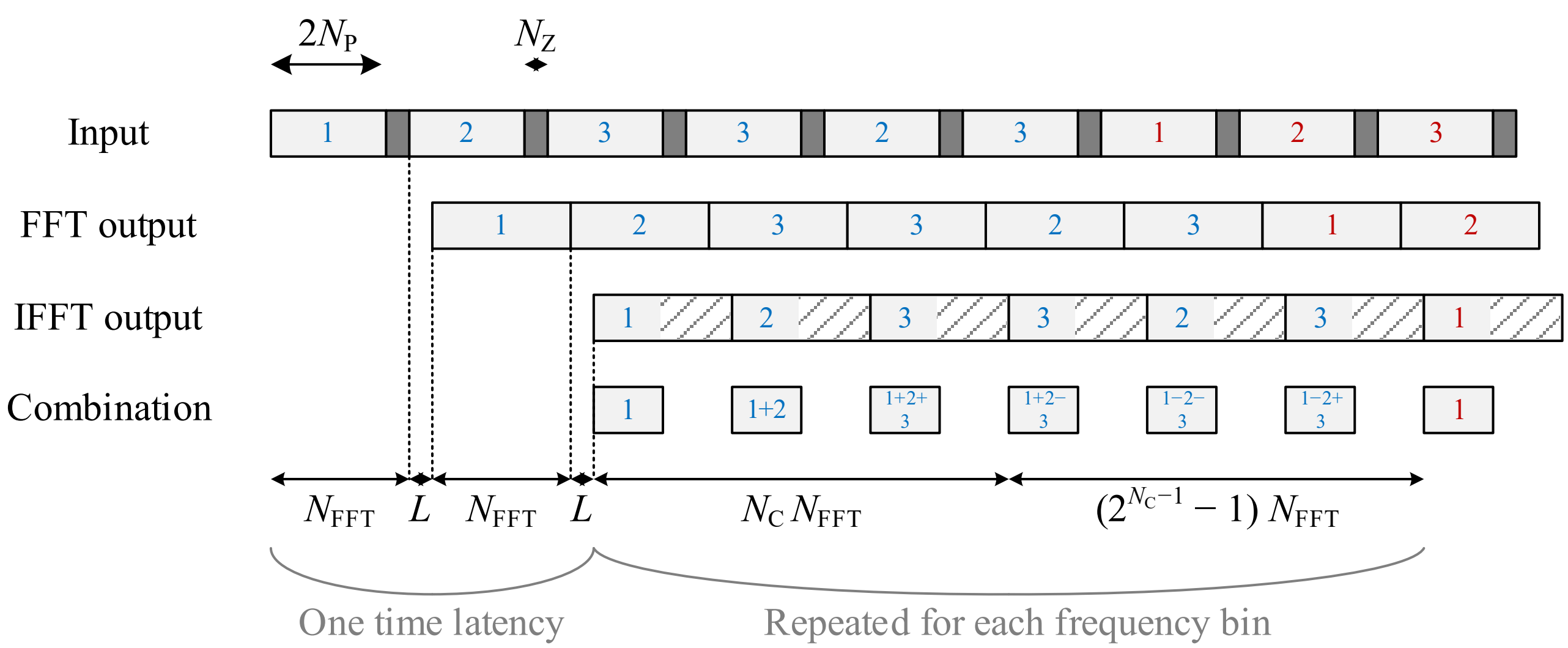

**Figure 7.8** Hardware timing for multi-hypothesis method, serial without intermediate memory, coherent only, $N_C = 3$ (**Figure 4.14**)

### *7.6.3. Memory requirements*

For the parallel coherent-only architecture of **Figure 4.12**, the memory requirement is

$$(N_C + 1)N_P B_{ADC} + N_{FFT,P} B_{FFT,P}/2 + 2M_{FFT,P} + 2^{N_C - 1} N_P B_C, \tag{41}$$

For the serial coherent-only architecture with intermediate memory of **Figure 4.13**, it is

$$(N_C + 1)N_P B_{ADC} + N_{FFT,P} B_{FFT,P}/2 + 2M_{FFT,P} + N_C N_P B_{FFT} + N_P B_C, \tag{42}$$

and without intermediate memory, as in **Figure 4.14**,

$$(N_{\mathrm{C}}+1)N_{\mathrm{P}}B_{\mathrm{ADC}}+N_{\mathrm{FFT,P}}B_{\mathrm{FFT,P}}/2+2M_{\mathrm{FFT,P}}+N_{\mathrm{P}}B_{\mathrm{C}}. \tag{43}$$

The corresponding hybrid architectures additionally require a non-coherent accumulator, giving

$$(N_{\mathrm{NC}}N_{\mathrm{C}}+1)N_{\mathrm{P}}B_{\mathrm{ADC}}+N_{\mathrm{FFT,P}}B_{\mathrm{FFT,P}}/2+2M_{\mathrm{FFT,P}}+2^{N_{\mathrm{C}}-1}N_{\mathrm{P}}B_{\mathrm{C}}+N_{\mathrm{P}}B_{\mathrm{NC}}, \tag{44}$$

for the parallel architecture of **Figure 4.15**,

$$(N_{\mathrm{NC}}N_{\mathrm{C}}+1)N_{\mathrm{P}}B_{\mathrm{ADC}}+N_{\mathrm{FFT,P}}B_{\mathrm{FFT,P}}/2+2M_{\mathrm{FFT,P}}+N_{\mathrm{C}}N_{\mathrm{P}}B_{\mathrm{FFT}}+N_{\mathrm{P}}B_{\mathrm{C}}+N_{\mathrm{P}}B_{\mathrm{NC}}, \tag{45}$$

for the serial architecture with intermediate memory of **Figure 4.16**, and

$$(N_{\mathrm{NC}}N_{\mathrm{C}}+1)N_{\mathrm{P}}B_{\mathrm{ADC}}+N_{\mathrm{FFT,P}}B_{\mathrm{FFT,P}}/2+2M_{\mathrm{FFT,P}}+N_{\mathrm{P}}B_{\mathrm{C}}+N_{\mathrm{P}}B_{\mathrm{NC}}, \tag{46}$$

for the serial architecture without intermediate memory of **Figure 4.17**.

## 7.7. Complexity for secondary code correlation

### *7.7.1. Software operation count for frequency-domain secondary-code correlation*

For software receivers, the frequency-domain secondary-code correlation architectures of **Figure 4.22** and **Figure 4.23** are considered. **TABLE 7.16** gives the operation count for partial secondary-code correlation, where $N_{\mathrm{C}} < N_{\mathrm{S}}$ but all $N_{\mathrm{S}}$ secondary-code delays are evaluated.

**TABLE 7.16** Software operation count for frequency-domain partial secondary-code correlation (**Figure 4.22**)

| Computation stage | Vector size | # real additions | # real multiplications | # repetitions |
|---|---|---|---|---|
| Local primary code FFT | $N_{\mathrm{FFT,P}}$ | $3N_{\mathrm{FFT,P}}\log_2(N_{\mathrm{FFT,P}})$ | $2N_{\mathrm{FFT,P}}\log_2(N_{\mathrm{FFT,P}})$ | 1 |
| Primary code correlation excluding local code FFT | in: $2N_{\mathrm{P}}$ out: $N_{\mathrm{P}}$ | $4N_{\mathrm{P}}+2N_{\mathrm{FFT,P}}$ $+6N_{\mathrm{FFT,P}}\log_2(N_{\mathrm{FFT,P}})$ | $8N_{\mathrm{P}}+4N_{\mathrm{FFT,P}}$ $+4N_{\mathrm{FFT,P}}\log_2(N_{\mathrm{FFT,P}})$ | $N_{\mathrm{FB}}N_{\mathrm{NC}}N_{\mathrm{C}}$ |
| Local secondary code FFT | $N_{\mathrm{FFT,S}}$ | $3N_{\mathrm{FFT,S}}\log_2(N_{\mathrm{FFT,S}})$ | $2N_{\mathrm{FFT,S}}\log_2(N_{\mathrm{FFT,S}})$ | 1 |
| Secondary code correlation excluding local code FFT | in: $N_{\mathrm{C}}$ out: $N_{\mathrm{S}}$ | $2N_{\mathrm{FFT,S}}$ $+6N_{\mathrm{FFT,S}}\log_2(N_{\mathrm{FFT,S}})$ | $4N_{\mathrm{FFT,S}}$ $+4N_{\mathrm{FFT,S}}\log_2(N_{\mathrm{FFT,S}})$ | $N_{\mathrm{FB}}N_{\mathrm{NC}}N_{\mathrm{P}}$ |
| Magnitude computation | $N_{\mathrm{S}}$ | $N_{\mathrm{S}}$ | $2N_{\mathrm{S}}$ | $N_{\mathrm{FB}}N_{\mathrm{NC}}N_{\mathrm{P}}$ |
| Non-coherent integration | $N_{\mathrm{S}}$ | $N_{\mathrm{S}}$ | 0 | $N_{\mathrm{FB}}N_{\mathrm{NC}}N_{\mathrm{P}}$ |
| Total | - | $3N_{\mathrm{FFT,P}}\log_2(N_{\mathrm{FFT,P}})$ $+3N_{\mathrm{FFT,S}}\log_2(N_{\mathrm{FFT,S}})$ $N_{\mathrm{FB}}N_{\mathrm{NC}}[N_{\mathrm{C}}(4N_{\mathrm{P}}+2N_{\mathrm{FFT,P}}$ $+6N_{\mathrm{FFT,P}}\log_2(N_{\mathrm{FFT,P}}))$ $+N_{\mathrm{P}}(2N_{\mathrm{FFT,S}}$ $+6N_{\mathrm{FFT,S}}\log_2(N_{\mathrm{FFT,S}})$ $+2N_{\mathrm{S}})]$ | $2N_{\mathrm{FFT,P}}\log_2(N_{\mathrm{FFT,P}})$ $+2N_{\mathrm{FFT,S}}\log_2(N_{\mathrm{FFT,S}})$ $N_{\mathrm{FB}}N_{\mathrm{NC}}[N_{\mathrm{C}}(8N_{\mathrm{P}}+4N_{\mathrm{FFT,P}}$ $+4N_{\mathrm{FFT,P}}\log_2(N_{\mathrm{FFT,P}}))$ $+N_{\mathrm{P}}(4N_{\mathrm{FFT,S}}$ $+4N_{\mathrm{FFT,S}}\log_2(N_{\mathrm{FFT,S}})$ $+2N_{\mathrm{S}})]$ | - |

The resulting total number of operations is

$$N_{\mathrm{FB}}N_{\mathrm{NC}}[N_{\mathrm{C}}(12N_{\mathrm{P}}+6N_{\mathrm{FFT,P}}+10N_{\mathrm{FFT,P}}\log_2(N_{\mathrm{FFT,P}}))+N_{\mathrm{P}}(4N_{\mathrm{S}}+6N_{\mathrm{FFT,S}}+10N_{\mathrm{FFT,S}}\log_2(N_{\mathrm{FFT,S}}))] \\ +5N_{\mathrm{FFT,P}}\log_2(N_{\mathrm{FFT,P}})+5N_{\mathrm{FFT,S}}\log_2(N_{\mathrm{FFT,S}}). \tag{47}$$

**TABLE 7.17** gives the corresponding operation count for full secondary-code correlation with $N_{\mathrm{C}} = kN_{\mathrm{S}}$.

**TABLE 7.17** Software operation count for frequency-domain full secondary-code correlation (**Figure 4.23**)

| Computation stage | Vector size | # real additions | # real multiplications | # repetitions |
|---|---|---|---|---|
| Local primary code FFT | $N_{\mathrm{FFT,P}}$ | $3N_{\mathrm{FFT,P}}\log_2(N_{\mathrm{FFT,P}})$ | $2N_{\mathrm{FFT,P}}\log_2(N_{\mathrm{FFT,P}})$ | 1 |
| Primary code correlation excluding local code FFT | in: $2N_\mathrm{P}$ out: $N_\mathrm{P}$ | $4N_\mathrm{P} + 2N_\mathrm{FFT,P} + 6N_\mathrm{FFT,P}\log_2(N_\mathrm{FFT,P})$ | $8N_\mathrm{P} + 4N_\mathrm{FFT,P} + 4N_\mathrm{FFT,P}\log_2(N_\mathrm{FFT,P})$ | $N_\mathrm{FB}N_\mathrm{NC}kN_\mathrm{S}$ |
| Local secondary code FFT | $N_\mathrm{FFT,S}$ | $3N_\mathrm{FFT,S}\log_2(N_\mathrm{FFT,S})$ | $2N_\mathrm{FFT,S}\log_2(N_\mathrm{FFT,S})$ | 1 |
| Secondary code correlation excluding local code FFT | in: $N_\mathrm{C}$ out: $N_\mathrm{S}$ | $2N_\mathrm{FFT,S} + 6N_\mathrm{FFT,S}\log_2(N_\mathrm{FFT,S})$ | $4N_\mathrm{FFT,S} + 4N_\mathrm{FFT,S}\log_2(N_\mathrm{FFT,S})$ | $N_\mathrm{FB}N_\mathrm{NC}kN_\mathrm{P}$ |
| Extra coherent integration | $N_\mathrm{S}$ | $2N_\mathrm{S}$ | 0 | $N_\mathrm{FB}N_\mathrm{NC}kN_\mathrm{P}$ |
| Magnitude computation | $N_\mathrm{S}$ | $N_\mathrm{S}$ | $2N_\mathrm{S}$ | $N_\mathrm{FB}N_\mathrm{NC}N_\mathrm{P}$ |
| Non-coherent integration | $N_\mathrm{S}$ | $N_\mathrm{S}$ | 0 | $N_\mathrm{FB}N_\mathrm{NC}N_\mathrm{P}$ |
| Total | - | $3N_\mathrm{FFT,P}\log_2(N_\mathrm{FFT,P}) + 3N_\mathrm{FFT,S}\log_2(N_\mathrm{FFT,S})$ $N_\mathrm{FB}N_\mathrm{NC}\big[kN_\mathrm{S}\big(4N_\mathrm{P} + 2N_\mathrm{FFT,P} + 6N_\mathrm{FFT,P}\log_2(N_\mathrm{FFT,P})\big) + kN_\mathrm{P}\big(2N_\mathrm{FFT,S} + 6N_\mathrm{FFT,S}\log_2(N_\mathrm{FFT,S}) + 2N_\mathrm{S}\big) + 2N_\mathrm{P}N_\mathrm{S}\big]$ | $2N_\mathrm{FFT,P}\log_2(N_\mathrm{FFT,P}) + 2N_\mathrm{FFT,S}\log_2(N_\mathrm{FFT,S})$ $N_\mathrm{FB}N_\mathrm{NC}\big[kN_\mathrm{S}\big(8N_\mathrm{P} + 4N_\mathrm{FFT,P} + 4N_\mathrm{FFT,P}\log_2(N_\mathrm{FFT,P})\big) + kN_\mathrm{P}\big(4N_\mathrm{FFT,S} + 4N_\mathrm{FFT,S}\log_2(N_\mathrm{FFT,S})\big) + 2N_\mathrm{S}N_\mathrm{P}\big]$ | - |

The resulting total number of operations is

$$N_\mathrm{FB}N_\mathrm{NC}\big[kN_\mathrm{S}\big(12N_\mathrm{P} + 6N_\mathrm{FFT,P} + 10N_\mathrm{FFT,P}\log_2(N_\mathrm{FFT,P})\big) + kN_\mathrm{P}\big(2N_\mathrm{S} + 6N_\mathrm{FFT,S} + 10N_\mathrm{FFT,S}\log_2(N_\mathrm{FFT,S})\big) + 4N_\mathrm{P}N_\mathrm{S}\big] + 5N_\mathrm{FFT,P}\log_2(N_\mathrm{FFT,P}) + 5N_\mathrm{FFT,S}\log_2(N_\mathrm{FFT,S}). \quad (48)$$

### *7.7.2. Hardware processing time for time-domain secondary-code correlation*

For hardware receivers, the time-domain architectures of **Figure 4.18** to **Figure 4.21** are considered. In the parallel architectures, all $N_\mathrm{S}$ secondary-code delays are evaluated simultaneously, whereas the serial architectures evaluate them sequentially.

For partial parallel secondary-code correlation of **Figure 4.18**, the processing time is

$$N_\mathrm{FB}N_\mathrm{NC}N_\mathrm{C}N_\mathrm{FFT,P}. \quad (49)$$

For full parallel secondary-code correlation of **Figure 4.19** with $N_\mathrm{C} = kN_\mathrm{S}$, it becomes

$$N_\mathrm{FB}N_\mathrm{NC}kN_\mathrm{S}N_\mathrm{FFT,P}. \quad (50)$$

For partial serial secondary-code correlation of **Figure 4.20**, the additional sequential search over the $N_\mathrm{S}$ delays gives

$$N_\mathrm{FB}N_\mathrm{NC}N_\mathrm{C}N_\mathrm{S}N_\mathrm{FFT,P}, \quad (51)$$

and for full serial secondary-code correlation of **Figure 4.21**,

$$N_\mathrm{FB}N_\mathrm{NC}kN_\mathrm{S}^2N_\mathrm{FFT,P}. \quad (52)$$

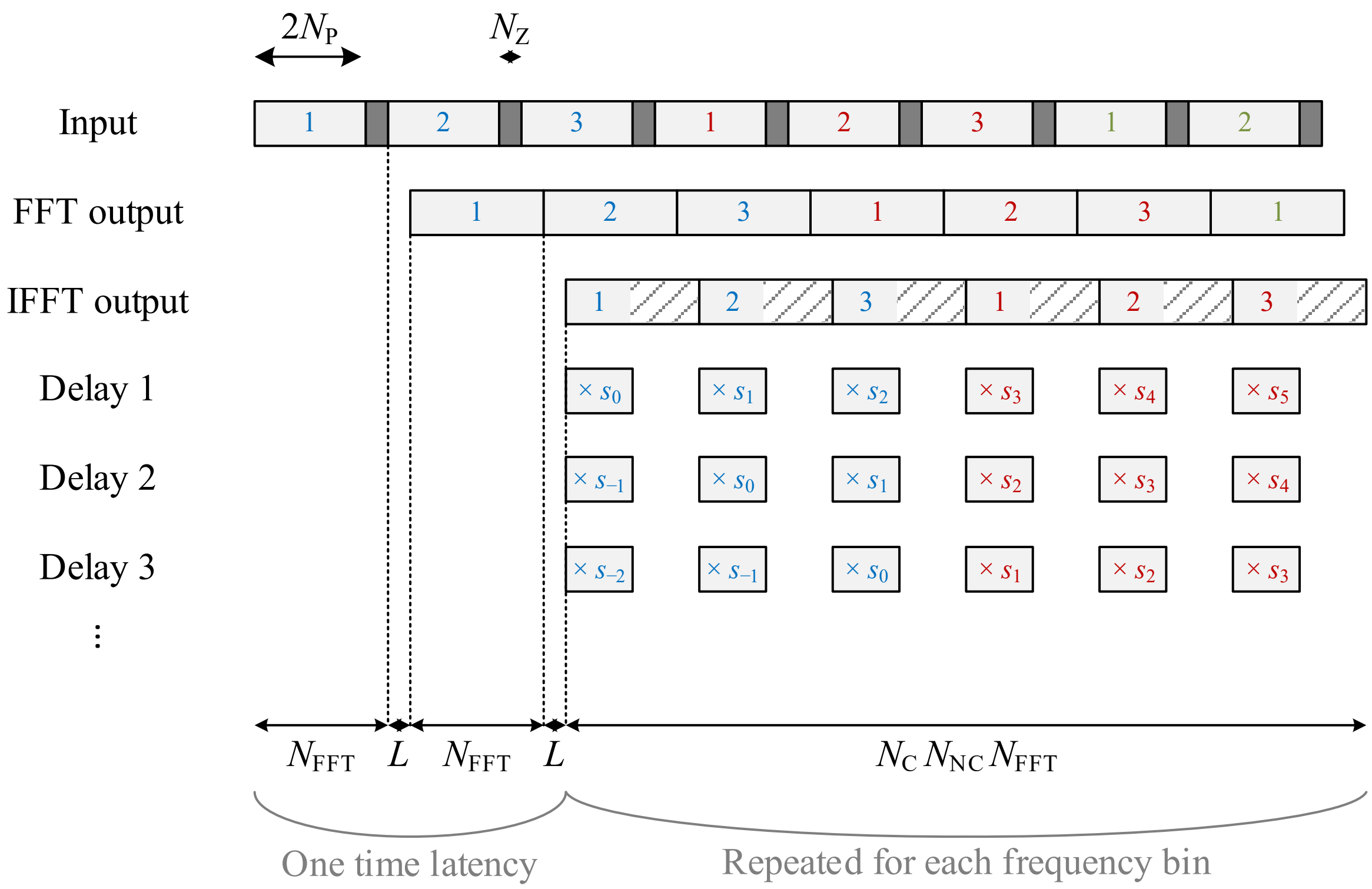

**Figure 7.9** Hardware timing for parallel partial secondary-code correlation, $N_C = 3$, $N_{NC} = 2$ (**Figure 4.18**)

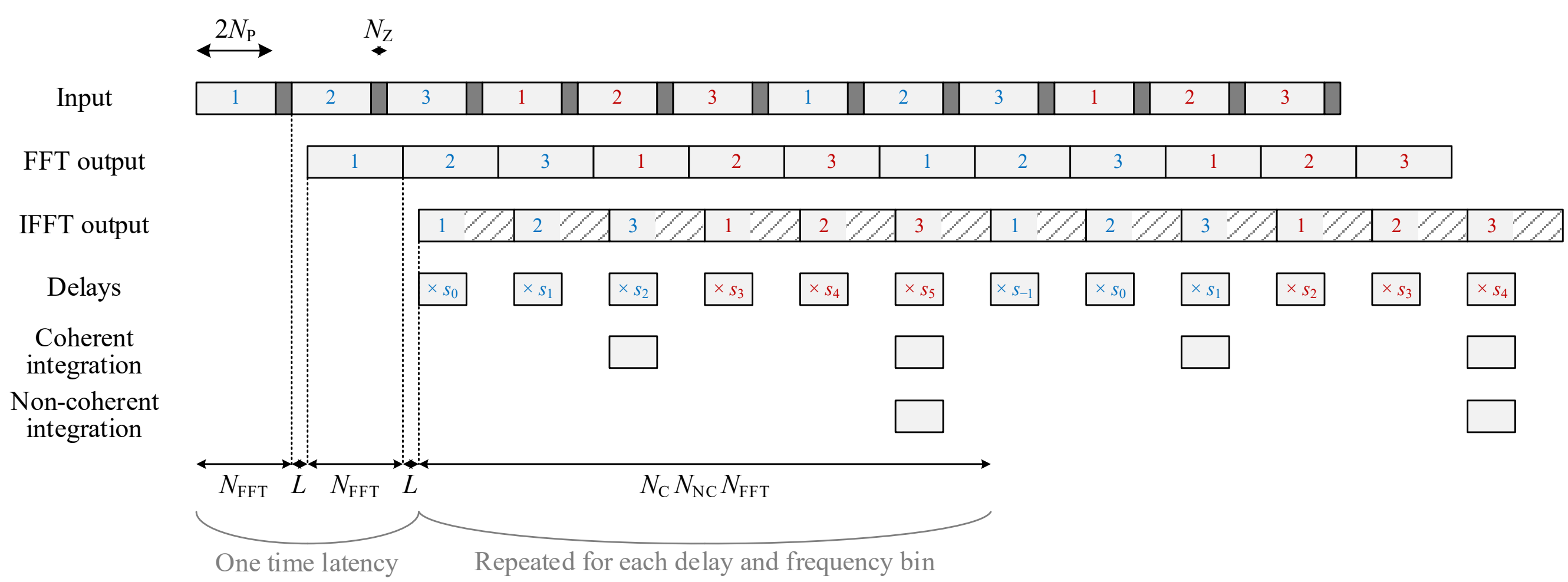

**Figure 7.10** Hardware timing for serial partial secondary-code correlation, $N_C = 3$, $N_{NC} = 2$ (**Figure 4.20**)

### *7.7.3. Memory requirements*

For partial parallel secondary-code correlation of **Figure 4.18**, the memory requirement is

$$\begin{cases} (N_C + 1)N_P B_{ADC} + N_{FFT,P}B_{FFT,P}/2 + 2M_{FFT,P} + N_S N_P B_C, & \text{if } N_{NC} = 1 \\ (N_{NC}N_C + 1)N_P B_{ADC} + N_{FFT,P}B_{FFT,P}/2 + 2M_{FFT,P} + N_S N_P (B_C + B_{NC}), & \text{if } N_{NC} > 1' \end{cases} \tag{53}$$

and for full parallel secondary-code correlation of **Figure 4.19**,

$$\begin{cases} (kN_S + 1)N_P B_{ADC} + N_{FFT,P}B_{FFT,P}/2 + 2M_{FFT,P} + N_S N_P B_C, & \text{if } N_{NC} = 1 \\ (N_{NC}kN_S + 1)N_P B_{ADC} + N_{FFT,P}B_{FFT,P}/2 + 2M_{FFT,P} + N_S N_P (B_C + B_{NC}), & \text{if } N_{NC} > 1' \end{cases} \tag{54}$$

For the corresponding serial architectures, the use of a single secondary-code delay accumulator reduces the memory requirement to

$$\begin{cases} (N_\mathrm{C}+1)N_\mathrm{P}B_\mathrm{ADC}+N_\mathrm{FFT,P}B_\mathrm{FFT,P}/2+2M_\mathrm{FFT,P}+N_\mathrm{P}B_\mathrm{C}, & \text{if } N_\mathrm{NC}=1 \\ (N_\mathrm{NC}N_\mathrm{C}+1)N_\mathrm{P}B_\mathrm{ADC}+N_\mathrm{FFT,P}B_\mathrm{FFT,P}/2+2M_\mathrm{FFT,P}+N_\mathrm{P}(B_\mathrm{C}+B_\mathrm{NC}), & \text{if } N_\mathrm{NC}>1' \end{cases} \quad (55)$$

for partial correlation in **Figure 4.20** and

$$\begin{cases} (kN_\mathrm{S}+1)N_\mathrm{P}B_\mathrm{ADC}+N_\mathrm{FFT,P}B_\mathrm{FFT,P}/2+2M_\mathrm{FFT,P}+N_\mathrm{P}B_\mathrm{C}, & \text{if } N_\mathrm{NC}=1 \\ (N_\mathrm{NC}kN_\mathrm{S}+1)N_\mathrm{P}B_\mathrm{ADC}+N_\mathrm{FFT,P}B_\mathrm{FFT,P}/2+2M_\mathrm{FFT,P}+N_\mathrm{P}(B_\mathrm{C}+B_\mathrm{NC}), & \text{if } N_\mathrm{NC}>1' \end{cases} \quad (56)$$

for full correlation in **Figure 4.21**.

For the frequency-domain software architectures, the intermediate primary-code correlation vectors and the FFT of the secondary code must additionally be stored. This gives

$$\begin{cases} (N_\mathrm{C}+1)N_\mathrm{P}B_\mathrm{ADC}+N_\mathrm{FFT,P}B_\mathrm{FFT,P}/2+N_\mathrm{C}N_\mathrm{P}B_\mathrm{FFT,P}+N_\mathrm{FFT,S}B_\mathrm{FFT,S}/2, & \text{if } N_\mathrm{NC}=1 \\ (N_\mathrm{NC}N_\mathrm{C}+1)N_\mathrm{P}B_\mathrm{ADC}+N_\mathrm{FFT,P}B_\mathrm{FFT,P}/2+N_\mathrm{C}N_\mathrm{P}B_\mathrm{FFT,P}+N_\mathrm{FFT,S}B_\mathrm{FFT,S}/2+N_\mathrm{S}N_\mathrm{P}B_\mathrm{NC}, & \text{if } N_\mathrm{NC}>1' \end{cases} \quad (57)$$

for partial secondary-code correlation in **Figure 4.22** and

$$\begin{cases} (kN_\mathrm{S}+1)N_\mathrm{P}B_\mathrm{ADC}+N_\mathrm{FFT,P}B_\mathrm{FFT,P}/2+N_\mathrm{S}N_\mathrm{P}B_\mathrm{FFT,P}+N_\mathrm{FFT,S}B_\mathrm{FFT,S}/2, & \text{if } k=1, N_\mathrm{NC}=1 \\ (kN_\mathrm{S}+1)N_\mathrm{P}B_\mathrm{ADC}+N_\mathrm{FFT,P}B_\mathrm{FFT,P}/2+N_\mathrm{S}N_\mathrm{P}B_\mathrm{FFT,P}+N_\mathrm{FFT,S}B_\mathrm{FFT,S}/2+N_\mathrm{S}N_\mathrm{P}B_\mathrm{C}, & \text{if } k>1, N_\mathrm{NC}=1 \\ (N_\mathrm{NC}kN_\mathrm{S}+1)N_\mathrm{P}B_\mathrm{ADC}+N_\mathrm{FFT,P}B_\mathrm{FFT,P}/2+N_\mathrm{S}N_\mathrm{P}B_\mathrm{FFT,P}+N_\mathrm{FFT,S}B_\mathrm{FFT,S}/2+N_\mathrm{S}N_\mathrm{P}B_\mathrm{NC}, & \text{if } k=1, N_\mathrm{NC}>1 \\ (N_\mathrm{NC}kN_\mathrm{S}+1)N_\mathrm{P}B_\mathrm{ADC}+N_\mathrm{FFT,P}B_\mathrm{FFT,P}/2+N_\mathrm{S}N_\mathrm{P}B_\mathrm{FFT,P}+N_\mathrm{FFT,S}B_\mathrm{FFT,S}/2+N_\mathrm{S}N_\mathrm{P}(B_\mathrm{C}+B_\mathrm{NC}), & \text{if } k>1, N_\mathrm{NC}>1 \end{cases} \quad (58)$$

for full secondary-code correlation in **Figure 4.23**.